\documentclass[aps,reprint,twocolumn,showpacs,preprintnumbers,superscriptaddress,amsmath,amssymb]{revtex4-2}
\usepackage{booktabs} 
\usepackage{graphicx}
\usepackage{siunitx}
\usepackage{xcolor}
\usepackage{amsmath,amssymb}
\usepackage[T1]{fontenc}
\usepackage{lmodern}
\usepackage[utf8]{inputenc}
\usepackage[english]{babel}
\usepackage{natbib}
\usepackage{hyperref}
\hypersetup{colorlinks=true,citecolor={blue},linkcolor={blue},urlcolor={blue}}
\usepackage{ulem}

\usepackage{amsfonts}
\usepackage{bm}
\usepackage{mathtools}
\usepackage{graphicx}
\usepackage{pst-poly,multido}
\usepackage{hyperref}

\newcommand{\hsn}[1]{{\color{black} #1}}
\newcommand{\highlight}[1]{{\color{black} #1}}

\renewcommand{\vec}[1]{\mathbf{#1}}

\DeclarePairedDelimiter\abs{\lvert}{\rvert}
\DeclarePairedDelimiter\bra{\langle}{\rvert}
\DeclarePairedDelimiter\ket{\lvert}{\rangle}
\DeclarePairedDelimiterX\braket[2]{\langle}{\rangle}{#1 \delimsize\vert #2}

\begin{document}

\title{Generalized Non-Hermitian Hamiltonian for Guided Resonances in Photonic Crystal Slabs}

\author{Viet Anh Nguyen}
\affiliation{Center for Environmental Intelligence, College of Engineering and Computer Science, VinUniversity, Gia Lam district, Hanoi 14000, Vietnam}
\author{Hung Son Nguyen}
\affiliation{Center for Environmental Intelligence, College of Engineering and Computer Science, VinUniversity, Gia Lam district, Hanoi 14000, Vietnam}

\author{Zhiyi Yuan}
\affiliation{Centre for OptoElectronics and Biophotonics (COEB), School of Electrical and Electronic Engineering, Nanyang Technological University, Singapore 639798}
\affiliation{ CNRS-International-NTU-Thales ResearchAlliance (CINTRA), IRL 3288, Singapore 637553}
\affiliation{Institute of Materials Research and Engineering, Agency for Science Technology and Research (A*STAR), 2 Fusionopolis Way, Singapore 138634}

\author{Dung Xuan Nguyen}
\affiliation{Center for Theoretical Physics of Complex Systems, Institute for Basic Science (IBS), Daejon, 34126, Republic of Korea}

\author{Cuong Dang}
\affiliation{Centre for OptoElectronics and Biophotonics (COEB), School of Electrical and Electronic Engineering, Nanyang Technological University, Singapore 639798}
\affiliation{ CNRS-International-NTU-Thales ResearchAlliance (CINTRA), IRL 3288, Singapore 637553}

\author{Son Tung Ha}
\affiliation{Institute of Materials Research and Engineering, Agency for Science Technology and Research (A*STAR), 2 Fusionopolis Way, Singapore 138634}

\author{Xavier Letartre}
\affiliation{Ecole Centrale de Lyon, CNRS, INSA Lyon, Université Claude Bernard Lyon 1, CPE Lyon, CNRS, INL, UMR5270, Ecully 69130, France}
\author{Quynh Le-Van}
\affiliation{Center for Environmental Intelligence, College of Engineering and Computer Science, VinUniversity, Gia Lam district, Hanoi 14000, Vietnam}
\author{Hai Son Nguyen}
\email{hai-son.nguyen@ec-lyon.fr}
\affiliation{Ecole Centrale de Lyon, CNRS, INSA Lyon, Université Claude Bernard Lyon 1, CPE Lyon, CNRS, INL, UMR5270, Ecully 69130, France}
\affiliation{Institut Universitaire de France (IUF), 75231 Paris, France}

\date{\today}

\begin{abstract}
\highlight{We develop a generalized non-Hermitian Hamiltonian formalism for guided resonances in photonic crystal slabs, derived directly from Maxwell’s equations} through a systematic guided-mode expansion. By expanding the electromagnetic fields over the complete mode basis of an unpatterned slab and systematically integrating out radiative Fabry--Pérot channels, we obtain the analytical operator structure of the Hamiltonian, which treats guided-mode coupling and radiation losses on equal footing. \hsn{The resulting Hamiltonian provides explicit expressions for both dispersive and radiative coupling terms in terms of modal overlap integrals and Fourier components of the permittivity modulation. For specific geometries, the Hamiltonian coefficients can be extracted from full-wave simulations}, enabling accurate modeling without phenomenological assumptions. As a case study, \highlight{we investigate hexagonal lattices with both preserved and broken $C_6$ symmetry, demonstrating predictive agreement for complex band structures, near-field distributions, and far-field polarization patterns}. In particular, the formalism \highlight{reproduces symmetry-protected bound states in the continuum (BICs) at the $\Gamma$ point, accidental off-$\Gamma$ BICs near the $\Gamma$ point, and the emergence of chiral exceptional points (EPs)}. It also captures the tunable behavior of eigenmodes near the $K$ point, including \highlight{Dirac-point shifts and the emergence of quasi-BICs or bandgap openings, depending on the nature of $C_6$ symmetry breaking}. We further demonstrate in the Appendix that \highlight{the same formalism extends naturally to other symmetry classes, including $C_2$ (1D grating) and $C_4$ (square lattice) photonic crystal slabs.}This approach enables predictive and efficient modeling of complex photonic resonances, revealing their topological and symmetry-protected characteristics in non-Hermitian systems.
\end{abstract}

\maketitle
\section{Introduction}

Understanding and engineering the resonant modes of photonic crystal (PhC) slabs~\cite{Johnson1999,Imada2002,Sakoda2001}—and more broadly, non-local metasurfaces composed of periodic subwavelength lattice elements~\cite{Malek2020,Overvig2022,Chen2025,Monticone2025}—is a central theme in modern nanophotonics, underpinning key applications in lasers, filters, sensors, and quantum optics. These systems exhibit rich physics due to their intrinsic non-Hermiticity, which arises from radiation leakage into the continuum and leads to complex-valued eigenfrequencies. \hsn{One of the most intriguing phenomena associated with non-Hermitian photonic systems is the emergence of bound states in the continuum (BICs) ~\cite{Hsu2016,Doeleman2018, Kang2023,MermetLyaudoz2023,Le2024}, which can be broadly categorized into two types: symmetry-protected BICs and accidental BICs. Symmetry-protected BICs occur at high-symmetry points in momentum space, where the confined electromagnetic modes are forbidden from coupling to outgoing radiation channels due to a symmetry mismatch. In contrast, accidental BICs arise away from these high-symmetry points as a result of destructive interference between multiple radiation pathways, and are therefore highly sensitive to geometric parameters of the structure. Another notable non-Hermitian feature is the presence of exceptional points (EPs)~\cite{Miri2019,Zhen2015,Zhou2018,Ferrier2022,Nasari2022,Nguyen2023}, where both eigenvalues and eigenmodes coalesce in momentum space, leading to non-trivial topological and spectral behaviors.} A unified and accurate modeling framework is thus essential for understanding and designing these non-Hermitian resonances in structured photonic media.

Guided resonances in PhC slabs—also referred to as quasi-guided modes—arise from the coupling between guided modes of an unpatterned dielectric slab and radiation continua induced by periodic modulation. These modes are characterized by complex eigenfrequencies encoding both the resonance frequencies and radiative losses. While full-wave numerical solvers (e.g., FEM, FDTD) can directly compute these quantities, they are computationally demanding for large parameter scans and often obscure the underlying physical mechanisms. Conversely, analytical or semi-analytical Hamiltonian approaches—such as temporal coupled-mode theory (TCMT)\cite{Fan2003,WonjooSuh2004} and phenomenological non-Hermitian models\cite{Zhen2015,Zhou2018,Ferrier2022,Letartre2022,Nguyen2023}—offer compact descriptions but typically start from assumed operator structures, which may restrict their applicability to specific designs or symmetry configurations.

A powerful intermediate approach is coupled-mode theory via permittivity perturbation, where the permittivity is written as $\epsilon(\mathbf{r}) = \epsilon_0(\mathbf{r}) + \Delta\epsilon(\mathbf{r})$, with $\epsilon_0(\mathbf{r})$ describing the unperturbed slab and $\Delta\epsilon(\mathbf{r})$ the periodic modulation. This formulation enables the eigenmodes of the full system to be expressed in terms of those of the homogeneous slab, yielding an effective Hamiltonian that captures both mode coupling and radiation leakage—particularly when extended to the complex frequency plane. This strategy dates back to distributed feedback laser theory~\cite{Streifer_1977,Kazarinov_1985} and has been extended to two-dimensional PhC slabs to describe phenomena such as band inversion, symmetry-protected BICs, and topological transitions in non-Hermitian photonics~\cite{Lee_2019,Nasari2023}. Complementary developments by Noda’s group provided effective mode-coupling models for various lattice geometries, offering insight into BIC formation and far-field radiation control~\cite{Liang2011,Liang_2012,Peng2012,Liang_2013,Yang_2014}, consistent with experimental observations from the MIT group~\cite{Hsu_2013}. However, a general and systematic formalism that includes the full mode basis of the slab and treats non-Hermiticity from the outset has remained unexplored.

\begin{figure}
\begin{center}
\includegraphics[width=1\linewidth]{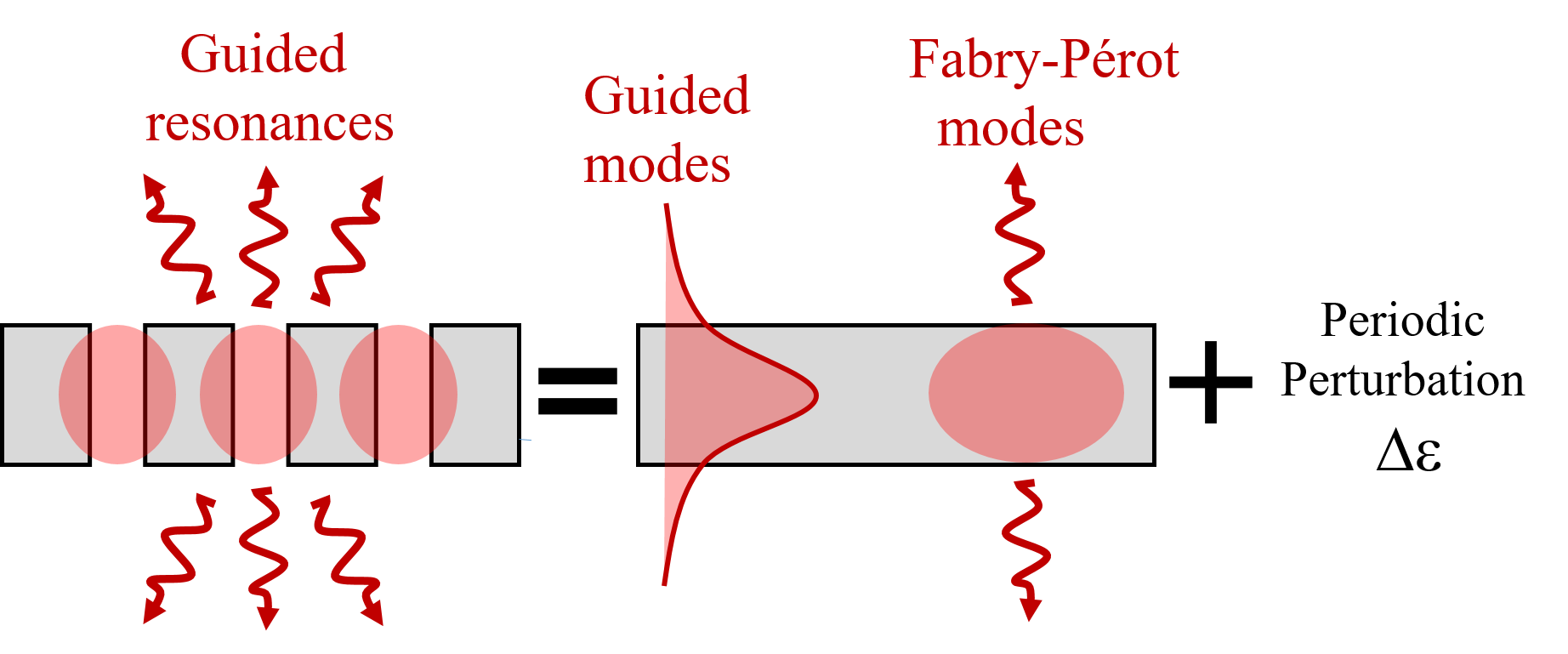}
\caption{\textbf{Generalized guided-mode expansion.} Guided resonances in a PhC slab result from periodic permittivity perturbations coupling guided modes and Fabry--Pérot modes of an unpatterned slab waveguide.}
\label{fig:couplingscheme}
\end{center}
\end{figure}

In this work, we \hsn{derive a generalized non-Hermitian Hamiltonian for guided resonances by expanding the fields onto} the complete set of eigenmodes from the unpatterned slab—including both guided and Fabry--Pérot modes(Fig.~\ref{fig:couplingscheme}). Starting from Maxwell’s equations, we systematically integrate out the Fabry--Pérot components to derive the analytical operator structure of a non-Hermitian Hamiltonian \hsn{—including diffractive coupling terms and radiative self-energy—expressed as overlap integrals between slab eigenmodes and the Fourier components of the permittivity modulation.} Our approach extends the guided-mode expansion method developed by Andreani and collaborators~\cite{Andreani_2006,Zanotti2024}, which constructs a Hermitian Hamiltonian restricted to guided modes and treats radiation perturbatively. In contrast, our method incorporates non-Hermiticity at the core, enabling accurate modeling of leaky modes and their complex interactions across the light cone.
In the numerical examples, the corresponding coupling coefficients are extracted from full-wave simulations to enable direct and quantitative comparison with complex band structures, quality factors, and far-field patterns. This approach avoids phenomenological assumptions about the Hamiltonian form while remaining computationally efficient and physically interpretable. 

To illustrate the efficiency and generality of the formalism, we apply it to hexagonal-lattice PhC slabs with both $C_6$-symmetric and symmetry-broken configurations near the $\Gamma$ and $K$ points of the Brillouin zone. \hsn{Beyond the hexagonal lattices discussed in the main text, we further demonstrate in the Appendix that the same formalism extends naturally to other symmetry classes, including $C_2$ (1D gratings) and $C_4$ (square lattices), confirming its generality across diverse photonic crystal geometries}. The derived non-Hermitian Hamiltonian not only reproduces the complex photonic band structure but also accurately captures near-field and far-field distributions, including polarization textures, in excellent agreement with full-wave finite-element simulations. The model predicts symmetry-protected BICs at the $\Gamma$ point (monopolar, hexapolar, and quadrupolar modes), accidental off-$\Gamma$ BICs from momentum-dependent destructive interference, and the emergence of chiral EP pairs in momentum space. It further describes eigenmode behavior near the $K$ point under $C_6$ symmetry breaking: in triangular lattices with broken $C_3$ symmetry, the Dirac point shifts away from $K$ while preserving degeneracy, with quasi-BIC features in the lower band; in honeycomb lattices with broken inversion symmetry, the Dirac degeneracy is lifted and a bandgap opens, with all modes becoming radiative. These results establish the generalized guided-mode expansion as a predictive and versatile framework for analyzing complex resonant phenomena and symmetry-governed radiation properties in non-Hermitian PhC slabs.

\section{Theoretical framework}

\subsection{Eigenmodes of PhC slabs}
PhC slabs are finite structures where light is confined vertically ($z$-direction) within a slab that consists of multilayers of high refractive index. The slab is periodically corrugated in the $x$-$y$ plane, and the vertical confinement is achieved by the refractive index contrast between the slab material and its surrounding lower refractive index environment (e.g., air and substrate). The in-plane periodicity gives rise to photonic band structures, while the vertical confinement allows the PhC slab to support fully guided modes that are confined by total internal reflection, useful for integrated photonic on-chip applications, and leaky modes (guided resonances) that interact with the radiation field in free space and are critical for applications such as light-emitting devices, sensors, or detectors. Guided resonances are specific to PhC slabs and introduce non-Hermitian physics. In this work, we focus solely on PhC slabs and sometimes refer to them as PhCs for brevity.\\  

The behavior of electromagnetic waves in PhC slabs is governed by Maxwell's equations. Here, for the sake of simplicity, we assume the materials are isotropic, lossless, non-magnetic, and non-dispersive, simplifying the permeability \(\mu=1\) and the permittivity to spatially varying constants \(\epsilon(\mathbf{r})\). For time-harmonic fields \(\mathbf{E}(\mathbf{r}, t) = \mathbf{E}(\mathbf{r}) e^{-i\omega t}\), the master equation for the electric field is given by:
\begin{equation}\label{eq:masterEq}
\nabla \times \left(\nabla \times \mathbf{E} \right) = \frac{\omega^2}{c^2} \epsilon(\mathbf{r}) \mathbf{E}.
\end{equation}
This eigenvalue equation describes the frequency \(\omega\) of electromagnetic waves as a function of the spatial variation of \(\epsilon(\mathbf{r})\), laying the foundation for the photonic band structure in PhCs. It is important to note that this equation is not a standard eigenvalue equation but a generalized one, requiring specific formulations for the inner product and orthogonality, which involve weighting by the permittivity \(\epsilon(\mathbf{r})\). Here the permittivity exhibits in-plane periodicity, satisfying $\epsilon(\mathbf{r})= \epsilon(\mathbf{r}+\mathbf{R})$ with \(\mathbf{R}\) being a lattice vector in the $xy$ plane. Since the periodicity exists in the in-plane directions ($x, y$) but not along $z$, the eigen Bloch modes are given by:
\begin{equation}
\mathbf{E}(\mathbf{r}) = \mathbf{u}_{\mathbf{k}\parallel}(\mathbf{r}) e^{i\mathbf{k}_\parallel \cdot \mathbf{r}_\parallel},
\end{equation}

where \(\mathbf{k}_\parallel = (k_x, k_y)\) is the in-plane wavevector, \(\mathbf{r}_\parallel= (x, y)\) are the in-plane spatial coordinates, and  \(\mathbf{u}_{\mathbf{k}_\parallel}(\mathbf{r})\) is a periodic function with the same periodicity as \(\epsilon(\mathbf{r})\). The eigenvalue problem yields the photonic band structure, \(\omega(\mathbf{k}_\parallel)\), which maps the allowed frequencies for each wavevector \(\mathbf{k}_\parallel\) within the first Brillouin zone.\\

The emergence of complex eigenvalues for guided resonances arises from their interaction with the radiative continuum. This coupling to the radiative continuum is a hallmark of non-Hermitian physics in PhC slabs. Mathematically, the eigenvalue problem for guided resonances can be written as $\hat{H} \mathbf{E} = \omega \mathbf{E}$, where the Hamiltonian $\hat{H}$ is non-Hermitian due to the inclusion of radiative losses. A non-Hermitian Hamiltonian is characterized by \(\hat{H} \neq \hat{H}^\dagger\), meaning it is not equal to its adjoint. This property introduces complex eigenvalues, where the imaginary part typically represents energy gain or loss in the system. In PhC slabs, the non-Hermitian nature of \(\hat{H}\) stems from the open-system interaction between guided resonances and the radiative continuum, which allows energy leakage. This is a direct departure from Hermitian physics, where systems are typically closed and do not interact with an external environment. The non-Hermitian framework enables the study of unique phenomena such as EPs and BICs, which are absent in purely Hermitian systems.\\

\subsection{Effective Hamiltonian from $\Delta\epsilon$ perturbation}
Before discussing the Hamiltonian of generalized guided modes expansion, we first introduce the common formalism of the coupled mode theory via $\Delta\epsilon$ perturbation.  The unperturbed system, characterized by \(\epsilon_0(\mathbf{r})\), has known eigenvalues \(\omega_n\) and eigenmodes \(\mathbf{E}_n(\mathbf{r})\): $\nabla \times \nabla \times \mathbf{E}_n = \frac{\omega_n^2}{c^2} \epsilon_0(\mathbf{r}) \mathbf{E}_n.$ Introducing \(\Delta \epsilon(\mathbf{r})\) perturbs these eigenmodes and eigenvalues, allowing us to write: $\mathbf{E}(\mathbf{r}) = \sum_n c_n \mathbf{E}_n(\mathbf{r}),$ where \(c_n\) are expansion coefficients that account for the perturbation. Substituting \(\mathbf{E}(\mathbf{r})\) into the perturbed master equation $\nabla \times \nabla \times \mathbf{E} = \frac{\omega^2}{c^2} \left[ \epsilon_0(\mathbf{r}) + \Delta \epsilon(\mathbf{r}) \right] \mathbf{E}$ and projecting onto the unperturbed basis modes \(\{\mathbf{E}_n(\mathbf{r})\}\) in the approximation $\omega\approx\omega_0$ to achieve an eigenvalue problem of $\omega$ instead of $\omega^2$, we obtain:
\begin{equation}
H_{\text{eff}} \mathbf{C} = \omega \mathbf{C},
\end{equation}
where \(\mathbf{C} = \{c_n\}\) is the vector of expansion coefficients. Here, the effective Hamiltonian \(H_{\text{eff}}\) includes the perturbative contributions from \(\Delta \epsilon(\mathbf{r})\):
\begin{equation}
[H_{\text{eff}}]_{nm} = \omega_n \delta_{nm} + \Delta H_{nm},
\end{equation}
with:
\begin{equation}
\Delta H_{nm} = \frac{\omega_0}{2} \int \mathbf{E}_n^* \cdot \Delta \epsilon(\mathbf{r}) \mathbf{E}_m \, d\mathbf{r}.
\end{equation}

\subsection{Field Expansion in the Generalized Guided mode Expansion}\label{sec:Field_Expansion}

In the generalized guided mode expansion, the unperturbed system corresponds to the unpatterned slab with a permittivity profile \(\epsilon_0(z)\), representing the zeroth Fourier component of \(\epsilon(\mathbf{r})\). The perturbation arises from the periodic modulation in the \(x\)-\(y\) plane, expressed as:
\begin{equation}
\Delta \epsilon(\mathbf{r}) = \sum_{\mathbf{G}} \epsilon_{\mathbf{G}}(z) e^{i\mathbf{G} \cdot \mathbf{r}_\parallel},
\end{equation}\label{eq:epsilon}
where \(\mathbf{G}\) denotes the reciprocal lattice vectors, and \(\mathbf{r}_\parallel = (x, y)\) represents the in-plane coordinates.

The eigenmodes of the unpatterned slab \(\epsilon_0(z)\) include the guided modes of the planar waveguide and the radiative Fabry-Pérot (FP) modes. In this work, we focus on the low-energy regime, where the wave vectors $\mathbf{G}n + \mathbf{k}\parallel$ primarily describe guided modes with evanescent out-of-plane components, while the wave vectors $\mathbf{k}_\parallel$ correspond to modes inside the light cone—i.e., radiative FP modes. The total electric field in the PhC slab can thus be expanded as:
\begin{equation}\label{eq:expansion_general_GME}
\mathbf{E} = \sum_n a_n \mathbf{E}_{n} + \sum_m b_m \mathbf{E}_{m,\text{FP}},
\end{equation}
where \(\mathbf{E}_{n} = \mathbf{u}_{n}(z) e^{i(\mathbf{G}_n + \mathbf{k}_\parallel) \cdot \mathbf{r}_\parallel}\) are the guided modes confined within the slab, having corresponding eigenvalues $\omega_{n}$. And \(\mathbf{E}_{m,\text{FP}} = \mathbf{u}_{m,\text{rad}}(z) e^{i\mathbf{k}_\parallel \cdot \mathbf{r}_\parallel}\) are the radiative FP modes propagating outside the slab, having corresponding eigenvalues $\tilde{\omega}_{m,\text{FP}} = \omega_{m,\text{FP}} - i\gamma_{m,\text{FP}}$. We note that in this study, we restrict ourselves to the low-energy regime, where only the zeroth-order FP modes contribute to radiation. However, the formalism can be naturally extended to higher-energy regimes by including FP modes associated with higher-order diffraction channels. We also focus exclusively on transverse electric (TE) guided modes. The same formalism can be applied to transverse magnetic (TM) guided modes by formulating the perturbation theory in terms of the magnetic field \(\mathbf{H}\), rather than the electric field \(\mathbf{E}\) as done here. We also neglect TE–TM coupling, which is justified in our system due to the strong spectral separation between TE and TM modes in high-contrast dielectric slabs.

It is important to note that despite the radiative Fabry-Pérot modes being quasinormal modes—characterized by complex eigenfrequencies due to their leaky nature, they can form a complete and orthogonal basis under specific conditions. The completeness and orthogonality of quasinormal modes in leaky optical cavities were rigorously established by Leung, Liu, and Young in 1994 \cite{Leung1994}. Their work demonstrated that, despite the non-Hermitian nature of such systems, a discrete set of quasinormal modes suffices to accurately describe the field distribution within the cavity. Therefore, the guided modes (\(\mathbf{E}_{n}\)) and radiative Fabry-Pérot modes (\(\mathbf{E}_{m,\text{FP}}\)) together provide a complete and orthogonal basis that ensures the expansion from Eq.~\eqref{eq:expansion_general_GME} is both accurate and comprehensive.\\

The perturbation \(\Delta \epsilon(\mathbf{r})\) introduces coupling between these guided and radiative modes. The coupling matrix elements are:

\begin{equation}\label{eq:coupling_element_GGME}
    \begin{split}
        \Delta H_{nm}^{\text{guided}} &= \frac{\omega_0}{2} \int \mathbf{E}_{n}^* \cdot \Delta \epsilon(\mathbf{r}) \mathbf{E}_{m} \, d\mathbf{r}  \quad \\ & = \frac{\omega_0}{2} \int \mathbf{u}_{n}^*(z) \cdot \epsilon_{\mathbf{G}_n - \mathbf{G}_m}(z) \mathbf{u}_{m}(z) \, dz,\\
        \Delta H_{nm}^{\text{guided-FP}} &= \frac{\omega_0}{2} \int \mathbf{E}_{n}^* \cdot \Delta \epsilon(\mathbf{r}) \mathbf{E}_{m,\text{FP}} \, d\mathbf{r}  \quad \\&=  \frac{\omega_0}{2} \int \mathbf{u}_{n}^*(z) \cdot \epsilon_{-\mathbf{G}_n}(z) \mathbf{u}_{\text{m,rad}}(z) \, dz,\\
        \Delta H_{nm}^{\text{FP-guided}} &= \frac{\omega_0}{2} \int \mathbf{E}_{n,\text{FP}}^* \cdot \Delta \epsilon(\mathbf{r}) \mathbf{E}_{m} \, d\mathbf{r} \quad \\&=  \frac{\omega_0}{2} \int \mathbf{u}_{\text{n,rad}}^*(z) \cdot \epsilon_{\mathbf{G}_m}(z) \mathbf{u}_{m}(z) \, dz.
    \end{split}
\end{equation}

These coupling terms correspond to three distinctive mechanisms:
\begin{itemize}
    \item \textbf{Guided-to-Guided Coupling}: The interaction between two guided modes is mediated by the Fourier component \(\Delta \epsilon_{\mathbf{G}_n - \mathbf{G}_m}(z)\), which matches their momentum difference. This term is crucial for photonic band structure modifications, such as band gaps.
    \item \textbf{Guided-to-FP Coupling}: The coupling of a guided mode to a radiative (Fabry–Pérot) mode depends on the Fourier component \(\Delta \epsilon_{-\mathbf{G}_n}(z)\). This term introduces radiative losses to guided modes, converting them into quasi-normal modes.
    \item \textbf{FP-to-Guided Coupling}: Radiative modes contribute to guided modes through the \(\Delta \epsilon_{\mathbf{G}_m}(z)\) term. This term describes how energy can leak from FP modes back into guided modes. This is the reversed mechanism of the Guided-to-FP Coupling, evidenced by $\Delta H_{nm}^{\text{guided-FP}} =\left(\Delta H_{nm}^{\text{FP-guided}}\right)^*$.
\end{itemize}
Note that as we restrict ourselves to the low-energy regime, where only the zeroth-order FP is involved, there is no FP-to-FP coupling.

\subsection{Effective Hamiltonian for Guided Modes}

The total Hamiltonian for the coupled system of guided and radiative modes can be written as:
\begin{equation}
    H =
    \begin{bmatrix}
    H_{\text{guided}} & H_{\text{guided-FP}} \\
    H_{\text{FP-guided}} & H_{\text{FP}}
    \end{bmatrix},
\end{equation}
which satisfies the eigenvalue problem:
\begin{equation}\label{eq:AB}
    H
    \begin{bmatrix}
    \mathbf{A} \\
    \mathbf{B}
    \end{bmatrix}
    =
    \omega
    \begin{bmatrix}
    \mathbf{A} \\
    \mathbf{B}
    \end{bmatrix},
\end{equation}
where \(\mathbf{A} = \{a_n\}\) and \(\mathbf{B} = \{b_m\}\) are the expansion coefficients of the total electric field \(\mathbf{E}\) in Eq.~\eqref{eq:expansion_general_GME}. From the second row of Eq.~\eqref{eq:AB}, we obtain a relation between the coefficients of the radiative modes and those of the guided modes $\mathbf{B} = \left( \omega - H_{\text{FP}} \right)^{-1} H_{\text{FP-guided}} \mathbf{A}$,
where \(\omega - H_{\text{FP}}\) is a diagonal matrix with entries \(\Delta_n + i\gamma_{n,\text{rad}}\), and \(\Delta_n = \omega - \omega_{n,\text{rad}}\) denotes the detuning from the radiative FP mode frequencies.

Substituting this relation into the first row of Eq.~\eqref{eq:AB} yields an effective operator acting on the guided-mode amplitudes:
\begin{equation}
    \hat{H} = H_{\text{guided}} + \Sigma(\omega),
\end{equation}
where \(\Sigma(\omega)\) is the self-energy term, given by $\Sigma(\omega) = H_{\text{guided-FP}} \, (\omega - H_{\text{FP}})^{-1} \, H_{\text{FP-guided}}$. This effective operator satisfies
\begin{equation}
    \hat{H} \mathbf{A} = \omega \mathbf{A},
\end{equation}
but the equation remains nonlinear due to the explicit \(\omega\)-dependence of the self-energy term \(\Sigma(\omega)\). Physically, this term encapsulates the modification of the guided-mode dynamics via coupling to the radiative Fabry–Pérot modes. \hsn{The imaginary part of \(\Sigma(\omega)\) captures the irreversible coupling to radiation and gives rise to the non-Hermitian character of the guided resonances. Its real part produces only a small dispersive frequency shift, which can be fully absorbed into the guided-mode Hamiltonian through the renormalization
$H_{\mathrm{guided}} \;\rightarrow\; H_{\mathrm{guided}} + \mathrm{Re}\,\Sigma$,
thereby shifting only the diagonal terms of $H_{\mathrm{guided}}$ without altering modal profiles, radiation rates, or topological properties. This procedure is standard in open electromagnetic and quantum systems—directly analogous to Lamb-shift corrections in QED, real-part self-energies in Green’s-function theory, resonance-frequency shifts in temporal coupled-mode theory, and on-site energy shifts in tight-binding models.

In the limit where the FP modes form a broadband and rapidly decaying reservoir (e.g., for thin or weakly confined slabs), the imaginary part of $\Sigma$ dominates over the real part. However, even when this strict quasi-continuum condition is not satisfied,  in practice, the diagonal parameters of $H_{\mathrm{guided}}$ in numerical examples are extracted from full-wave simulations, so any dispersive contribution from $\mathrm{Re}\,\Sigma$ is naturally absorbed into these extracted coefficients. This makes the approximation robust.  In the concrete examples presented in Sec.3, this renormalization simply corresponds to a small adjustment of the  guided-mode resonance frequencies $\omega_{\Gamma}$ and $\omega_{K}$. Retaining only the imaginary part of the self-energy therefore yields a compact closed-form} non-Hermitian effective Hamiltonian acting on the guided-mode subspace:
\begin{equation}
    \hat{H} = H_{\text{guided}} 
    \underbrace{-i\,H_{\text{guided-FP}}\,\text{Im}\left(H_{\text{FP}}^{-1}\right) H_{\text{FP-guided}}}_{H_{\text{rad}}}. \label{eq:Hamiltonian_construction}
\end{equation}

This is the non-Hermitian effective Hamiltonian we set out to derive. It captures both the dispersive properties of the guided modes and their radiative losses through coupling to the continuum. Importantly, the approximation eliminates the $\omega$-dependence in the self-energy, so that $\hat{H}$ now defines a standard linear eigenvalue problem. 

Moreover, within the quasi-continuum regime of FP modes, the radiation amplitudes \(\mathbf{B}\)—corresponding to the far-field leakage—can be directly computed from the eigenvector \(\mathbf{A}\) of \(\hat{H}\) using:
\begin{equation}\label{eq:Bapprox}
    \mathbf{B} \approx -i\,\text{Im}\left(H_{\text{FP}}^{-1}\right) H_{\text{FP-guided}} \mathbf{A}.
\end{equation}

\subsection{Compact expression and physical meaning of the non-Hermitian Hamiltonian}\label{sec:compact_expression}

The expressions of the coupling terms of $\hat{H}$, given in Eq.~\eqref{eq:coupling_element_GGME}, are simplified by summing out the polarization cross products (see Appendix ~\ref{sec:sumingout}). We now discuss the compact expression and physical meaning of each of these terms. 
\begin{itemize}

\item The first term $H_{\text{guided}}$ represents the unperturbed guided modes and the diffractive couplings between them. The diagonal elements correspond to the frequencies of guided modes in the unpatterned slab: 
\begin{equation}
H^{\text{guided}}_{nn}=\omega_{n}.\label{eq:guided_nn}
\end{equation}
The off-diagonal elements describe direct coupling between guided modes via diffractive mechanism:
\begin{equation}
   H^\text{guided}_{nm} =\vec{p}_n\cdot\vec{p}_m U_{nm}, \label{eq:diff_nm}
\end{equation}
    where $\vec{p}_n$, $\vec{p}_m$ are the polarization vector of the guided mode $\mathbf{E}_n$,$\mathbf{E}_m$ respectively; and the coupling strength $U_{nm}$ is given by:
    \begin{equation}
         U_{nm}=\frac{\omega_0}{2}\int u_{n}^* \cdot \epsilon_{\mathbf{G}_n - \mathbf{G}_m} \cdot u_{m}\, dz.\label{eq:U_nm}
    \end{equation}

\item The second term $\hat{H}_\text{rad}$ represents the radiative losses and radiative coupling of the guided modes. The diagonal elements introduce imaginary components in the eigenfrequencies of the guided modes due to coupling with radiative modes:
    \begin{equation}
    H_{nn}^{\text{rad}} = -i\sum_l \gamma_n^{(l)} \label{eq:rad_nn}
    \end{equation}
    Here $\gamma_n^{(l)}$ represents the radiative losses of the guided mode $n$ via the radiative Faby-Pérot mode $l$, and the expression of $\gamma_n^{(l)}$ is given by: 
    \begin{equation}
         \gamma_n^{(l)}=\frac{\omega_0^2\abs{\int u_{n}^*\cdot \epsilon_{-\mathbf{G}_n}\cdot u_{l,\text{rad}} \, dz}^2 }{4c^4\gamma_{l,\text{rad}}} \label{eq:gamma_l} 
    \end{equation}
    The off-diagonal elements describe indirect coupling between guided modes mediated by radiative modes:
    \begin{equation}
          H_{nm}^{\text{rad}} = -i\vec{p}_n\cdot\vec{p}_m\sum_l \sqrt{\gamma_n^{(l)}\gamma_m^{(l)}}e^{i(\phi_n^{(l)}-\phi_m^{(l)})}. \label{eq:rad_nm}
    \end{equation}
    where coupling phase $\phi_n^{(l)}$ is given by:
    \begin{equation}
         \phi_n^{(l)}=arg\left(\int u_{n}^*\cdot \epsilon_{-\mathbf{G}_n}\cdot u_{l,\text{rad}} \, dz\right)\label{eq:phase_n}
    \end{equation}
\end{itemize}

It is worth emphasizing that the coupling terms expressed in Eqs.\eqref{eq:U_nm} and \eqref{eq:rad_nm}, derived here from first-principles Maxwell equations via the guided-mode expansion, correspond directly to the so-called diffractive coupling and radiative coupling often introduced phenomenologically in the literature on 1D photonic gratings~\cite{Lu2020,Letartre2022} and 2D PhC slabs~\cite{MermetLyaudoz2023,Do2025} . In particular, Eq.\eqref{eq:rad_nm}, which captures the off-diagonal elements of the radiative loss operator, is in excellent agreement with the inter-mode coupling terms found in the radiative Hamiltonian of temporal coupled-mode theory applied to non-orthogonal resonators~\cite{WonjooSuh2004}. This highlights the consistency between the full-wave modal expansion approach and reduced-order models, and provides a rigorous microscopic foundation for the coupling coefficients often assumed in heuristic or fitted models.\\

Furthermore, we highlight that the radiative losses of the guided mode $n$ via the radiative channel $l$ is governed by the overlap integral  $\int u_{n}^*\cdot \epsilon_{-\mathbf{G}_n}\cdot u_{l,\text{rad}} \, dz$, as shown in Eq.~\eqref{eq:gamma_l}. It can be accidentally suppressed if the overlap integral is zero, leading to the suppression of a radiative channel for a given guided mode. In particular, for some particular design, it is possible to render null the radiative losses of every guided modes in the spectral window of interest. In such a configuration, the eigenmode is evidently lossless. This corresponds to an accidental BIC configuration, well documented in the literature, whose emergence is extremely sensitive to the slab thickness, effective index, and requires a vertical symmetry design~\cite{Blanchard2014,Ovcharenko2020,Contractor20}.

\subsection{From eigenmodes to nearfield pattern and farfield pattern of Bloch modes}
 
 The eigenmodes of the system are obtained by diagonalizing the effective non-Hermitian Hamiltonian $\hat{H}$ through the characteristic equation $det(H-\omega)=0$. Due to the non-Hermitian nature of $\hat{H}$, its eigenvalues $\Omega_n$  are generally complex, and the corresponding eigenvectors $\ket{\Omega_n}$ describe the resonant modes of the structure.If the imaginary part of an eigenvalue vanishes, i.e. $\text{Im}(\Omega_n) = 0$, the corresponding eigenmode $\ket{\Omega_n}$ is lossless. Such states are known as BICs. Conversely, if $\text{Im}(\Omega_n) < 0$, the mode exhibits radiative loss and is referred to as a leaky mode. The eigenvector $\mathbf{A}_n=(a_{n_1},a_{n_2},a_{n_3},..)$ associated with $\ket{\Omega_n}$  defines the near-field spatial distribution of the electric field:
 \begin{equation}
  \begin{split}
        \vec{E}_{\ket{\Omega_n}}&=\sum_{m} a_{nm}\vec{E_m}\\
           &=\sum_{m} a_{nm}u_{m}(z) e^{i(\mathbf{G}_m+\vec{k}_\parallel)\cdot \mathbf{r}_\parallel}\vec{p_m} 
  \end{split}\label{eq:NF_E}
 \end{equation}
 A similar superposition principle applies to the magnetic field. Due to the TE nature of the guided modes, the magnetic field is predominantly polarized along the z-direction. The corresponding near-field profile of the magnetic field can thus be expressed as a scalar field:
\begin{equation}
     \vec{H}_{\ket{\Omega_n}}\propto \sum_{m} a_{nm}u_m(z)e^{i\vec{G_m}\vec{r}_\parallel}\hat{\vec{z}} \label{eq:NF_H}
\end{equation}
The far-field radiation pattern is governed by the FP components in the field expansion of Eq.~\eqref{eq:expansion_general_GME}. For a given eigenmode $\ket{\Omega_n}$, the FP coefficients $\mathbf{B}_n = (b_{n_1}, b_{n_2}, b_{n_3}, \dots)$ can be obtained from the eigenvector $\mathbf{A}_n$ using Eq.~\eqref{eq:Bapprox}. The resulting far-field electric field is given by:
 \begin{equation}
   \begin{split}
\vec{E}^{\text{farfield}}_{\ket{\Omega_n}}&=\sum_{m} b_{nm}\vec{E_{m,\text{FP}}}\\
&\propto e^{i(\vec{k}_\parallel\vec{r}_\parallel+k_z.z)}\sum_{m} a_{nm}\alpha_m\vec{p_m} 
  \end{split}\label{eq:FF_E}
 \end{equation}
where $\alpha_m = \sum_l \sqrt{\gamma_m^{(l)}}, e^{i\phi_m^{(l)}}$ accounts for the radiation amplitude and phase into each far-field channel $l$.
We note a direct correspondence between the near-field and far-field expressions in Eqs.~\eqref{eq:NF_E} and \eqref{eq:FF_E}: the far-field radiation is derived from the near-field by replacing the mode profiles $u_m(z) , e^{i\vec{G}m \cdot \vec{r}\parallel}$ with radiating plane wave components $e^{i k_z z}, \vec{p}_m$. This procedure effectively represents the folding of guided Bloch modes into the first Brillouin zone and their subsequent coupling to the radiation continuum.

The polarization texture—including polarization orientation, ellipticity, and topological charge of polarization singularities—can be readily computed from the far-field electric field (see Appendix ~\ref{sec:appendix_polarization}).
 \begin{figure}[!ht]
\begin{center}
\includegraphics[width=1\linewidth]{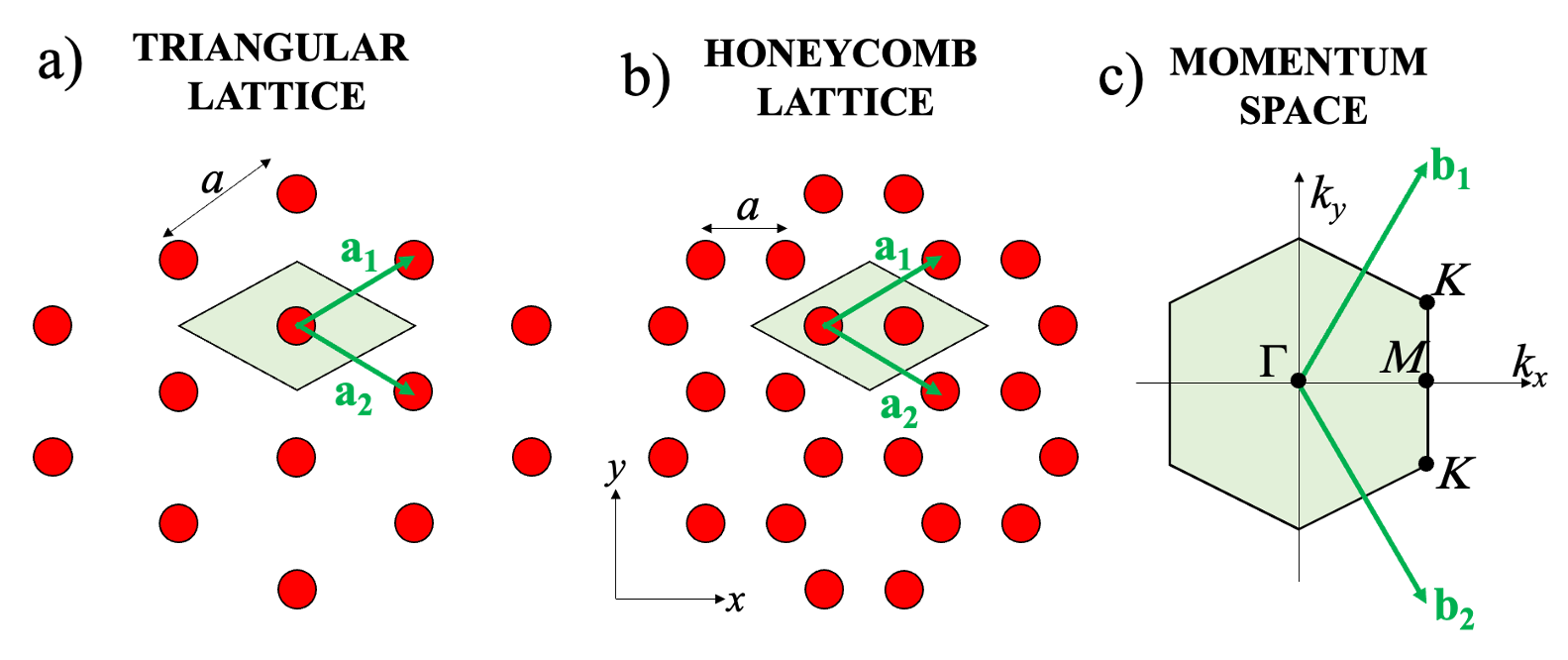}
\caption{\textbf{Geometry of hexagonal lattices}. a) Triangular lattice. b) Honeycomb lattice. c) First Brillouin zone in momentum space.}
\label{fig:honeycomb_lattice}
\end{center}
\end{figure}
 \begin{figure}[!ht]
\begin{center}
\includegraphics[width=1\linewidth]{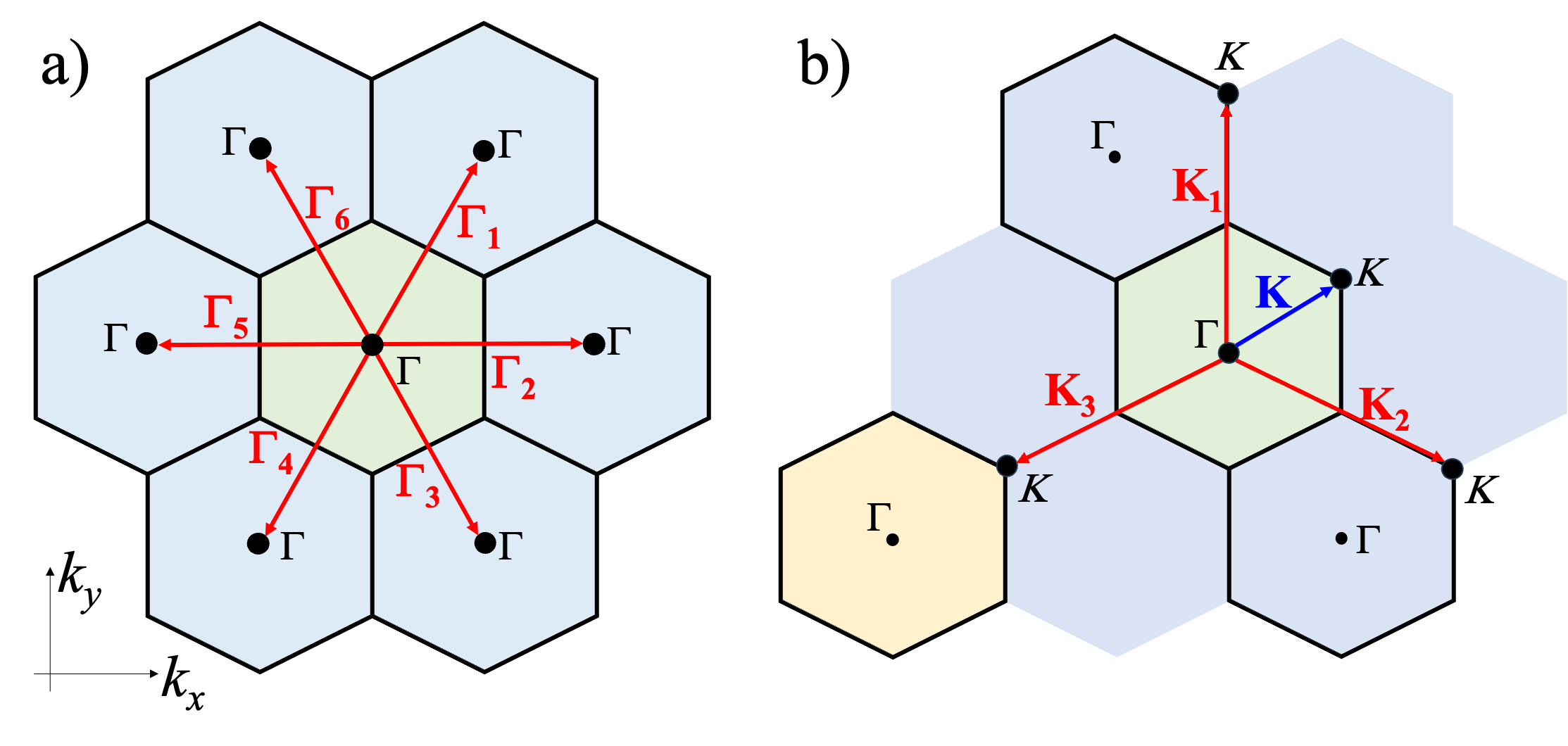}
\caption{\textbf{Guided mode basis.} a) Eigenmodes operating at the $\Gamma$ point and above the light cone are described by six guided modes $\ket{\Gamma_n}$ of wave vector $\vec{\Gamma_n}$, with $n=1\rightarrow 6$. b) Eigenmodes operating at the $K$ point and above the light cone are described by three guided modes $\ket{K_n}$ of wave vector $\vec{K_n}$, with $n=1\rightarrow 3$. The green region indicates the first Brillouin zone, the blue regions indicate the second Brillouin zone, and the yellow region indicates the third Brillouin zone.}
\label{fig:BZs}
\end{center}
\end{figure}
\section{Non-Hermitian Hamiltonian versus Numerical Simulations}
\subsection{Hexagonal lattices with $C_6$ symmetry}
\subsubsection{System description}
We now apply our general non-Hermitian Hamiltonian to described guided resonances in the vicinity of the $\Gamma$ and $K$ point of a PhC slab with hexagonal lattices. \hsn{The case of grating and square lattices are presented in the Appendix \ref{sec:appendix_Applicability}}. The PhC slab of consideration is of a hexagonal lattice with $C_6$ symmetry, for example, a triangular lattice with a single circular hole in the unit cell (see Fig.~\ref{fig:honeycomb_lattice}a), or a honeycomb lattice with two identical circular holes (see Fig.~\ref{fig:honeycomb_lattice}b). The unit vectors are given by $\vec{a_1}=\Lambda\left(\frac{\sqrt{3}}{2}, \frac{1}{2}\right)$ and $ \vec{a_2}=\Lambda\left(\frac{\sqrt{3}}{2},-\frac{1}{2}\right)$ with $\Lambda=a$ for the triangular lattice, and $\Lambda=\sqrt{3}a$ for the honeycomb lattice.

In the momentum space, the high symmetry points are $K$, $K'$, and $M$ at the edge of the Brillouin zone, and $\Gamma$ at the centre of the Brillouin zone (see Fig.~\ref{fig:honeycomb_lattice}c). The corresponding unit vectors in momentum space are given by: $\vec{b_1}=b\left(\frac{1}{2}, \frac{\sqrt{3}}{2}\right),$ and $
\vec{b_2}=b\left(\frac{1}{2}, -\frac{\sqrt{3}}{2}\right)$, with $b=\frac{4\pi}{\sqrt{3}a}$ for the triangular lattice, and $b=\frac{4\pi}{3a}$ for the honeycomb lattice. We note that $|\vec{b_1}|=|\vec{b_2}|= b$ is the distance between the $\Gamma$ point of the first Brillouin zone (BZ) to the six closest $\Gamma$ points of the neighbor (i.e. second) BZs  (see Fig.~\ref{fig:BZs}a).  
\subsection{Guided Mode Basis}
In this work, we only focus on photonic modes above the line cone. In this region, the lowest photonic bands at the $\Gamma$ point is described by the basis consisting of guided modes operating at the six $\Gamma$ points of the second BZs(see Fig.~\ref{fig:BZs}a). These correspond to six guided modes $\ket{\Gamma_n}$, with $n=1\rightarrow 6$, that both originate from the fundamental TE-guided mode of the slab $\epsilon_0(z)$ and have wavevector $\vec{\Gamma_n}$ (see Fig.~\ref{fig:BZs}a).

On the other hand, the lowest photonic bands at the $K$ point is described by the basis consisting of guided modes operating in three $K$ points, two of them are in the second BZ and the third one is in the third BZ (see Fig.~\ref{fig:BZs}b). These correspond to three guided modes $\ket{K_n}$ with $n=1\rightarrow 3$, that both originate from the fundamental TE-guided mode of the slab $\epsilon_0(z)$ and have wavevector $\vec{K_n}$ (see Fig.~\ref{fig:BZs}b).  In a general way,  $\vec{\Gamma_n}=n_1\vec{b_1} + n_2\vec{b_2}$ and $\vec{K_n}=n_1\vec{b_1} + n_2\vec{b_2} + \vec{K} $ with $\vec{K}=b\left(\frac{1}{2}, \frac{\sqrt{3}}{4}\right)$ is the position of the $K$ point in the first BZ. The couples $(n_1,n_2)$ are given by: 
\begin{equation}
    (n_1,n_2)|_{\Gamma}=\begin{cases}(1,0)\\(1,1)\\(0,1)\\(-1,0)\\(-1,-1)\\(0,-1)\end{cases} \text{and}\,\,\,
    (n_1,n_2)|_{K}=\begin{cases}(0,-1)\\(0,1)\\(-2,-1)\end{cases}\label{eq:couples_nm}
\end{equation}
for $\ket{\Gamma_1}\rightarrow\ket{\Gamma_6}$ and $\ket{K_1}\rightarrow\ket{K_3}$ respectively. We note that $\abs{\vec{\Gamma_n}}=b$ and $\abs{\vec{K_n}}=\frac{2}{\sqrt{3}}b$. Explicit expressions of $\vec{\Gamma_n}$ and $\vec{K_n}$ are provided in the Appendix ~\ref{appendix:wavevector}.\\

The six guided modes $\ket{\Gamma_n}$ have the same energy $\omega_\Gamma$ at the $\Gamma$ point, and the three guided modes $\ket{K_n}$ have the same energy $\omega_K$ at the $K$ point. Adding an in-plane momentum $\vec{k}=(k\cos\varphi,k\sin\varphi)$ with $|\vec{k}|\ll b$ , the propagation vector of $\ket{\Gamma_n}$ and $\ket{K_n}$ becomes $\vec{\Gamma_n} + \vec{k}$ and $\vec{K_n} + \vec{k}$ respectively. Thus the energy  of $\ket{\Gamma_n}$ and  $\ket{K_n}$ at $\vec{k}$ are given by $\omega_{\ket{\Gamma_n}}(\vec{k}) \approx \omega_\Gamma+ v_\Gamma\frac{\vec{\Gamma_n}\cdot\vec{k}}{\abs{\vec{\Gamma_n}}}$ and $
    \omega_{\ket{K_n}}(\vec{k}) \approx \omega_K + v_K\frac{\vec{K_n}\cdot\vec{k}}{\abs{\vec{K_n}}}$ where $v_\Gamma$ and $v_K$ are the group velocities of the guided mode at $\Gamma$ and $K$ respectively. Explicit expressions of $\omega_{\ket{\Gamma_n}}(\vec{k})$ and $\omega_{\ket{K_n}}(\vec{k})$ are provided in the Appendix ~\ref{appendix:energy}. Due to the TE nature, our guided modes exhibit in-plane polarization that are perpendicular to their wavevector. Thus $\vec{p_{\ket{\Gamma_n}}}\cdot\left(\vec{\Gamma_n}+\vec{k}\right)=\vec{p_{\ket{K_n}}}\cdot\left(\vec{K_n}+\vec{k}\right)=0$. In the limit of $k\ll b$, the previous condition is approximately reduced to $\vec{p_{\ket{\Gamma_n}}}\cdot\vec{\Gamma_n}=\vec{p_{\ket{K_n}}}\cdot\vec{K_n}=0$. Thus one may easily obtain $\vec{p_{\ket{\Gamma_n}}}$ and $\vec{p_{\ket{K_n}}}$ (see Appendix.~\ref{appendix:polar}).

Finally, since $\ket{\Gamma_n}$ belongs to the same planar waveguide mode, they exhibit the same vertical confinement profile $u_\Gamma(z)$. The same, all three $\ket{K_n}$ exhibit the same vertical confinement profile $u_K(z)$. We note that each guided mode is associated with a pair $(n_1,n_2)$ that corresponds to a Bloch vector $\vec{G_n}=n_1\vec{b_1}+n_2\vec{b_2}$. Therefore, the electric field of our two basis are given by:
\begin{equation}
    \begin{split}
    \vec{E_{\ket{\Gamma_n}}}&=u_\Gamma(z)e^{i(\vec{G_n}+\vec{k}_\parallel)\cdot\vec{r}_\parallel}\vec{p_{\ket{G_n}}}\\
        \vec{E_{\ket{K_n}}}&=u_K(z)e^{i(\vec{G_n}+\vec{k}_\parallel)\cdot\vec{r}_\parallel}\vec{p_{\ket{K_n}}}
    \end{split}
\end{equation}
Here the momentum $\vec{k}_\parallel$ within the first BZ is given by $\vec{k}$ for $\ket{\Gamma_n}$ and $\vec{K}+\vec{k}$ for $\ket{K_n}$.

\subsection{Effective Hamiltonian}
We now apply the general expressions derived in Section~\ref{sec:compact_expression} to the two mode bases $\vec{\Gamma_n}$ and $\vec{K_n}$. The diagonal elements of the guided-mode Hamiltonian $H_\text{guided}$ are simply the dispersion relations of the corresponding modes:
\begin{equation}
    \begin{split}
        H^{\text{guided}(\Gamma)}_{nn}(\vec{k})&=\omega_{\ket{\Gamma_n}}(\vec{k})\\
        H^{\text{guided}(K)}_{nn}(\vec{k})&=\omega_{\ket{K_n}}(\vec{k})
    \end{split}
\end{equation}
The off-diagonal elements, corresponding to diffractive coupling between guided modes, take the form:
\begin{equation}
    \begin{split}
        H^{\text{guided}(\Gamma)}_{nm}&=\hat{U}^{(\Gamma)}_{nm}=\vec{p_{\ket{G_n}}}\cdot\vec{p_{\ket{G_m}}}U^{(\Gamma)}_{nm}\\
        H^{\text{guided}(K)}_{nm}&=\hat{U}^{(K)}_{nm}=\vec{p_{\ket{K_n}}}\cdot\vec{p_{\ket{K_m}}}U^{(K)}_{nm}
    \end{split}
\end{equation}
where the coupling coefficients are defined as:
    \begin{equation}
    \begin{split}
         U^{\Gamma}_{nm}=&\frac{\omega_\Gamma}{2}\int \abs{u_{\Gamma}(z)}^2 \epsilon_{n_1-m_1,n_2-m_2}(z)\, dz\\
         U^{K}_{nm}=&\frac{\omega_K}{2}\int \abs{u_{K}(z)}^2 \epsilon_{n_1-m_1,n_2-m_2}(z)\, dz
    \end{split}   
    \end{equation}\label{eq:U_nm_Gamma}

Here, $\epsilon_{n_1 - m_1, n_2 - m_2}$ denotes the Fourier component $\epsilon_{\vec{G_n} - \vec{G_m}}$ of the in-plane permittivity modulation $\Delta \epsilon(\vec{r}_\parallel)$.
Thanks to the $C_6$ symmetry of the lattice and using the specific mode indices $(n,m)$ given in Eq.~\eqref{eq:couples_nm}, along with the polarization vectors $\vec{p}_{\ket{G_n}}$ and $\vec{p}_{\ket{K_n}}$ defined in Eq.~\eqref{eq:pol_guidedmodes}, the coupling matrices acquire the simplified form:
\begin{equation}
    \begin{cases}       \hat{U}^{\Gamma}_{12}&=\hat{U}^{\Gamma}_{23}=\hat{U}^{\Gamma}_{34}=\hat{U}^{\Gamma}_{45}=\hat{U}^{\Gamma}_{56}=\hat{U}^{\Gamma}_{61}\equiv V\\
        \hat{U}^{\Gamma}_{13}&=\hat{U}^{\Gamma}_{24}=\hat{U}_{35}=\hat{U}^{\Gamma}_{46}=\hat{U}^{\Gamma}_{51}=\hat{U}^{\Gamma}_{62}\equiv W\\
        \hat{U}^{\Gamma}_{14}&=\hat{U}^{\Gamma}_{25}=\hat{U}^{\Gamma}_{36}\equiv U\\
        \hat{U}^{K}_{12}&=\hat{U}^{K}_{23}=\hat{U}^{K}_{31}\equiv T
    \end{cases}
\end{equation}
The couplings $U$, $V$, $W$, and $T$ are all real-valued and governed by specific Fourier components of $\Delta \epsilon$:
\begin{equation*}
\begin{cases}
    V &: \epsilon_{1,0}=\epsilon_{1,1}=\epsilon_{0,1}=\epsilon_{-1,0}=\epsilon_{-1,-1}=\epsilon_{0,-1}\\
    W &: \epsilon_{1,-1}  = \epsilon_{2,1}  = \epsilon_{1,2} = \epsilon_{-1,1}=\epsilon_{-2,-1}=\epsilon_{-1,-2}\\
    U &: \epsilon_{2,0}=\epsilon_{2,2}=\epsilon_{0,2}\\
    T &: \epsilon_{0,2}=\epsilon_{2,0}=\epsilon_{-2,-2}
\end{cases} 
\end{equation*}

The $C_6$ symmetry also simplifies the structure of the radiative Hamiltonian $\hat{H}_{\text{rad}}$. The diagonal elements are given by:
 \begin{equation}
 \begin{split}
     \hat{H}_{nn}^{\text{rad}(\Gamma)} &= -i\gamma_0 \\
     \hat{H}_{nn}^{\text{rad}(K)} &= 
     \begin{cases}
         -i\gamma_1 \quad\text{ for } n=1,2\\
         -i\gamma_2 \quad\text{ for } n=3
     \end{cases}
 \end{split}    
    \end{equation}
where $\gamma_0$, $\gamma_1$, and $\gamma_2$ represent the radiative loss rates for the modes at $\Gamma$ and $K$, respectively, under the assumption of a single radiation channel. These are governed by:
\begin{equation*}
  \begin{cases}
    \gamma_0 : \epsilon_{1,0}=\epsilon_{1,1}=\epsilon_{0,1}=\epsilon_{-1,0}=\epsilon_{-1,-1}=\epsilon_{0,-1}\\
    \gamma_1 : \epsilon_{0,1}=\epsilon_{0,-1}\\
    \gamma_2 : \epsilon_{-2,-1}\\
\end{cases}  
\end{equation*}

The off-diagonal elements are expressed as:
\begin{equation}
 \begin{split}
     \hat{H}_{nm}^{\text{rad}(\Gamma)} &= -i\vec{p_{\ket{G_n}}}\cdot\vec{p_{\ket{G_m}}}\gamma_0 \\
     \hat{H}_{nm}^{\text{rad}(K)} &= i\vec{p_{\ket{K_n}}}\cdot\vec{p_{\ket{K_m}}}\sqrt{\hat{H}_{nn}^{\text{rad}(K)}\hat{H}_{mm}^{\text{rad}(K)}}
 \end{split}    
    \end{equation}

Finally, the full non-Hermitian effective Hamiltonians for guided resonances near the $\Gamma$ and $K$ points are constructed as:
\begin{widetext}
\begin{equation}
\begin{aligned}
    &\hat{H}_\Gamma(\vec{k})=\omega_\Gamma +\begin{pmatrix}
        v_\Gamma k\cos{\left(\varphi-\frac{\pi}{3}\right)} &V & W & U & W & V\\
        V&v_\Gamma k\cos\varphi& V & W & U & W\\
        W & V &v_\Gamma k\cos{\left(\varphi+\frac{\pi}{3}\right)} &V & W & U\\
        U & W & V&-v_\Gamma k\cos{\left(\varphi-\frac{\pi}{3}\right)} &V & W\\
        W & U & W & V&-v_\Gamma k\cos\varphi&V\\
        V & W & U & W & V& -v_\Gamma k\cos{\left(\varphi+\frac{\pi}{3}\right)}\\
    \end{pmatrix} \\
    &  
    - i\gamma_0\begin{pmatrix}
        1 & \frac{1}{2} & -\frac{1}{2} & -1 & -\frac{1}{2} & \frac{1}{2}\\
        \frac{1}{2} & 1 & \frac{1}{2} & -\frac{1}{2} & -1 & -\frac{1}{2}\\
        -\frac{1}{2} &  \frac{1}{2} & 1 & \frac{1}{2} & -\frac{1}{2} & -1\\
        -1&-\frac{1}{2} &  \frac{1}{2} & 1 & \frac{1}{2} & -\frac{1}{2}\\
        -\frac{1}{2}&-1&-\frac{1}{2} &  \frac{1}{2} & 1 & \frac{1}{2}\\
        \frac{1}{2}&-\frac{1}{2}&-1&-\frac{1}{2} &  \frac{1}{2} & 1  \end{pmatrix}
     \label{eq:H_gamma}
    \end{aligned}
    \end{equation}
and
\begin{equation}
    \hat{H}_K(\vec{k})=\omega_K+ \begin{pmatrix}
        v_K k\sin{\varphi} & T & T \\
        T &-v_K k\sin{\left(\varphi-\frac{\pi}{3}\right)}& T \\
        T & T &-v_K k\sin{\left(\varphi+\frac{\pi}{3}\right)}  
    \end{pmatrix}  
    - i\begin{pmatrix}
        \gamma_1 & -\frac{1}{2}\gamma_1 & -\frac{1}{2}\sqrt{\gamma_1\gamma_2} \\
        -\frac{1}{2}\gamma_1 & \gamma_1 & -\frac{1}{2}\sqrt{\gamma_1\gamma_2}\\
        -\frac{1}{2}\sqrt{\gamma_1\gamma_2} &  -\frac{1}{2}\sqrt{\gamma_1\gamma_2} & \gamma_2         
    \end{pmatrix}\\
      \label{eq:H_K}
\end{equation}
\end{widetext}
\subsection{Eigenmodes at $\Gamma$: emergence of symmetry-protected BICs}\label{sec:Eigenmode at Gamma}
At the $\Gamma$ point,  $\hat{H}_\Gamma(\vec{k})$ can be diagonalized analytically, yielding six eigenmodes $\ket{\Omega_{n=1,\dots,6}^{(\Gamma)}}$ with  eigenvalues:
\begin{equation}
    \begin{cases}
        \Omega_1^{(\Gamma)}(\vec{k}=0) = \omega_\Gamma + U + 2V + 2W \\
        \Omega_2^{(\Gamma)}(\vec{k}=0) = \omega_\Gamma + U - V - W \\
        \Omega_3^{(\Gamma)}(\vec{k}=0) = \omega_\Gamma + U - V - W \\
        \Omega_4^{(\Gamma)}(\vec{k}=0) = \omega_\Gamma - U - 2V + 2W \\
        \Omega_5^{(\Gamma)}(\vec{k}=0) = \omega_\Gamma - U + V - W + 3i\gamma_0 \\
        \Omega_6^{(\Gamma)}(\vec{k}=0) = \omega_\Gamma - U + V - W + 3i\gamma_0
    \end{cases}
    \label{eq:values_k=0}
\end{equation}
From these expressions, we identify four BICs at the $\Gamma$ point: $\ket{\Omega_1^{(\Gamma)}}$ and $\ket{\Omega_4^{(\Gamma)}}$ are non-degenerate, while $\ket{\Omega_2^{(\Gamma)}}$ and $\ket{\Omega_3^{(\Gamma)}}$ form a doubly degenerate pair. The remaining two modes, $\ket{\Omega_5^{(\Gamma)}}$ and $\ket{\Omega_6^{(\Gamma)}}$, are leaky modes and also form a degenerate pair.
\begin{figure}[h]
\centering
\includegraphics[width=1\linewidth]{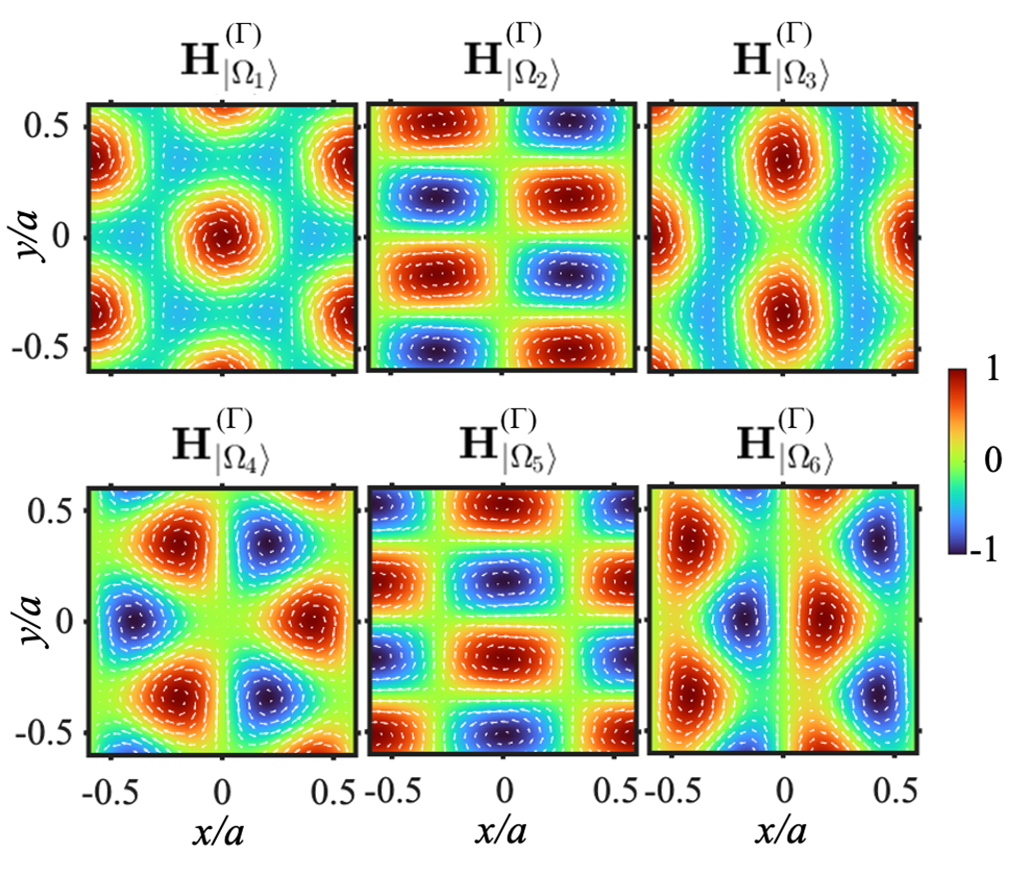}
\caption{\textbf{General near-field patterns.} Calculated magnetic near-field profiles $\vec{H}_{\ket{\Omega_n}}^{(\Gamma)}$ for the eigenmodes at the $\Gamma$ point. Arrows represent the electric field vectors $\vec{E}_{\ket{\Omega_n}}^{\Gamma}$.}
\label{fig:NF_Gamma}
\end{figure}

The corresponding eigenvectors, non-normalized, at $\vec{k} = 0$ are:
\begin{equation}
    \begin{cases}
        \mathbf{A}_1^{(\Gamma)} = (1, 1, 1, 1, 1, 1) \\
        \mathbf{A}_2^{(\Gamma)} = (-1, 0, 1, -1, 0, 1) \\
        \mathbf{A}_3^{(\Gamma)} = (-\frac{1}{2}, 1, -\frac{1}{2}, -\frac{1}{2}, 1, -\frac{1}{2}) \\
        \mathbf{A}_4^{(\Gamma)} = (-1, 1, -1, 1, -1, 1) \\
        \mathbf{A}_5^{(\Gamma)} = (1, 0, -1, -1, 0, 1) \\
        \mathbf{A}_6^{(\Gamma)} = (\frac{1}{2}, 1, -\frac{1}{2}, \frac{1}{2}, -1, -\frac{1}{2})
    \end{cases}
    \label{eq:eigmodes_k=0}
\end{equation}
Using Eqs.~\eqref{eq:NF_E} and~\eqref{eq:NF_H}, the near-field distributions $\vec{E}^{(\Gamma)}_{\ket{\Omega_n}}$ and $\vec{H}^{(\Gamma)}_{\ket{\Omega_n}}$ can be computed analytically. Remarkably, these spatial field patterns are fully determined by the symmetry of the eigenvectors and are independent of the specific values of the coupling parameters $U$, $V$, and $W$. Figure~\ref{fig:NF_Gamma} presents the calculated magnetic near-field profiles. Based on the spatial symmetry of these modes, we assign:
\begin{equation*}
    \begin{cases}
    \ket{\Omega_1^{(\Gamma)}}\text{: magnetic monopolar mode,}\\
    \ket{\Omega_{2,3}^{(\Gamma)}}\text{: magnetic quadripolar modes,}\\
    \ket{\Omega_4^{(\Gamma)}}\text{: magnetic hexapolar mode,}\\
    \ket{\Omega_{5,6}^{(\Gamma)}}\text{: magnetic dipolar modes.}
    \end{cases}
\end{equation*}
These modal patterns can also be classified according to the irreducible representations of the $C_6$ point group symmetry~\cite{Sakoda2001}, offering a clear group-theoretical interpretation of their polarization textures. We emphasize that the general near-field patterns predicted by our effective Hamiltonian model are in perfect agreement with those reported in the literature using full-wave finite-difference time-domain (FDTD) simulations~\cite{Imada2002}, for the six photonic modes at the $\Gamma$ point of a triangular lattice. This agreement not only validates the accuracy of our model but also underscores its ability to capture the essential physics of PhC slabs with high symmetry.

Moreover, using Eq.~\eqref{eq:FF_E}, the radiation pattern of these six photonic modes can be computed.  One may confirm that the farfield radiation of the monopolar mode corresponds to a polarization singularity of topological charge 1, while the topological charge of the hexapolar mode is -2, and the degenerated quadipolar modes are pinned at a polarization singularity of topological charge -2. 

Interestingly, an accidental degeneracy between the quadripolar modes and the hexapolar mode occurs when the condition $ W = \frac{2U + V}{3} $ is satisfied. Under this configuration, the resulting triple degeneracy at the $\Gamma$ point consists of one quadratic band and a pair of Dirac cones. This accidental degeneracy, involving three BICs, is particularly relevant for applications such as zero-refractive-index metamaterials~\cite{Minkov2018,Dong2021} and scalable lasing~\cite{Contractor2022,Kant2024}, both of which benefit from lossless Dirac cones at the $\Gamma$ point.

\subsection{ Eigenmodes at $K$: symmetry-protected Dirac dispersion}

In the vicinity of the $K$ point, the non-Hermitian Hamiltonian $\hat{H}_K(\vec{k})$ can be approximately diagonalized in the regime \( T \gg \gamma_1, \gamma_2 \). This yields three eigenmodes $\ket{\Omega_{n=1,2,3}^{(K)}}$ with eigenvalues:

\begin{equation}
    \begin{cases}
        \Omega_1^{(K)}(\vec{k}) &= \omega_K - T + v_K k \sin\left(\varphi + \frac{\pi}{3} \right) - i\frac{3\gamma_1}{2} \\
        \Omega_2^{(K)}(\vec{k}) &= \omega_K - T - v_K k \sin\left(\varphi + \frac{\pi}{3} \right) - i\frac{ \left( \sqrt{\gamma_1} + 2\sqrt{\gamma_2} \right)^2 }{6} \\
        \Omega_3^{(K)}(\vec{k}) &= \omega_K + 2T - i\frac{ \left( \sqrt{\gamma_1} - \sqrt{\gamma_2} \right)^2 }{3} + \mathcal{O}(k^2)
    \end{cases}
    \label{eq:values_K}
\end{equation}

The corresponding eigenvectors at $\vec{k} = 0$ are:
\begin{equation}
    \begin{cases}
        \mathbf{A}_1^{(K)} = \left(\frac{-2(1+\sin\varphi)}{1+\sin\varphi - \sqrt{3}\cos\varphi}, \frac{2(1+\sin\varphi)}{1+\sin\varphi - \sqrt{3}\cos\varphi}-1, 1\right) \\
        \mathbf{A}_2^{(K)} = \left(\frac{-2(1-\sin\varphi)}{1-\sin\varphi + \sqrt{3}\cos\varphi}, \frac{2(1-\sin\varphi)}{1-\sin\varphi + \sqrt{3}\cos\varphi}-1, 1\right) \\
        \mathbf{A}_3^{(K)} = (1, 1, 1) 
    \end{cases}
    \label{eq:eigmodes_k=K}
\end{equation}
The first two modes, $\ket{\Omega_1^{(K)}}$ and $\ket{\Omega_2^{(K)}}$, are degenerate at \( \vec{k} = 0 \) and split linearly with \( k \), forming a Dirac cone centered at the $K$ point. This Dirac cone is robust against variations in $\gamma_1$, $\gamma_2$, and $T$, as long as the condition \( T \gg \gamma_1, \gamma_2 \) holds.

\begin{figure*}[!ht]
\centering
\includegraphics[width=1\linewidth]{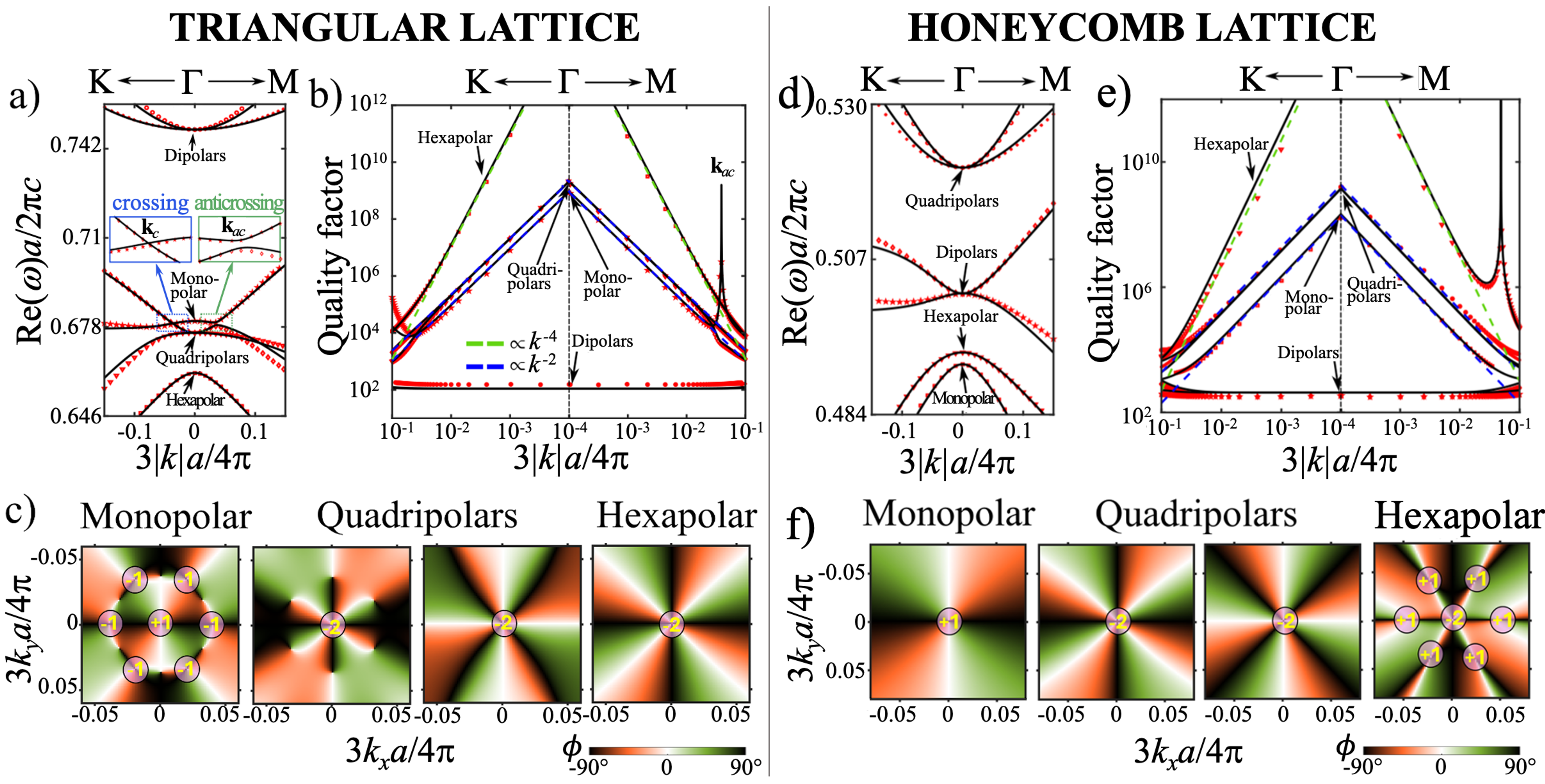}
\caption{\textbf{Eigenmodes near $\Gamma$ for triangular and honeycomb lattice design.}  
a,d) Real part of the photonic band energies as a function of the in-plane wavevector. The zoomed-in insets highlight the crossing along $\Gamma K$(blue box) and anticrossing $\Gamma M$ (green box) between the third and fourth bands of the triangular lattice. b,e) Quality factors of the photonic bands. Green and blue dashed lines indicate reference curves proportional to \(1/k^4\) and \(1/k^2\), respectively. Red scatters represent numerically simulated photonic bands, while black lines show their corresponding analytical fitting using the effective theory. c,f) Far-field polarization textures (i.e., the orientation of radiated polarization) of the photonic bands hosting monopolar, quadrupolar, and hexapolar modes. }
\label{fig:Figure_5}
\end{figure*}
\subsection{Effective Theory vs. Numerical Simulations near the $\Gamma$ Point}

In this section, we focus on the eigenmodes in the vicinity of the $\Gamma$ point, where various symmetry-protected and accidental BICs and EPs can emerge. The analysis of modes near the $K$ point—characterized by distinct degeneracy lifting and topological transitions under $C_6$ symmetry breaking—will be presented in the next section.

The simulated structures consist of air hole arrays in dielectric slabs with two lattice geometries: triangular and honeycomb. In both designs, the lattice constant is $a = 400$~nm, the air hole diameter is $D = 0.35a$, and the slab thickness is $h = 100$~nm. The refractive index of the slab is $n = 2.0$, and the structures are embedded in air.
\subsubsection{Complex Band Structures and Symmetry-Protected BICs}

We first show that the photonic band structures near the $\Gamma$ point are accurately described by the effective non-Hermitian Hamiltonians introduced in Eq.~\eqref{eq:H_gamma}. To validate this theory, we perform full-wave simulations using the finite element method (FEM) implemented in \textsc{COMSOL Multiphysics}. Floquet boundary conditions are applied in the in-plane directions, while perfectly matched layers (PMLs) along $z$ model radiation into the far field. Complex eigenfrequencies $\Omega_n(\vec{k})$ are computed for a dense sampling of $\vec{k}$-points near the $\Gamma$ point, and are fitted using the analytical eigenvalues of the effective Hamiltonians. The corresponding fitting parameters are listed in Appendix.~\ref{appendix:fitting_para}.

Figures~\ref{fig:Figure_5}a and~\ref{fig:Figure_5}d present the real parts of the eigenfrequencies for the triangular and honeycomb lattices, respectively. The band structures show excellent agreement between numerical simulations and the analytical \hsn{calculation}s, both for the real and imaginary parts of the eigenfrequencies, as evidenced by the quality factors plotted in Figs.~\ref{fig:Figure_5}b and~\ref{fig:Figure_5}e. In particular, the expected scaling laws for BICs are recovered: $Q \propto 1/k^{2q}$, where $q$ is the topological charge. The hexapolar mode with $q = -2$ follows $Q \propto 1/k^4$, while the monopolar ($q = +1$) and each of the two quadrupolar modes ($q = -1$) exhibit $Q \propto 1/k^2$, fully consistent with theoretical \hsn{calculation}s.

\subsubsection{Accidental off-$\Gamma$ BICs}

While the presence of symmetry-protected BICs at the $\Gamma$ point is independent of the lattice details—as discussed in Sec.~\ref{sec:Eigenmode at Gamma} and confirmed above for both geometries—accidental off-$\Gamma$ BICs can emerge in specific bands depending on the lattice design.

In the triangular lattice, off-$\Gamma$ BICs are found in the monopolar mode band at $3|k|a/4\pi = 0.0391$ along the $\Gamma \rightarrow M$ direction (Fig.~\ref{fig:Figure_5}b). In the honeycomb lattice, off-$\Gamma$ BICs appear in the hexapolar mode band at $3|k|a/2\pi = 0.0488$ (Fig.~\ref{fig:Figure_5}e). In both cases, their positions and properties are accurately predicted by the effective model, confirming that the mechanism of accidental destructive interference is fully captured by our generalized guided-mode expansion  framework.

Interestingly, in the triangular lattice, the off-$\Gamma$ BIC occurs near an anticrossing between the third (quadrupolar) and fourth (monopolar) bands, as highlighted in the zoom-in inset of Fig.~\ref{fig:Figure_5}a. The corresponding quality factor profiles in Fig.~\ref{fig:Figure_5}b reveal a clear loss exchange: at the anticrossing wavevector $\vec{k}_{ac}$, the quality factor of the monopolar band increases by five orders of magnitude, while that of the quadrupolar band drops sharply. This strongly suggests that the off-$\Gamma$ BIC arises from two-band Friedrich–Wintgen interference.

By contrast, the off-$\Gamma$ BIC in the honeycomb lattice does not coincide with any visible anticrossing in Fig.~\ref{fig:Figure_5}d. Furthermore, the increase in quality factor for the hexapolar band is not accompanied by a corresponding drop in any nearby band (Fig.~\ref{fig:Figure_5}e), indicating that the BIC arises from multi-band interference and cannot be reduced to a two-mode interaction picture.

The far-field polarization textures (see Appendix~\ref{sec:appendix_polarization} for details) associated with the monopolar, quadrupolar, and hexapolar modes are presented in Figs.~\ref{fig:Figure_5}c and~\ref{fig:Figure_5}f for the triangular and honeycomb lattices, respectively. The expected topological charges—$+1$ for the monopolar, $-2$ for the hexapolar, and $-2$ total for the twofold-degenerate quadrupolar modes—are clearly observed, confirming that each quadrupolar mode carries charge $-1$. Additionally, six off-$\Gamma$ topological charges corresponding to accidental BICs are identified along the $\Gamma \rightarrow M$ path, with their positions and bands depending on the lattice type. These features, in agreement with both simulations and analytical \hsn{calculation}s of quality factors, further demonstrate that our effective theory captures not only the complex eigenfrequencies but also the far-field polarization topology.

\subsubsection{Emergence of Chiral Exceptional Points}
\begin{figure}[h!]
\begin{center}
\includegraphics[width=1\linewidth]{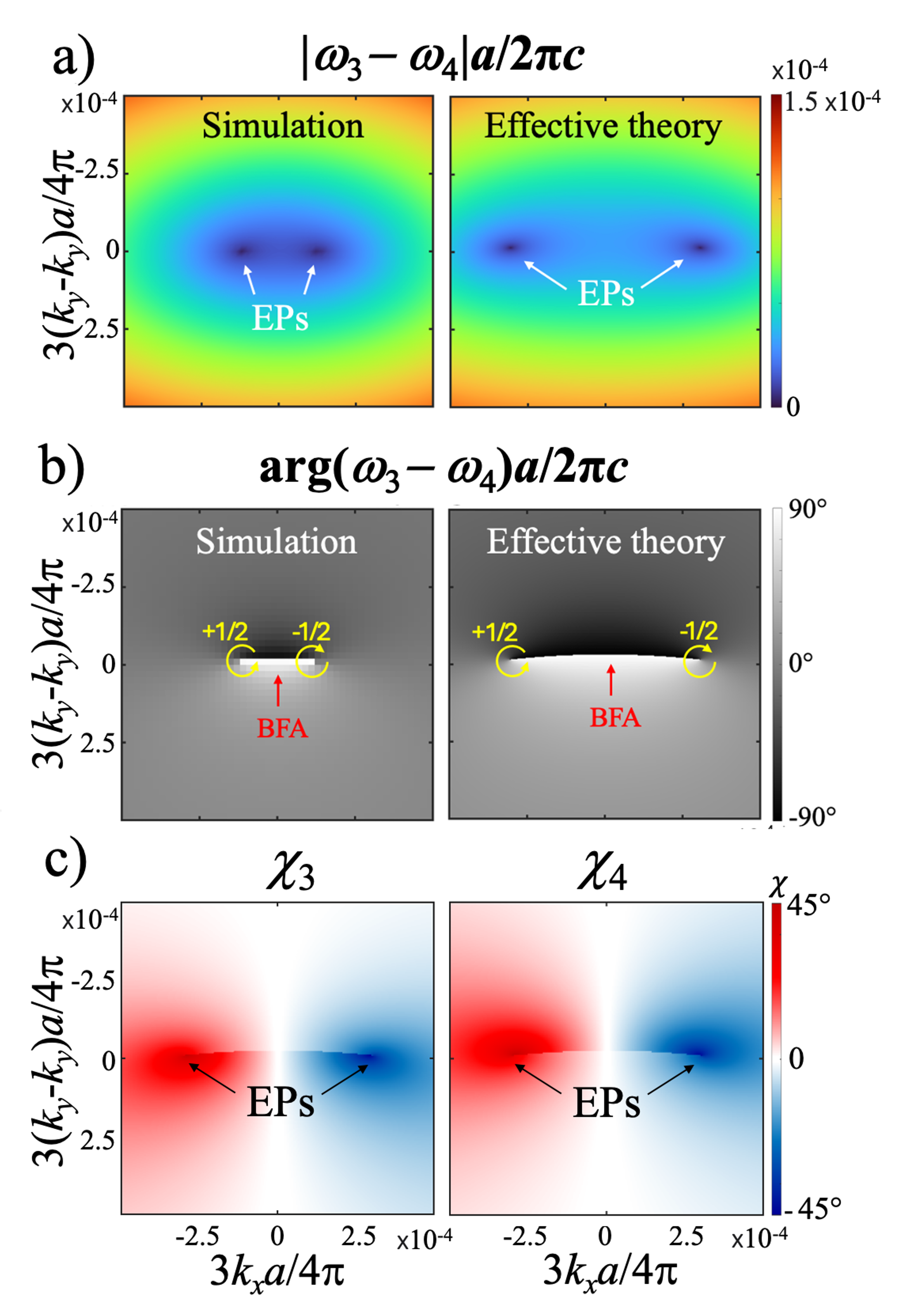}
\caption{\textbf{Chiral EPs in triangular lattice design.} a,b) Amplitude (a) and argument (b) of the complex gap between the third and fourth band in the vicinity of the crossing point of Fig.~\ref{fig:Figure_5}a. Left panels are results obtained from numerical simulations, while right panels are from the analytical models. c) \hsn{Ellipticity of the far-field polarization} of the third and fourth bands in the vicinity of the crossing point $\vec{k_c}=(0,k_c)$ of Fig.~\ref{fig:Figure_5}a. }
\label{fig:EP}
\end{center}
\end{figure}
Beyond BICs, the effective theory also successfully predicts the emergence of EPs in the PhC slabs. As pointed out in Ref.~\cite{frau2025}, EPs are expected to appear near crossings of bands with opposite symmetry. To identify possible EPs, we examine the band structures along high-symmetry directions. In the triangular lattice, the third and fourth bands cross along the $\Gamma \rightarrow K$ direction (Fig.~\ref{fig:Figure_5}a), suggesting the emergence of EPs in their vicinity. Due to the $C_6$ symmetry, there are six equivalent crossing points. Without loss of generality, we focus on $\vec{k}_c = (0, k_c)$. To probe the EPs, we map the amplitude and argument of the complex gap between the third and fourth bands around $\vec{k}_c$. As shown in Figs.~\ref{fig:EP}a and~\ref{fig:EP}b, two EPs are clearly identified by the vanishing of the gap amplitude and the presence of a singularity in the phase. The winding number of each EP, $w = \pm \frac{1}{2}$, is computed from the gap argument as $w = \frac{1}{2\pi} \oint_\mathcal{C} d\vec{k} \cdot \nabla_{\vec{k}} \arg\left[\omega_4(\vec{k}) - \omega_3(\vec{k})\right]$.

The phase map of the gap also reveals a bulk Fermi arc (BFA) connecting the two EPs~\cite{Zhou2018,frau2025}, characterized by a $\pi$ jump in the argument (Fig.~\ref{fig:EP}b), marking a degeneracy of the real parts of the eigenfrequencies. The agreement between the effective theory and numerical simulations for both the gap amplitude and phase confirms the robustness of our model in capturing non-Hermitian degeneracies and their topological features.

Finally, we compute the \hsn{ellipticity of the far-field polarization} (see Appendix~\ref{sec:appendix_polarization} for details) of the third and fourth bands near $\vec{k}_c$. As shown in Fig.~\ref{fig:EP}c, the two EPs exhibit opposite handedness in their polarization textures, confirming their chiral nature. To the best of our knowledge, this is the first demonstration of \textit{chiral EPs} in a triangular lattice without explicit symmetry breaking.

\begin{figure}[h!]
\begin{center}
\includegraphics[width=0.8\linewidth]{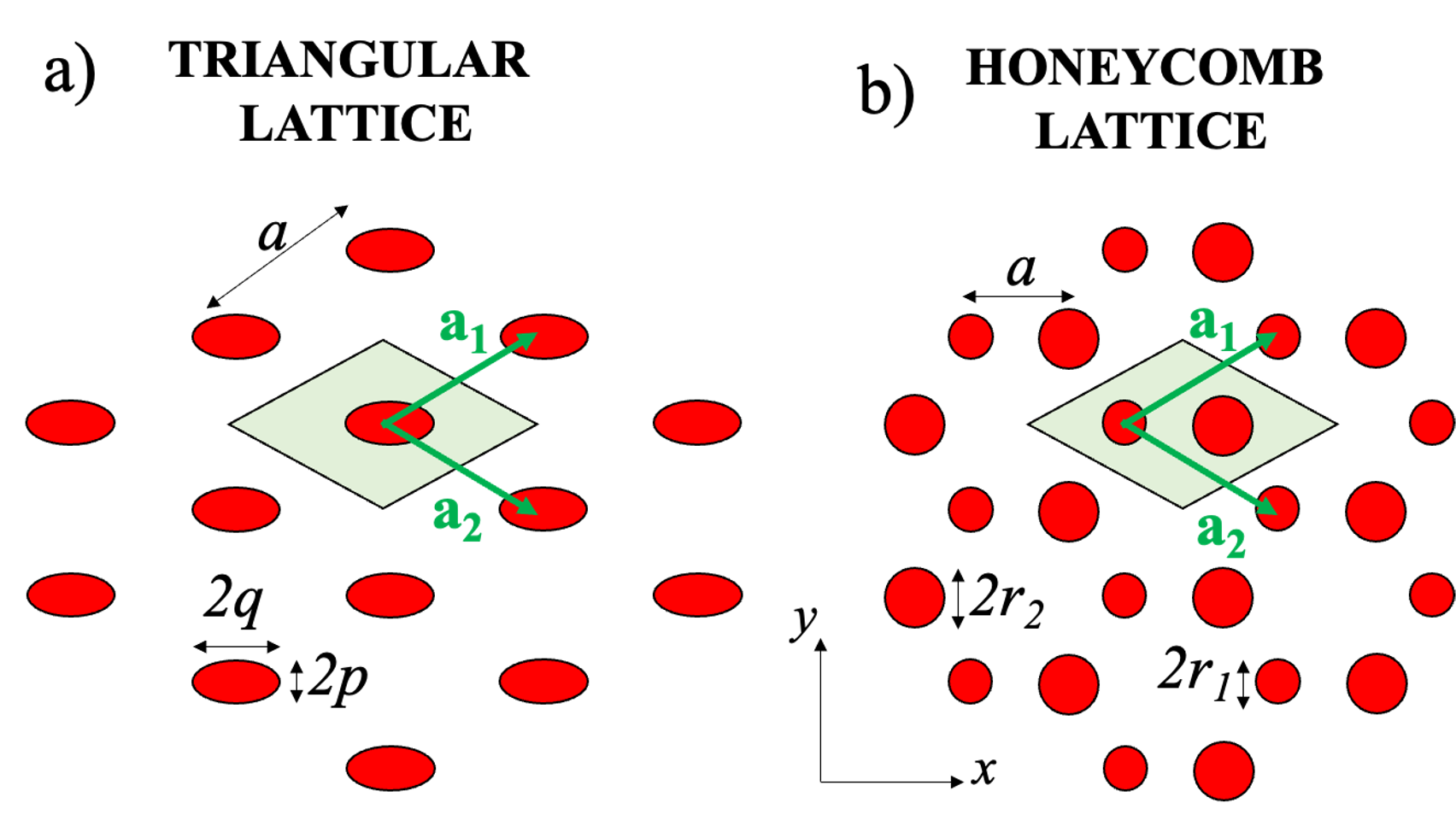}
\caption{\textbf{Hexagonal lattices with broken $C_6$ symmetry.} a) Triangular lattice with elliptical holes. b) Honeycomb lattice with two circular holes of different sizes.}
\label{fig:brokenC6}
\end{center}
\end{figure}

\subsection{Hexagonal lattices with broken $C_6$ symmetry}

The $C_6$ symmetry is broken either by using elliptical holes instead of circular ones in a triangular lattice, or by using two circular holes of different sizes in a honeycomb lattice (see Fig.~\ref{fig:brokenC6}). In general, breaking the $C_6$ symmetry lifts the degeneracy of the quadrupolar and dipolar modes at the $\Gamma$ point, as well as the Dirac point degeneracy at the $K$ point. However, depending on the specific geometry of the symmetry breaking, the form of the effective Hamiltonian will differ. In this section, we focus specifically on the band structure in the vicinity of the $K$ point, where the lifting of Dirac degeneracies gives rise to rich topological and non-Hermitian phenomena~\cite{Dong2016,Guo2020,Xue2021}.

\subsubsection{Modification of the effective Hamiltonian}

For the case of triangular lattice with elliptical holes, when the elliptical holes are aligned along the $x$- or $y$-axis (see Fig.~\ref{fig:brokenC6}a), the two mirror symmetries $x \rightarrow -x$ and $y \rightarrow -y$, corresponding to $C_2$ operations, are preserved. As a result, all Fourier components of $\Delta\epsilon(\vec{r}_\parallel)$ remain real-valued, leading to real-valued coupling coefficients $U^{(K)}_{nm}$, as defined in Eq.~\eqref{eq:U_nm}. However, the $C_3$ symmetry is broken, and the effective Hamiltonian near the $K$ point now becomes:

\begin{align}
    \begin{split}
    \hat{H}_K^{(1)}(\vec{k})=\omega_K - i\begin{pmatrix}
        \gamma_1 & -\frac{1}{2}\sqrt{\gamma_1\gamma_2} & -\frac{1}{2}\sqrt{\gamma_1\gamma_3} \\
        -\frac{1}{2}\sqrt{\gamma_1\gamma_2} & \gamma_2 & -\frac{1}{2}\sqrt{\gamma_2\gamma_3}\\
        -\frac{1}{2}\sqrt{\gamma_1\gamma_3} &  -\frac{1}{2}\sqrt{\gamma_2\gamma_3} & \gamma_3         
    \end{pmatrix}
    \\ + \begin{pmatrix}
        v_K k\sin{\varphi} & T_1 & T_2 \\
        T_1 &-v_K k\sin{\left(\varphi-\frac{\pi}{3}\right)}& T_1 \\
        T_2 & T_1 &-v_K k\sin{\left(\varphi+\frac{\pi}{3}\right)}  
    \end{pmatrix} \\ &
    \label{eq:H_K1}
    \end{split}
\end{align}

On the other hand, for the case of honeycomb lattice with different hole sizes,  the $C_3$ symmetry is preserved, but the difference in hole sizes breaks the mirror symmetry $x \rightarrow -x$ (see Fig.~\ref{fig:brokenC6}b), while the symmetry $y \rightarrow -y$ remains. Consequently, the Fourier components of $\Delta\epsilon(\vec{r}_\parallel)$ can be complex, resulting in complex coupling strengths $U^{(K)}_{nm}$ with  $\hat{U}^{K}_{mn} = \left(\hat{U}^{K}_{nm}\right)^*$, being complex-valued as a direct consequence of the broken mirror symmetry. Therefore, the effective Hamiltonian near the $K$ point is given by:

\begin{align}
    \begin{split}
    \hat{H}_K^{(1)}(\vec{k})=\omega_K - i\begin{pmatrix}
        \gamma_1 & -\frac{1}{2}\gamma_1 & -\frac{1}{2}\sqrt{\gamma_1\gamma_2} \\
        -\frac{1}{2}\gamma_1 & \gamma_1 & -\frac{1}{2}\sqrt{\gamma_1\gamma_2}\\
        -\frac{1}{2}\sqrt{\gamma_1\gamma_2} &  -\frac{1}{2}\sqrt{\gamma_1\gamma_2} & \gamma_2         
    \end{pmatrix} 
    \\ + \begin{pmatrix}
        v_K k\sin{\varphi} & T & T^* \\
        T^* &-v_K k\sin{\left(\varphi-\frac{\pi}{3}\right)}& T \\
        T & T^* &-v_K k\sin{\left(\varphi+\frac{\pi}{3}\right)}  
    \end{pmatrix} \\ &
    \label{eq:H_K2}
    \end{split}
\end{align}

\subsubsection{Band Structure: Effective Theory vs. Numerical Simulations near the $K$ point}

\begin{figure}[h!]
 \begin{center}
\includegraphics[width=1\linewidth]{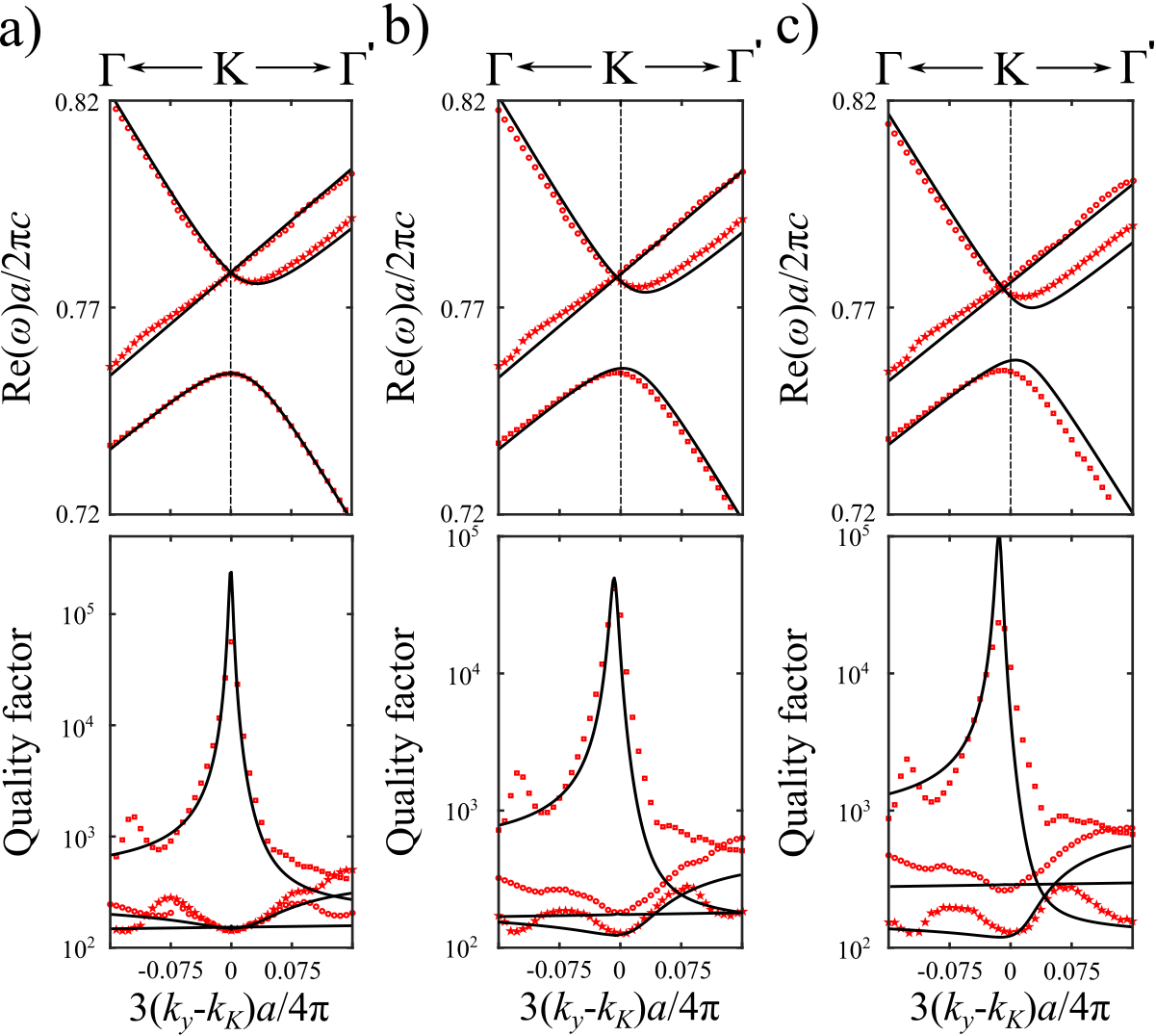}
\caption{\textbf{Triangular lattice with elliptical holes}. Red scatters represent numerically simulated photonic bands, while black lines show their corresponding analytical fitting using the effective theory. The upper panel shows the band structures near the $K$ point, while the lower panel depicts the corresponding quality factor for each band. (a) $p = q = 60$ nm. (b) $p = 50$ nm and $q = 70$ nm. (c) $p = 40$ nm and $q = 80$ nm.}
\label{fig:figure_9}
\end{center}
\end{figure}

To investigate the role of $C_3$ symmetry, we designed a structure consisting of a triangular lattice of elliptical air holes with lattice constant $a = 440$~nm, slab thickness $h = 180$~nm, and refractive index $n = 2.02$, placed on a glass substrate with $n = 1.46$. The elliptical holes are defined by their semi-axes $p$ and $q$, allowing controlled breaking of higher-order rotational symmetries. The results of three representative cases are presented in Fig.~\ref{fig:figure_9}, showing both the band structure and the associated quality factors.

\begin{figure}[h!]
\begin{center}
\includegraphics[width=1\linewidth]{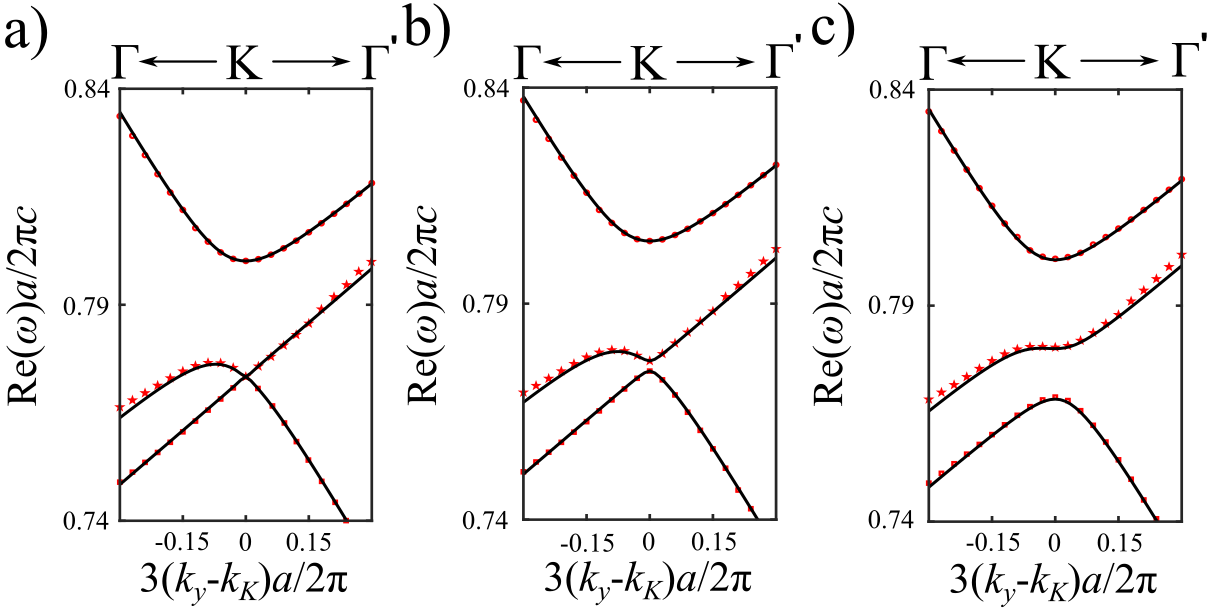}
\caption{\textbf{ Honeycomb lattice bands with different hole sizes.} 
Red scatters represent numerically simulated photonic bands, while black lines show their corresponding analytical fitting using the effective theory.
(a) $r_{1} = r_{2} = 50$ nm. (b) $r_{1} = 50$ nm and $r_{2} = 55$ nm. (c) $r_{2} = 40$ nm and $r_{2} = 60$ nm.}
\label{fig:figure10}
\end{center}
\end{figure}

In Fig.~\ref{fig:figure_9}a, we consider the high-symmetry case where $p = q = 60$~nm, which preserves $C_6$ symmetry. In this configuration, a Dirac point is formed at the $K$ point by the crossing of two upper bands, while the lowest band remains isolated. Interestingly, this lowest band exhibits a pronounced quasi-BIC character: its quality factor reaches a sharp peak (exceeding $10^5$) precisely at the momentum corresponding to the Dirac point. We then break the $C_3$ symmetry while preserving inversion symmetry ($C_2$) by elongating one semi-axis and reducing the other. Specifically, in Fig.~\ref{fig:figure_9}b, we increase $p$ by 10~nm and decrease $q$ by 10~nm. The Dirac point is no longer pinned to the high-symmetry $K$ point but shifts along the $K \rightarrow \Gamma$ direction, appearing at $3(k_{y}-k_K)a/4\pi = -0.0075$. Notably, the quasi-BIC peak in the lowest band follows this shift, indicating that the momentum-space location of maximal radiation suppression remains locked to the displaced Dirac crossing. This trend becomes more pronounced as the symmetry breaking increases. In Fig.~\ref{fig:figure_9}c, the semi-axes differ by 40~nm, and the Dirac point moves further to $3k_{y}a/4\pi = -0.015$. Owing to the preserved $C_2$ symmetry, this displacement is symmetric: if the major axis were instead aligned along the $y$-direction, the shift would occur in the opposite direction. These observations confirm that while the Dirac degeneracy persists due to inversion symmetry, its location in momentum space is no longer protected by $C_3$ symmetry and becomes tunable through geometry.

Crucially, the momentum-dependent complex eigenfrequencies obtained from full-wave simulations are in excellent quantitative agreement with the \hsn{calculation}s of the effective non-Hermitian Hamiltonian. This confirms that our analytical model faithfully captures both the band dispersion and the quasi-BIC behavior induced by symmetry breaking.

To investigate the role of inversion symmetry ($C_2$) in honeycomb lattices, we consider a slab similar to the previous cases, but with a reduced lattice constant of $a = 400$~nm and air holes of different radii $r_1$ and $r_2$. When inversion symmetry is preserved (i.e., $r_1 = r_2 = 50$~nm), the structure exhibits a Dirac point at the $K$ point, as shown in Fig.~\ref{fig:figure10}a, consistent with the symmetry-protected degeneracy of the honeycomb lattice. However, when inversion symmetry is broken by introducing a small size asymmetry between the two sublattices (e.g., $r_1 = 50$~nm and $r_2 = 55$~nm), the Dirac point degeneracy is lifted, and a bandgap opens at the $K$ point, as seen in Fig.~\ref{fig:figure10}b. This gap becomes significantly larger with stronger symmetry breaking. In Fig.~\ref{fig:figure10}c, a larger contrast between $r_1$ and $r_2$ results in a pronounced gap, demonstrating how geometric perturbations directly control the topological features of the band structure. Once again, the \hsn{calculation}s of the effective non-Hermitian Hamiltonian show excellent quantitative agreement with the full-wave numerical simulations, confirming the accuracy and robustness of the theoretical model.\\

Compared to the case of triangular lattices with elliptical holes, where breaking $C_3$ symmetry (while preserving inversion symmetry $C_2$) causes the Dirac point to shift in momentum space without lifting the degeneracy, the honeycomb lattice exhibits a qualitatively different response: breaking inversion symmetry directly opens a bandgap at the $K$ point. This contrast underscores the distinct roles of $C_2$ and $C_3$ symmetries in protecting Dirac points. Moreover, the energetic ordering and radiative properties of the bands also differ significantly between the two cases. In the triangular lattice, the singly degenerate band with quadratic dispersion lies below the Dirac point and exhibits a quasi-BIC character, with strongly suppressed radiation losses at the Dirac momentum. In contrast, for the honeycomb lattice, this quadratic band lies above the Dirac point, and all three bands near the $K$ point exhibit significant radiation losses. No quasi-BIC behavior is observed in this case, reflecting the absence of symmetry protection and destructive interference mechanisms that suppress radiation. These differences further highlight how lattice geometry and symmetry breaking govern both the topological and radiative characteristics of the photonic band structure.

\hsn{\subsection{Parameter retrieval and computational efficiency}
To determine the parameters of the analytical Hamiltonian, full-wave simulations are required only at a single high-symmetry point (e.g., $\Gamma$ or $K$). From the complex eigenfrequencies at this point, we extract all model coefficients—including the coupling parameters $U$, $V$, $W$, the diagonal frequencies $\omega_{\Gamma}$ and $\omega_{K}$, and the radiative loss rate $\gamma_{0}$ (numerical values of parameters used in the results are reported in Appendix \ref{appendix:fitting_para}. Once these coefficients are known, the effective Hamiltonian reproduces the full complex band structure in the vicinity of the high-symmetry point, including both the radiative linewidths and the far-field polarization textures, through instantaneous matrix diagonalization.

In our examples, subtle non-Hermitian features such as off-$\Gamma$ bound states in the continuum and chiral exceptional points were first revealed by the analytical Hamiltonian. Only after their approximate locations were identified did we refine our full-wave simulations—using significantly increased mesh density and finer $k$-space sampling—to confirm these features numerically. This highlights the predictive power of the analytical model and its ability to guide full-wave solvers toward the relevant regions of parameter space.

For a representative hexagonal-lattice structure, a full-wave FEM sweep ($\sim 50{,}000$ mesh elements) required approximately five hours on a standard desktop computer (AMD Ryzen 7 processor, 3.3 GHz; RAM 16 GB) to evaluate 1000-points and 7 frequency samples. In contrast, once the Hamiltonian parameters were extracted, the analytical model generated the corresponding results within a fraction of a second on the same hardware. This demonstrates the substantial computational advantage of the proposed framework, particularly for broad parameter scans or for exploring high-$Q$ resonances.}

\section{Conclusion and Perspectives}

In this work, we have developed a general and systematic formalism for modeling complex resonances in PhC slabs within a non-Hermitian framework. Starting from Maxwell’s equations, we derive an effective non-Hermitian Hamiltonian by expanding the electromagnetic fields onto the complete set of guided and radiative modes of an unpatterned slab. This approach provides a unified and physically grounded alternative to earlier phenomenological models that have been applied to periodic photonic structures such as gratings~\cite{Lu2020, Nguyen2024}, square~\cite{Do2025}, and rectangular lattices~\cite{MermetLyaudoz2023}.  We illustrated the effectiveness of our approach through a case study on hexagonal PhC slabs, under both preserved and broken $C_6$ symmetry. The effective Hamiltonian accurately reproduces complex band structures, near-field mode profiles, and far-field polarization textures, in excellent agreement with full-wave simulations. These results demonstrate that the formalism reliably captures how lattice symmetry and geometry govern radiation and resonance properties.

This framework paves the way for designing non-local metasurfaces with controlled radiation losses, enabling applications in high-$Q$ lasers, filters, and sensors. It also provides a powerful tool for exploring topological photonics in open systems, including bulk–radiation correspondence~\cite{Yin2025,yuan2025} and non-Hermitian effects such as EPs and spectral degeneracies~\cite{Ferrier2022}. Future extensions to multilayer slabs, moiré superlattices, or aperiodic structures will further broaden its scope, enabling the study of flatbands~\cite{Dong2021,Nguyen2022,Saadi2025} and other exotic radiative phenomena~\cite{Zhang2023,Ni2024,Gromyko2024}.\\

\section*{Acknowledgements} 
The authors thank Xinyi Yuan and Grazia Salerno for fruitful discussions. This research is funded by Vietnam National Foundation for Science and Technology Development (NAFOSTED) under grant number 103.03-2024.16,  by the French National Research Agency (ANR) under the project POLAROID (ANR-24-CE24-7616-01), by the Ministry of Education, Singapore, under its AcRF Tier 1 grant (RG140/23) and by the Institute for Basic Science in Korea through the Project IBS-R024-D1.

\onecolumngrid
\section*{Appendix}
\appendix
\section{Summing out the polarization products}\label{sec:sumingout}
We can simplify the expression of the coupling terms of $H$, given in Eq. ~\eqref{eq:coupling_element_GGME}, by summing out the polarization cross products. To do so, we write separately the polarization and the amplitude of the periodic function $\mathbf{u}_{n}(\mathbf{r})$ and $\mathbf{u}_{m,\text{rad}}(\mathbf{r})$. While $\mathbf{u}_{n}(z)=u_{n}(z)\vec{p}_{n}$, we note that  $\mathbf{u}_{m,\text{rad}}(z)$ are counted twice, once for $u_{m,\text{rad}}(z)\mathbf{\hat{x}}$  and one for $ u_{m,\text{rad}}(z)\mathbf{\hat{y}.}$ As consequence, we obtain:

\begin{align}
     &\int \mathbf{u}_{n}^* \cdot \epsilon_{\mathbf{G}_n - \mathbf{G}_m} \cdot\mathbf{u}_{m}\, dz \; = \;\vec{p}_n\vec{p}_m\int u_{n}^* \cdot \epsilon_{\mathbf{G}_n - \mathbf{G}_m} \cdot u_{m}\, dz,\\
     &\sum_{l}{\frac{|\int \mathbf{u}_{n}^* \cdot \epsilon_{-\mathbf{G}_l} \cdot\mathbf{u}_{l,\text{rad}} \, dz|^2 } {\gamma_{l,rad}}} 
     = \sum_{l}{\frac{|\int u_{n}^*\cdot \epsilon_{-\mathbf{G}_l}\cdot u_{l,\text{rad}} \, dz|^2 } {\gamma_{l,rad}}},\\
    &\sum_{l}{\frac{\int \mathbf{u}_{n}^* \cdot \epsilon_{-\mathbf{G}_n} \cdot \mathbf{u}_{l,\text{rad}} \, dz \int \mathbf{u}_{l,\text{rad}}^* \cdot \epsilon_{\mathbf{G}_m}  \cdot\mathbf{u}_{m} \, dz}{\gamma_{l,rad}}}
         =\vec{p}_n\vec{p}_m \sum_{l}{\frac{\int u_{n}^*\cdot \epsilon_{-\mathbf{G}_n}\cdot u_{l,\text{rad}} \, dz \int u_{l,\text{rad}}^* \cdot \epsilon_{\mathbf{G}_m}\cdot u_{m} \, dz} {\gamma_{l,rad}}} 
\end{align}

\section{Wavectors of the $\Gamma$ points and $K$ points for guided modes inside the light cone}\label{appendix:wavevector}
The six $\Gamma$ points corresponding to $\ket{\Gamma_n}$, and the three $K$ points corresponding to  $\ket{K_n}$,shown in Fig.~\ref{fig:BZs}, are of wavevectors:
\begin{equation}
\begin{cases} 
    \vec{G_1}&=\vec{b_1} = b\left(\frac{1}{2}, \frac{\sqrt{3}}{2} \right)\\
    \vec{G_2}&=\vec{b_1}+\vec{b_2} = b\left(1, 0 \right)\\
    \vec{G_3}&=\vec{b_2}= b\left(\frac{1}{2}, -\frac{\sqrt{3}}{2} \right)\\
    \vec{G_4}&=-\vec{b_1}= b\left(-\frac{1}{2}, -\frac{\sqrt{3}}{2} \right)\\
    \vec{G_5}&=-\vec{b_1}-\vec{b_2} = b\left(-1, 0 \right)\\
    \vec{G_6}&=-\vec{b_2}= b\left(-\frac{1}{2}, -\frac{\sqrt{3}}{2} \right)
\end{cases} \quad\quad
\text{and} \quad\quad
\begin{cases} 
    \vec{K_1}&=-\vec{b_2}+\vec{K} = b\left(0, \frac{2}{\sqrt{3}} \right)\\
    \vec{K_2}&=\vec{b_2}+\vec{K} = b\left(1,-\frac{1}{\sqrt{3}} \right)\\
    \vec{K_3}&=-2\vec{b_1}-\vec{b_2}+\vec{K}= b\left(-1,-\frac{1}{\sqrt{3}} \right)
\end{cases}  
\label{eq:Gm}
\end{equation}

\section{Energy of $\ket{\Gamma_n}$ and $\ket{K_n}$ out of $\Gamma$ and $K$ points}\label{appendix:energy}
Using the approximation:
\begin{equation}
\begin{split}
      \omega_{\ket{\Gamma_n}}(\vec{k}) &=v_\Gamma\left(\abs{\vec{\Gamma_n}+\vec{k}}-\abs{\vec{\Gamma_n}}\right)\approx \omega_\Gamma+ v_\Gamma\frac{\vec{\Gamma_n}\cdot\vec{k}}{\abs{\vec{\Gamma_n}}}, \\
    \omega_{\ket{K_n}}(\vec{k})  &=v_K\left(\abs{\vec{K_n}+\vec{k}}-\abs{\vec{K_n}}\right)\approx \omega_K + v_K\frac{\vec{K_n}\cdot\vec{k}}{\abs{\vec{K_n}}}
\end{split}
\label{eq:approximation}
 \end{equation}

We obtain:
\begin{equation}
      \begin{cases}
        \omega_{\ket{G_1}}(\vec{k})&\simeq\omega_\Gamma + v_\Gamma k\cos{\left(\varphi-\frac{\pi}{3}\right)}\\
        \omega_{\ket{G_2}}(\vec{k})&\simeq\omega_\Gamma + v_\Gamma k\cos\varphi\\
        \omega_{\ket{G_3}}(\vec{k})&\simeq\omega_\Gamma + v_\Gamma k\cos{\left(\varphi+\frac{\pi}{3}\right)}\\
        \omega_{\ket{G_4}}(\vec{k})&\simeq\omega_\Gamma - v_\Gamma k\cos{\left(\varphi-\frac{\pi}{3}\right)}\\
        \omega_{\ket{G_5}}(\vec{k})&\simeq\omega_\Gamma - v_\Gamma k\cos\varphi\\
        \omega_{\ket{G_6}}(\vec{k})&\simeq\omega_\Gamma - v_\Gamma k\cos{\left(\varphi+\frac{\pi}{3}\right)}
    \end{cases} \quad\quad
\text{and} \quad\quad
\begin{cases}
        \omega_{\ket{K_1}}(\vec{k})&\simeq\omega_K + v_K k\sin\varphi\\
        \omega_{\ket{K_2}}(\vec{k})&\simeq\omega_K - v_K k\sin{\left(\varphi-\frac{\pi}{3}\right)}\\
        \omega_{\ket{K_3}}(\vec{k})&\simeq\omega_K - v_K k\sin{\left(\varphi+\frac{\pi}{3}\right)}
    \end{cases}
\label{eq:omega_GmKm_explicit} 
\end{equation}

\section{Polarization of $\ket{\Gamma_n}$ and $\ket{K_n}$}\label{appendix:polar}
Using the condition $\vec{p_{\ket{\Gamma_n}}}\cdot\vec{\Gamma_n}=\vec{p_{\ket{K_n}}}\cdot\vec{K_n}=0$ and Eq.\eqref{eq:Gm}, we obtain:
\begin{equation}
  \begin{cases}
    \vec{p_{\ket{G_1}}}&=\left(-\frac{\sqrt{3}}{2}, \frac{1}{2} \right)\\
    \vec{p_{\ket{G_2}}}&=\left(0, 1 \right)\\
    \vec{p_{\ket{G_3}}}&=\left(\frac{\sqrt{3}}{2}, \frac{1}{2} \right)\\
    \vec{p_{\ket{G_4}}}&=\left(\frac{\sqrt{3}}{2}, -\frac{1}{2} \right)\\
    \vec{p_{\ket{G_5}}}&=\left(0, -1 \right)\\
    \vec{p_{\ket{G_6}}}&=\left(-\frac{\sqrt{3}}{2}, -\frac{1}{2} \right)
\end{cases}\quad\quad
\text{and} \quad\quad
\begin{cases}
    \vec{p_{\ket{K_1}}}&=\left(-1, 0 \right)\\
    \vec{p_{\ket{K_2}}}&=\left(\frac{1}{2},\frac{\sqrt{3}}{2} \right)\\
    \vec{p_{\ket{K_3}}}&=\left(\frac{1}{2},-\frac{\sqrt{3}}{2} \right)
\end{cases}\label{eq:pol_guidedmodes}
\end{equation} 

\section{Polarization Texture of the Farfield}\label{sec:appendix_polarization}

Using the effective non-Hermitian Hamiltonian, the farfield electric field of the eigenmodes is computed via Eq.~\eqref{eq:FF_E}. For a given eigenmode, the farfield electric field is expressed as  $\vec{E}^{\text{farfield}}(\vec{k}) = E_x(\vec{k})\,\hat{\mathbf{x}} + E_y(\vec{k})\,\hat{\mathbf{y}}$, where $E_x(\vec{k})$ and $E_y(\vec{k})$ are the complex field components in the Cartesian basis. The polarization orientation, defined as the angle $\phi$ between $\vec{E}^{\text{farfield}}(\vec{k})$ and the $\hat{\mathbf{x}}$ axis, is given by~\cite{Dennis2002}:
\begin{equation}\label{eq:phi}
    \tan 2\phi = \frac{2\,\text{Re}(E_{x}^*E_{y})}{|E_{x}|^2 - |E_{y}|^2}
\end{equation}
The ellipticity of the polarization, characterized by the angle $\chi$, is given by~\cite{Dennis2002}:
\begin{equation}\label{eq:chi}
    \sin 2\chi = \frac{2\,\text{Im}(E_{x}^*E_{y})}{|E_{x}|^2 + |E_{y}|^2}
\end{equation}

In the presence of a polarization singularity at $\vec{k} = \vec{k}_0$, the associated topological charge is defined as the winding number of the polarization orientation $\phi(\vec{k})$ around a closed contour $\mathcal{C}$ encircling $\vec{k}_0$:
\begin{equation}\label{eq:topological_charge_polarization}
   q = \frac{1}{2\pi} \oint_\mathcal{C} d\vec{k} \cdot \nabla_{\vec{k}} \phi(\vec{k})
\end{equation}

\begin{figure}[h!]
 \begin{center}
\includegraphics[width=1\linewidth]{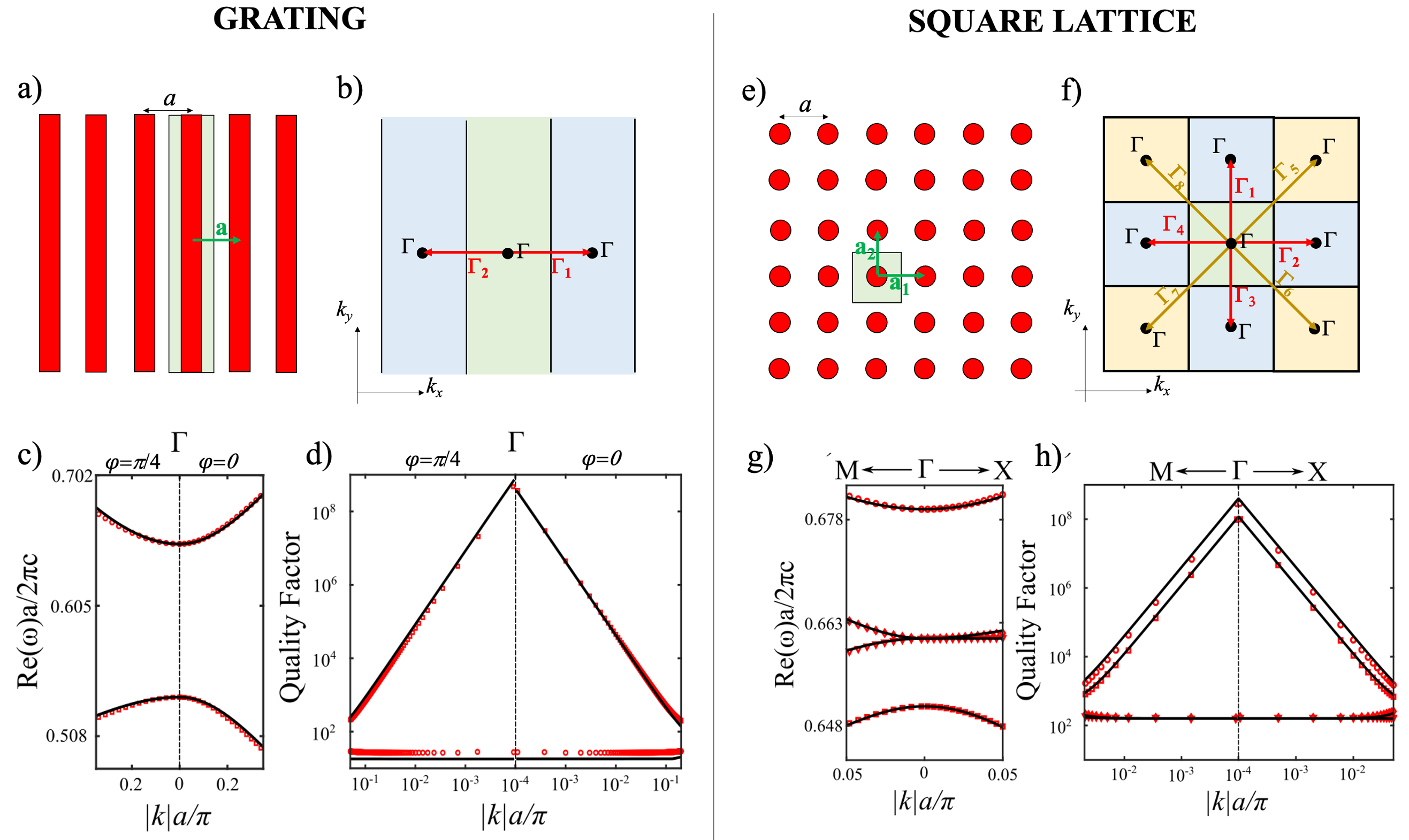}
\caption{\hsn{\textbf{Grating and square lattice}. 
a,e) Geometries of the two photonic crystal slabs. 
b,f) Guided-mode bases for the eigenmodes at the $\Gamma$ point above the light cone: two guided modes $\ket{\Gamma_{1,2}}$ for the 1D grating, and four guided modes $\ket{\Gamma_{1\rightarrow4}}$ for the square lattice. The green region indicates the first Brillouin zone, the blue regions indicate the second Brillouin zone, and the yellow region indicates the third Brillouin zone. 
c,g) Real parts of the photonic band energies as functions of the in-plane wavevector. 
d,h) Quality factors of the photonic bands. Red markers show the numerically simulated complex eigenfrequencies, while black lines show the corresponding analytical bands obtained from the effective non-Hermitian Hamiltonian.}}
\label{app_fig}
\end{center}
\end{figure}

\hsn{\section{Non-Hermitian Hamiltonian for 1D Grating and Square-Lattice Photonic Crystal Slabs}
\label{sec:appendix_Applicability}

To demonstrate the broad applicability of the generalized guided-mode expansion framework, we extend our analysis to two additional photonic crystal slab geometries: a one-dimensional (1D) grating (Fig.~\ref{app_fig}a) and a two-dimensional (2D) square lattice (Fig.~\ref{app_fig}b). These structures possess distinct point-group symmetries—$C_2$ for the grating and $C_4$ for the square lattice—thereby providing complementary test cases to the triangular and honeycomb lattices discussed in the main text. Applying the same formalism to these different symmetry classes allows us to verify the generality and robustness of the effective non-Hermitian Hamiltonian approach.

\subsection{Grating}

For a one-dimensional photonic grating, the lowest photonic bands at the $\Gamma$ point can be described by a basis consisting of two counter-propagating guided modes located at the two equivalent $\Gamma$ points of the second Brillouin zone (see Fig.~\ref{app_fig}b). These modes, denoted $\ket{\Gamma_1}$ and $\ket{\Gamma_2}$, both originate from the fundamental TE-guided mode of the unpatterned slab $\epsilon_0(z)$ and possess in-plane wavevectors $\vec{\Gamma}_1$ and $\vec{\Gamma}_2$.

Applying the general expressions derived in Sec.~\ref{sec:compact_expression}, and using the $C_2$ symmetry of the unit cell, we obtain the effective non-Hermitian Hamiltonian in the vicinity of the $\Gamma$ point:

\begin{align}
\hat{H}_{\Gamma}(\vec{k})=\omega_{\Gamma}
+ 
\begin{pmatrix}
v_{\Gamma}\!\left[k\cos{\varphi}+\frac{(k\sin{\varphi})^{2}}{2}\right] & U \\
U & v_{\Gamma}\!\left[k\cos(\pi-{\varphi})+\frac{(k\sin{\varphi})^{2}}{2}\right]
\end{pmatrix}
-i\gamma_0 
\begin{pmatrix}
1 & -1\\
-1 & 1
\end{pmatrix}.
\label{eq:H_K_Grating}
\end{align}

Here, the diffractive coupling $U$ is governed by the second-order Fourier components $\epsilon_{2}=\epsilon_{-2}$ of $\Delta\epsilon$, while the radiative loss rate $\gamma_0$ is governed by the first-order components $\epsilon_{1}=\epsilon_{-1}$.

Diagonalizing Eq.~\eqref{eq:H_K_Grating} yields two eigenmodes with eigenvalues:
\begin{equation}
\Omega_1^{(\Gamma)} = \omega_{\Gamma} + U, \qquad
\Omega_2^{(\Gamma)} = \omega_{\Gamma} - U - 2i\gamma_0,
\label{H_Grating:values_Gamma}
\end{equation}
and normalized eigenvectors:
\begin{equation}
\mathbf{A}_1^{(\Gamma)} = \tfrac{1}{\sqrt{2}}(1,1), \qquad
\mathbf{A}_2^{(\Gamma)} = \tfrac{1}{\sqrt{2}}(-1,1).
\label{H_Grating:eigmodes_k=0}
\end{equation}

These results are consistent with Refs.~\cite{Lee_2019,Lu2020}, where the two lowest band edges correspond to a symmetry-protected BIC (dark mode) and a bright mode. The sign of $U$, controlled by the sign of $\epsilon_2$ (e.g., via filling fraction), determines whether band inversion occurs.

\subsection{Square lattice}

For the square lattice, the lowest photonic bands near the $\Gamma$ point arise from four guided modes associated with the four $\Gamma$ points of the second Brillouin zone (see Fig.~\ref{app_fig}f). These modes form two pairs of counter-propagating states, $(\ket{\Gamma_1},\ket{\Gamma_3})$ and $(\ket{\Gamma_2},\ket{\Gamma_4})$, each derived from the fundamental TE-guided mode of the slab.

For a unit cell with $C_4$ symmetry, the effective non-Hermitian Hamiltonian in the basis $\{\ket{\Gamma_1},\dots,\ket{\Gamma_4}\}$ takes the form:

\begin{align}
\hat{H}_{\Gamma}(\vec{k})=\omega_{\Gamma}
+
\begin{pmatrix}
v_{\Gamma}k\cos(\varphi-\frac{\pi}{2}) & V & W & V \\
V & v_{\Gamma}k\cos(\varphi) & V & W \\
W & V & -v_{\Gamma}k\cos(\varphi-\frac{\pi}{2}) & V \\
V & W & V & -v_{\Gamma}k\cos(\varphi)
\end{pmatrix}
-i\gamma_0
\begin{pmatrix}
1 & 0 & -1 & 0 \\
0 & 1 & 0 & -1 \\
-1 & 0 & 1 & 0 \\
0 & -1 & 0 & 1
\end{pmatrix}.
\label{eq:H_Gamma_Square}
\end{align}

The diffractive coupling $V$ between counter-propagating guided modes is governed by the second-order Fourier components $\epsilon_{\pm 2,0}$ and $\epsilon_{0,\pm 2}$, while the radiative loss $\gamma_0$ is governed by $\epsilon_{\pm 1,0}$ and $\epsilon_{0,\pm 1}$.

In contrast, the coupling $W$ between orthogonally propagating modes vanishes at first order due to orthogonal polarization, but appears via second-order processes involving guided modes at the four $\Gamma$ points of the third Brillouin zone ($n=5\rightarrow 8$), consistent with Ref.~\cite{Liang2011}:
\begin{equation}
W = 
\sum_{n=5}^{8}
\bra{\Gamma_1}\Delta\epsilon\ket{\Gamma_n}
\bra{\Gamma_n}\Delta\epsilon\ket{\Gamma_2}
=
\sum_{n=5}^{8}
\bra{\Gamma_3}\Delta\epsilon\ket{\Gamma_n}
\bra{\Gamma_n}\Delta\epsilon\ket{\Gamma_4}.
\end{equation}

This indicates that $W$ depends on combinations of first- and higher-order Fourier components such as 
$\epsilon_{\pm 1,\pm 3}$ and $\epsilon_{\pm 3,\pm 1}$.

Diagonalizing Eq.~\eqref{eq:H_Gamma_Square} yields four eigenmodes:
\begin{equation}
\begin{aligned}
\Omega_1^{(\Gamma)} &= \omega_{\Gamma} - 2V + W,\\
\Omega_2^{(\Gamma)} &= \omega_{\Gamma} + 2V + W,\\
\Omega_3^{(\Gamma)} &= \omega_{\Gamma} - W - 2i\gamma_0,\\
\Omega_4^{(\Gamma)} &= \omega_{\Gamma} - W - 2i\gamma_0,
\end{aligned}
\label{H_Square:values_Gamma}
\end{equation}
with corresponding normalized eigenvectors:
\begin{equation}
\mathbf{A}_1^{(\Gamma)}=(-1,1,-1,1),\qquad
\mathbf{A}_2^{(\Gamma)}=(1,1,1,1),\qquad
\mathbf{A}_3^{(\Gamma)}=(0,-1,0,1),\qquad
\mathbf{A}_4^{(\Gamma)}=(-1,0,1,0).
\label{H_Square:eigmodes_k=0}
\end{equation}

These results agree with previous analyses of square-lattice photonic crystal slabs~\cite{Liang2011,yuan2025}, where the band edges consist of two degenerate bright modes (here $\ket{\Omega_3^{(\Gamma)}}$ and $\ket{\Omega_4^{(\Gamma)}}$) and two symmetry-protected BICs corresponding to monopolar and quadrupolar field profiles (here $\ket{\Omega_2^{(\Gamma)}}$ and $\ket{\Omega_1^{(\Gamma)}}$).

\subsection{Effective Theory vs Numerical Simulations near the $\Gamma$ Point}

We consider dielectric slabs ($n=2.5$) patterned into a 1D grating or a square lattice, with lattice constant $a=300$ nm and slab thickness $h=100$ nm. The grating consists of air grooves of width $w=0.3a$, while the square lattice consists of circular air holes of diameter $D=0.4a$.

Figures~\ref{app_fig}(c,d) and \ref{app_fig}(g,h) present the complex band structures obtained from full-wave simulations and from the analytical Hamiltonian. The agreement is excellent for both the real and imaginary parts of the eigenfrequencies, demonstrating that the generalized Hamiltonian framework extends naturally to other lattice geometries beyond the triangular and honeycomb cases discussed in the main text.

These examples confirm that the proposed effective Hamiltonian is broadly applicable to diverse photonic lattices and accurately captures both dispersion and radiation characteristics.}

\section{Fitting Parameters}\label{appendix:fitting_para}

All parameters are normalized to the lattice constant. They were used to fit the data shown in Fig.~\ref{fig:Figure_5}, Fig.~\ref{fig:figure_9}, Fig.~\ref{fig:figure10} and Fig.~\ref{app_fig}, with axis units specified in each respective figure. The corresponding fitting parameters are listed in the tables below:

\vspace{1em}
\noindent\textbf{1. Fitting parameters for photonic modes near the $\Gamma$ point (see Fig.~\ref{fig:Figure_5}):}
\begin{center}
\begin{tabular}{l*{6}{c}}
 & $\omega_{\Gamma}$ & $v_{\Gamma}$ & $V$ & $W$ & $U$ & $\gamma_{0}$\\ \midrule
Triangular lattice & 0.6968 & 0.3228 & 1.54E-2 & -1.4E-2 & -2.13E-2 & 2.26E-3 \\
Honeycomb lattice & 0.5035 & 0.2033 & -3.33E-3 & -6.18E-3 & 5.83E-3 & 3.87E-4 \\
\end{tabular}
\end{center}

\vspace{1em}
\noindent\textbf{2. Fitting parameters for photonic modes near the $K$ point in triangular lattices with elliptical holes (see Fig.~\ref{fig:figure_9}):}
\begin{center}
\begin{tabular}{l*{7}{c}}
 & $\omega_{K}$ & $v_{K}$ & $T_{1}$ & $T_{2}$ & $\gamma_{1}$ & $\gamma_{2}$ & $\gamma_{3}$\\ \midrule
$p = q = 60$ nm & 0.7704 & -0.3333 & -8.14E-3 & -8.14E-3 & 1.75E-3 & 1.75E-3 & 1.75E-3 \\
$p = 50$ nm, $q = 70$ nm & 0.7699 & -0.3333 & -6.9E-3 & -8.11E-3 & 2.4E-3 & 1.3E-3 & 1.7E-3 \\
$p = 40$ nm, $q = 80$ nm & 0.7686 & -0.3333 & -4.81E-3 & -7.4E-3 & 2.8E-3 & 8E-4 & 1E-3 \\
\end{tabular}
\end{center}

\vspace{1em}
\noindent\textbf{3. Fitting parameters for photonic modes near the $K$ point in honeycomb lattices with varying hole sizes (see Fig.~\ref{fig:figure10}):}
\begin{center}
\begin{tabular}{l*{5}{c}}
 & $\omega_{K}$ & $v_{K}$ & $T$ & $\gamma_{1}$ & $\gamma_{2}$ \\ \midrule
$r_{1} = r_{2} = 50$ nm & 0.7823 & -0.1666 & 8.94E-3 & 7.3E-4 & 2.43E-3 \\
$r_{1} = 50$ nm, $r_{2} = 70$ nm & 0.7852 & -0.1666 & 9.64E-3 + $i$8E-4 & 1.2E-3 & 1.2E-5 \\
$r_{1} = 40$ nm, $r_{2} = 80$ nm & 0.783 & -0.1666 & 8.8E-3 + $i$3.4E-3 & 1.2E-3 & 1.2E-5 \\
\end{tabular}
\end{center}

\hsn{\vspace{1em}
\noindent\textbf{4. Fitting parameters for photonic modes near the $\Gamma$ point in 1D grating (see Fig.~\ref{app_fig})}:
\begin{center}
\begin{tabular}{l*{4}{c}}
$\omega_{\Gamma}$ & $v_{\Gamma}$ & $U$ & $\gamma$ \\ \midrule
0.5939 & 0.218 & -0.0568 & 0.0194 \\
\end{tabular}
\end{center}

\vspace{1em}
\noindent\textbf{5. Fitting parameters for photonic modes near the $\Gamma$ point in square lattice (see Fig.~\ref{app_fig})}:
\begin{center}
\begin{tabular}{l*{5}{c}} 
$\omega_{\Gamma}$ & $v_{\Gamma}$ & $V$ & $W$ &$\gamma$  \\ \midrule
0.663 & 0.16 & -0.0069 & -0.00216 & 0.0021 \\
\end{tabular}
\end{center}}

\twocolumngrid
\bibliography{bib} 

\begin{thebibliography}{56}%
\makeatletter
\providecommand \@ifxundefined [1]{%
 \@ifx{#1\undefined}
}%
\providecommand \@ifnum [1]{%
 \ifnum #1\expandafter \@firstoftwo
 \else \expandafter \@secondoftwo
 \fi
}%
\providecommand \@ifx [1]{%
 \ifx #1\expandafter \@firstoftwo
 \else \expandafter \@secondoftwo
 \fi
}%
\providecommand \natexlab [1]{#1}%
\providecommand \enquote  [1]{``#1''}%
\providecommand \bibnamefont  [1]{#1}%
\providecommand \bibfnamefont [1]{#1}%
\providecommand \citenamefont [1]{#1}%
\providecommand \href@noop [0]{\@secondoftwo}%
\providecommand \href [0]{\begingroup \@sanitize@url \@href}%
\providecommand \@href[1]{\@@startlink{#1}\@@href}%
\providecommand \@@href[1]{\endgroup#1\@@endlink}%
\providecommand \@sanitize@url [0]{\catcode `\\12\catcode `\$12\catcode `\&12\catcode `\#12\catcode `\^12\catcode `\_12\catcode `\%12\relax}%
\providecommand \@@startlink[1]{}%
\providecommand \@@endlink[0]{}%
\providecommand \url  [0]{\begingroup\@sanitize@url \@url }%
\providecommand \@url [1]{\endgroup\@href {#1}{\urlprefix }}%
\providecommand \urlprefix  [0]{URL }%
\providecommand \Eprint [0]{\href }%
\providecommand \doibase [0]{https://doi.org/}%
\providecommand \selectlanguage [0]{\@gobble}%
\providecommand \bibinfo  [0]{\@secondoftwo}%
\providecommand \bibfield  [0]{\@secondoftwo}%
\providecommand \translation [1]{[#1]}%
\providecommand \BibitemOpen [0]{}%
\providecommand \bibitemStop [0]{}%
\providecommand \bibitemNoStop [0]{.\EOS\space}%
\providecommand \EOS [0]{\spacefactor3000\relax}%
\providecommand \BibitemShut  [1]{\csname bibitem#1\endcsname}%
\let\auto@bib@innerbib\@empty
\bibitem [{\citenamefont {Johnson}\ \emph {et~al.}(1999)\citenamefont {Johnson}, \citenamefont {Fan}, \citenamefont {Villeneuve}, \citenamefont {Joannopoulos},\ and\ \citenamefont {Kolodziejski}}]{Johnson1999}%
  \BibitemOpen
  \bibfield  {author} {\bibinfo {author} {\bibfnamefont {S.~G.}\ \bibnamefont {Johnson}}, \bibinfo {author} {\bibfnamefont {S.}~\bibnamefont {Fan}}, \bibinfo {author} {\bibfnamefont {P.~R.}\ \bibnamefont {Villeneuve}}, \bibinfo {author} {\bibfnamefont {J.~D.}\ \bibnamefont {Joannopoulos}},\ and\ \bibinfo {author} {\bibfnamefont {L.~A.}\ \bibnamefont {Kolodziejski}},\ }\bibfield  {title} {\bibinfo {title} {Guided modes in photonic crystal slabs},\ }\href {https://doi.org/10.1103/physrevb.60.5751} {\bibfield  {journal} {\bibinfo  {journal} {Physical Review B}\ }\textbf {\bibinfo {volume} {60}},\ \bibinfo {pages} {5751–5758} (\bibinfo {year} {1999})}\BibitemShut {NoStop}%
\bibitem [{\citenamefont {Imada}\ \emph {et~al.}(2002)\citenamefont {Imada}, \citenamefont {Chutinan}, \citenamefont {Noda},\ and\ \citenamefont {Mochizuki}}]{Imada2002}%
  \BibitemOpen
  \bibfield  {author} {\bibinfo {author} {\bibfnamefont {M.}~\bibnamefont {Imada}}, \bibinfo {author} {\bibfnamefont {A.}~\bibnamefont {Chutinan}}, \bibinfo {author} {\bibfnamefont {S.}~\bibnamefont {Noda}},\ and\ \bibinfo {author} {\bibfnamefont {M.}~\bibnamefont {Mochizuki}},\ }\bibfield  {title} {\bibinfo {title} {Multidirectionally distributed feedback photonic crystal lasers},\ }\bibfield  {journal} {\bibinfo  {journal} {Physical Review B}\ }\textbf {\bibinfo {volume} {65}},\ \href {https://doi.org/10.1103/physrevb.65.195306} {10.1103/physrevb.65.195306} (\bibinfo {year} {2002})\BibitemShut {NoStop}%
\bibitem [{\citenamefont {Sakoda}(2001)}]{Sakoda2001}%
  \BibitemOpen
  \bibfield  {author} {\bibinfo {author} {\bibfnamefont {K.}~\bibnamefont {Sakoda}},\ }\bibinfo {title} {Photonic crystal slabs},\ in\ \href {https://doi.org/10.1007/978-3-662-14324-7_8} {\emph {\bibinfo {booktitle} {Optical Properties of Photonic Crystals}}}\ (\bibinfo  {publisher} {Springer Berlin Heidelberg},\ \bibinfo {year} {2001})\ p.\ \bibinfo {pages} {177–188}\BibitemShut {NoStop}%
\bibitem [{\citenamefont {Malek}\ \emph {et~al.}(2020)\citenamefont {Malek}, \citenamefont {Overvig}, \citenamefont {Shrestha},\ and\ \citenamefont {Yu}}]{Malek2020}%
  \BibitemOpen
  \bibfield  {author} {\bibinfo {author} {\bibfnamefont {S.~C.}\ \bibnamefont {Malek}}, \bibinfo {author} {\bibfnamefont {A.~C.}\ \bibnamefont {Overvig}}, \bibinfo {author} {\bibfnamefont {S.}~\bibnamefont {Shrestha}},\ and\ \bibinfo {author} {\bibfnamefont {N.}~\bibnamefont {Yu}},\ }\bibfield  {title} {\bibinfo {title} {Active nonlocal metasurfaces},\ }\href {https://doi.org/10.1515/nanoph-2020-0375} {\bibfield  {journal} {\bibinfo  {journal} {Nanophotonics}\ }\textbf {\bibinfo {volume} {10}},\ \bibinfo {pages} {655–665} (\bibinfo {year} {2020})}\BibitemShut {NoStop}%
\bibitem [{\citenamefont {Overvig}\ and\ \citenamefont {Alù}(2022)}]{Overvig2022}%
  \BibitemOpen
  \bibfield  {author} {\bibinfo {author} {\bibfnamefont {A.}~\bibnamefont {Overvig}}\ and\ \bibinfo {author} {\bibfnamefont {A.}~\bibnamefont {Alù}},\ }\bibfield  {title} {\bibinfo {title} {Diffractive nonlocal metasurfaces},\ }\bibfield  {journal} {\bibinfo  {journal} {Laser and Photonics Reviews}\ }\textbf {\bibinfo {volume} {16}},\ \href {https://doi.org/10.1002/lpor.202100633} {10.1002/lpor.202100633} (\bibinfo {year} {2022})\BibitemShut {NoStop}%
\bibitem [{\citenamefont {Chen}\ \emph {et~al.}(2025)\citenamefont {Chen}, \citenamefont {Fleury}, \citenamefont {Seppecher}, \citenamefont {Hu},\ and\ \citenamefont {Wegener}}]{Chen2025}%
  \BibitemOpen
  \bibfield  {author} {\bibinfo {author} {\bibfnamefont {Y.}~\bibnamefont {Chen}}, \bibinfo {author} {\bibfnamefont {R.}~\bibnamefont {Fleury}}, \bibinfo {author} {\bibfnamefont {P.}~\bibnamefont {Seppecher}}, \bibinfo {author} {\bibfnamefont {G.}~\bibnamefont {Hu}},\ and\ \bibinfo {author} {\bibfnamefont {M.}~\bibnamefont {Wegener}},\ }\bibfield  {title} {\bibinfo {title} {Nonlocal metamaterials and metasurfaces},\ }\href {https://doi.org/10.1038/s42254-025-00829-1} {\bibfield  {journal} {\bibinfo  {journal} {Nature Reviews Physics}\ }\textbf {\bibinfo {volume} {7}},\ \bibinfo {pages} {299–312} (\bibinfo {year} {2025})}\BibitemShut {NoStop}%
\bibitem [{\citenamefont {Monticone}\ \emph {et~al.}(2025)\citenamefont {Monticone}, \citenamefont {Mortensen}, \citenamefont {Fern{\'a}ndez-Dom{\'i}nguez},\ and\ \citenamefont {et~al}}]{Monticone2025}%
  \BibitemOpen
  \bibfield  {author} {\bibinfo {author} {\bibfnamefont {F.}~\bibnamefont {Monticone}}, \bibinfo {author} {\bibfnamefont {N.}~\bibnamefont {Mortensen}}, \bibinfo {author} {\bibfnamefont {A.}~\bibnamefont {Fern{\'a}ndez-Dom{\'i}nguez}},\ and\ \bibinfo {author} {\bibnamefont {et~al}},\ }\bibfield  {title} {{\selectlanguage {English}\bibinfo {title} {Nonlocality in photonic materials and metamaterials: roadmap}},\ }\href {https://doi.org/10.1364/ome.559374} {\bibfield  {journal} {\bibinfo  {journal} {Optical Materials Express}\ }\textbf {\bibinfo {volume} {15}},\ \bibinfo {pages} {1544} (\bibinfo {year} {2025})},\ \bibinfo {note} {publisher Copyright: Journal {\textcopyright} 2025.}\BibitemShut {Stop}%
\bibitem [{\citenamefont {Hsu}\ \emph {et~al.}(2016)\citenamefont {Hsu}, \citenamefont {Zhen}, \citenamefont {Stone}, \citenamefont {Joannopoulos},\ and\ \citenamefont {Soljačić}}]{Hsu2016}%
  \BibitemOpen
  \bibfield  {author} {\bibinfo {author} {\bibfnamefont {C.~W.}\ \bibnamefont {Hsu}}, \bibinfo {author} {\bibfnamefont {B.}~\bibnamefont {Zhen}}, \bibinfo {author} {\bibfnamefont {A.~D.}\ \bibnamefont {Stone}}, \bibinfo {author} {\bibfnamefont {J.~D.}\ \bibnamefont {Joannopoulos}},\ and\ \bibinfo {author} {\bibfnamefont {M.}~\bibnamefont {Soljačić}},\ }\bibfield  {title} {\bibinfo {title} {Bound states in the continuum},\ }\bibfield  {journal} {\bibinfo  {journal} {Nature Reviews Materials}\ }\textbf {\bibinfo {volume} {1}},\ \href {https://doi.org/10.1038/natrevmats.2016.48} {10.1038/natrevmats.2016.48} (\bibinfo {year} {2016})\BibitemShut {NoStop}%
\bibitem [{\citenamefont {Doeleman}\ \emph {et~al.}(2018)\citenamefont {Doeleman}, \citenamefont {Monticone}, \citenamefont {den Hollander}, \citenamefont {Al{\`u}},\ and\ \citenamefont {Koenderink}}]{Doeleman2018}%
  \BibitemOpen
  \bibfield  {author} {\bibinfo {author} {\bibfnamefont {H.~M.}\ \bibnamefont {Doeleman}}, \bibinfo {author} {\bibfnamefont {F.}~\bibnamefont {Monticone}}, \bibinfo {author} {\bibfnamefont {W.}~\bibnamefont {den Hollander}}, \bibinfo {author} {\bibfnamefont {A.}~\bibnamefont {Al{\`u}}},\ and\ \bibinfo {author} {\bibfnamefont {A.~F.}\ \bibnamefont {Koenderink}},\ }\bibfield  {title} {\bibinfo {title} {Experimental observation of a polarization vortex at an optical bound state in the continuum},\ }\href {https://doi.org/10.1038/s41566-018-0177-5} {\bibfield  {journal} {\bibinfo  {journal} {Nature Photonics}\ }\textbf {\bibinfo {volume} {12}},\ \bibinfo {pages} {397} (\bibinfo {year} {2018})}\BibitemShut {NoStop}%
\bibitem [{\citenamefont {Kang}\ \emph {et~al.}(2023)\citenamefont {Kang}, \citenamefont {Liu}, \citenamefont {Chan},\ and\ \citenamefont {Xiao}}]{Kang2023}%
  \BibitemOpen
  \bibfield  {author} {\bibinfo {author} {\bibfnamefont {M.}~\bibnamefont {Kang}}, \bibinfo {author} {\bibfnamefont {T.}~\bibnamefont {Liu}}, \bibinfo {author} {\bibfnamefont {C.~T.}\ \bibnamefont {Chan}},\ and\ \bibinfo {author} {\bibfnamefont {M.}~\bibnamefont {Xiao}},\ }\bibfield  {title} {\bibinfo {title} {Applications of bound states in the continuum in photonics},\ }\href {https://doi.org/10.1038/s42254-023-00642-8} {\bibfield  {journal} {\bibinfo  {journal} {Nature Reviews Physics}\ }\textbf {\bibinfo {volume} {5}},\ \bibinfo {pages} {659–678} (\bibinfo {year} {2023})}\BibitemShut {NoStop}%
\bibitem [{\citenamefont {Mermet-Lyaudoz}\ \emph {et~al.}(2023)\citenamefont {Mermet-Lyaudoz}, \citenamefont {Symonds}, \citenamefont {Berry}, \citenamefont {Drouard}, \citenamefont {Chevalier}, \citenamefont {Trippé-Allard}, \citenamefont {Deleporte}, \citenamefont {Bellessa}, \citenamefont {Seassal},\ and\ \citenamefont {Nguyen}}]{MermetLyaudoz2023}%
  \BibitemOpen
  \bibfield  {author} {\bibinfo {author} {\bibfnamefont {R.}~\bibnamefont {Mermet-Lyaudoz}}, \bibinfo {author} {\bibfnamefont {C.}~\bibnamefont {Symonds}}, \bibinfo {author} {\bibfnamefont {F.}~\bibnamefont {Berry}}, \bibinfo {author} {\bibfnamefont {E.}~\bibnamefont {Drouard}}, \bibinfo {author} {\bibfnamefont {C.}~\bibnamefont {Chevalier}}, \bibinfo {author} {\bibfnamefont {G.}~\bibnamefont {Trippé-Allard}}, \bibinfo {author} {\bibfnamefont {E.}~\bibnamefont {Deleporte}}, \bibinfo {author} {\bibfnamefont {J.}~\bibnamefont {Bellessa}}, \bibinfo {author} {\bibfnamefont {C.}~\bibnamefont {Seassal}},\ and\ \bibinfo {author} {\bibfnamefont {H.~S.}\ \bibnamefont {Nguyen}},\ }\bibfield  {title} {\bibinfo {title} {Taming friedrich–wintgen interference in a resonant metasurface: Vortex laser emitting at an on-demand tilted angle},\ }\href {https://doi.org/10.1021/acs.nanolett.2c04936} {\bibfield  {journal} {\bibinfo  {journal} {Nano Letters}\ }\textbf {\bibinfo {volume} {23}},\ \bibinfo {pages} {4152–4159}
  (\bibinfo {year} {2023})}\BibitemShut {NoStop}%
\bibitem [{\citenamefont {Le}\ \emph {et~al.}(2024)\citenamefont {Le}, \citenamefont {Bouteyre}, \citenamefont {Kheir-Aldine}, \citenamefont {Dubois}, \citenamefont {Cueff}, \citenamefont {Berguiga}, \citenamefont {Letartre}, \citenamefont {Viktorovitch}, \citenamefont {Benyattou},\ and\ \citenamefont {Nguyen}}]{Le2024}%
  \BibitemOpen
  \bibfield  {author} {\bibinfo {author} {\bibfnamefont {N.~D.}\ \bibnamefont {Le}}, \bibinfo {author} {\bibfnamefont {P.}~\bibnamefont {Bouteyre}}, \bibinfo {author} {\bibfnamefont {A.}~\bibnamefont {Kheir-Aldine}}, \bibinfo {author} {\bibfnamefont {F.}~\bibnamefont {Dubois}}, \bibinfo {author} {\bibfnamefont {S.}~\bibnamefont {Cueff}}, \bibinfo {author} {\bibfnamefont {L.}~\bibnamefont {Berguiga}}, \bibinfo {author} {\bibfnamefont {X.}~\bibnamefont {Letartre}}, \bibinfo {author} {\bibfnamefont {P.}~\bibnamefont {Viktorovitch}}, \bibinfo {author} {\bibfnamefont {T.}~\bibnamefont {Benyattou}},\ and\ \bibinfo {author} {\bibfnamefont {H.~S.}\ \bibnamefont {Nguyen}},\ }\bibfield  {title} {\bibinfo {title} {Super bound states in the continuum on a photonic flatband: Concept, experimental realization, and optical trapping demonstration},\ }\bibfield  {journal} {\bibinfo  {journal} {Physical Review Letters}\ }\textbf {\bibinfo {volume} {132}},\ \href {https://doi.org/10.1103/physrevlett.132.173802}
  {10.1103/physrevlett.132.173802} (\bibinfo {year} {2024})\BibitemShut {NoStop}%
\bibitem [{\citenamefont {Miri}\ and\ \citenamefont {Alù}(2019)}]{Miri2019}%
  \BibitemOpen
  \bibfield  {author} {\bibinfo {author} {\bibfnamefont {M.-A.}\ \bibnamefont {Miri}}\ and\ \bibinfo {author} {\bibfnamefont {A.}~\bibnamefont {Alù}},\ }\bibfield  {title} {\bibinfo {title} {Exceptional points in optics and photonics},\ }\bibfield  {journal} {\bibinfo  {journal} {Science}\ }\textbf {\bibinfo {volume} {363}},\ \href {https://doi.org/10.1126/science.aar7709} {10.1126/science.aar7709} (\bibinfo {year} {2019})\BibitemShut {NoStop}%
\bibitem [{\citenamefont {Zhen}\ \emph {et~al.}(2015)\citenamefont {Zhen}, \citenamefont {Hsu}, \citenamefont {Igarashi}, \citenamefont {Lu}, \citenamefont {Kaminer}, \citenamefont {Pick}, \citenamefont {Chua}, \citenamefont {Joannopoulos},\ and\ \citenamefont {Soljačić}}]{Zhen2015}%
  \BibitemOpen
  \bibfield  {author} {\bibinfo {author} {\bibfnamefont {B.}~\bibnamefont {Zhen}}, \bibinfo {author} {\bibfnamefont {C.~W.}\ \bibnamefont {Hsu}}, \bibinfo {author} {\bibfnamefont {Y.}~\bibnamefont {Igarashi}}, \bibinfo {author} {\bibfnamefont {L.}~\bibnamefont {Lu}}, \bibinfo {author} {\bibfnamefont {I.}~\bibnamefont {Kaminer}}, \bibinfo {author} {\bibfnamefont {A.}~\bibnamefont {Pick}}, \bibinfo {author} {\bibfnamefont {S.-L.}\ \bibnamefont {Chua}}, \bibinfo {author} {\bibfnamefont {J.~D.}\ \bibnamefont {Joannopoulos}},\ and\ \bibinfo {author} {\bibfnamefont {M.}~\bibnamefont {Soljačić}},\ }\bibfield  {title} {\bibinfo {title} {Spawning rings of exceptional points out of dirac cones},\ }\href {https://doi.org/10.1038/nature14889} {\bibfield  {journal} {\bibinfo  {journal} {Nature}\ }\textbf {\bibinfo {volume} {525}},\ \bibinfo {pages} {354–358} (\bibinfo {year} {2015})}\BibitemShut {NoStop}%
\bibitem [{\citenamefont {Zhou}\ \emph {et~al.}(2018)\citenamefont {Zhou}, \citenamefont {Peng}, \citenamefont {Yoon}, \citenamefont {Hsu}, \citenamefont {Nelson}, \citenamefont {Fu}, \citenamefont {Joannopoulos}, \citenamefont {Soljačić},\ and\ \citenamefont {Zhen}}]{Zhou2018}%
  \BibitemOpen
  \bibfield  {author} {\bibinfo {author} {\bibfnamefont {H.}~\bibnamefont {Zhou}}, \bibinfo {author} {\bibfnamefont {C.}~\bibnamefont {Peng}}, \bibinfo {author} {\bibfnamefont {Y.}~\bibnamefont {Yoon}}, \bibinfo {author} {\bibfnamefont {C.~W.}\ \bibnamefont {Hsu}}, \bibinfo {author} {\bibfnamefont {K.~A.}\ \bibnamefont {Nelson}}, \bibinfo {author} {\bibfnamefont {L.}~\bibnamefont {Fu}}, \bibinfo {author} {\bibfnamefont {J.~D.}\ \bibnamefont {Joannopoulos}}, \bibinfo {author} {\bibfnamefont {M.}~\bibnamefont {Soljačić}},\ and\ \bibinfo {author} {\bibfnamefont {B.}~\bibnamefont {Zhen}},\ }\bibfield  {title} {\bibinfo {title} {Observation of bulk fermi arc and polarization half charge from paired exceptional points},\ }\href {https://doi.org/10.1126/science.aap9859} {\bibfield  {journal} {\bibinfo  {journal} {Science}\ }\textbf {\bibinfo {volume} {359}},\ \bibinfo {pages} {1009–1012} (\bibinfo {year} {2018})}\BibitemShut {NoStop}%
\bibitem [{\citenamefont {Ferrier}\ \emph {et~al.}(2022)\citenamefont {Ferrier}, \citenamefont {Bouteyre}, \citenamefont {Pick}, \citenamefont {Cueff}, \citenamefont {Dang}, \citenamefont {Diederichs}, \citenamefont {Belarouci}, \citenamefont {Benyattou}, \citenamefont {Zhao}, \citenamefont {Su}, \citenamefont {Xing}, \citenamefont {Xiong},\ and\ \citenamefont {Nguyen}}]{Ferrier2022}%
  \BibitemOpen
  \bibfield  {author} {\bibinfo {author} {\bibfnamefont {L.}~\bibnamefont {Ferrier}}, \bibinfo {author} {\bibfnamefont {P.}~\bibnamefont {Bouteyre}}, \bibinfo {author} {\bibfnamefont {A.}~\bibnamefont {Pick}}, \bibinfo {author} {\bibfnamefont {S.}~\bibnamefont {Cueff}}, \bibinfo {author} {\bibfnamefont {N.}~\bibnamefont {Dang}}, \bibinfo {author} {\bibfnamefont {C.}~\bibnamefont {Diederichs}}, \bibinfo {author} {\bibfnamefont {A.}~\bibnamefont {Belarouci}}, \bibinfo {author} {\bibfnamefont {T.}~\bibnamefont {Benyattou}}, \bibinfo {author} {\bibfnamefont {J.}~\bibnamefont {Zhao}}, \bibinfo {author} {\bibfnamefont {R.}~\bibnamefont {Su}}, \bibinfo {author} {\bibfnamefont {J.}~\bibnamefont {Xing}}, \bibinfo {author} {\bibfnamefont {Q.}~\bibnamefont {Xiong}},\ and\ \bibinfo {author} {\bibfnamefont {H.}~\bibnamefont {Nguyen}},\ }\bibfield  {title} {\bibinfo {title} {Unveiling the enhancement of spontaneous emission at exceptional points},\ }\bibfield  {journal} {\bibinfo  {journal} {Physical Review Letters}\
  }\textbf {\bibinfo {volume} {129}},\ \href {https://doi.org/10.1103/physrevlett.129.083602} {10.1103/physrevlett.129.083602} (\bibinfo {year} {2022})\BibitemShut {NoStop}%
\bibitem [{\citenamefont {Nasari}\ \emph {et~al.}(2022)\citenamefont {Nasari}, \citenamefont {Lopez-Galmiche}, \citenamefont {Lopez-Aviles}, \citenamefont {Schumer}, \citenamefont {Hassan}, \citenamefont {Zhong}, \citenamefont {Rotter}, \citenamefont {LiKamWa}, \citenamefont {Christodoulides},\ and\ \citenamefont {Khajavikhan}}]{Nasari2022}%
  \BibitemOpen
  \bibfield  {author} {\bibinfo {author} {\bibfnamefont {H.}~\bibnamefont {Nasari}}, \bibinfo {author} {\bibfnamefont {G.}~\bibnamefont {Lopez-Galmiche}}, \bibinfo {author} {\bibfnamefont {H.~E.}\ \bibnamefont {Lopez-Aviles}}, \bibinfo {author} {\bibfnamefont {A.}~\bibnamefont {Schumer}}, \bibinfo {author} {\bibfnamefont {A.~U.}\ \bibnamefont {Hassan}}, \bibinfo {author} {\bibfnamefont {Q.}~\bibnamefont {Zhong}}, \bibinfo {author} {\bibfnamefont {S.}~\bibnamefont {Rotter}}, \bibinfo {author} {\bibfnamefont {P.}~\bibnamefont {LiKamWa}}, \bibinfo {author} {\bibfnamefont {D.~N.}\ \bibnamefont {Christodoulides}},\ and\ \bibinfo {author} {\bibfnamefont {M.}~\bibnamefont {Khajavikhan}},\ }\bibfield  {title} {\bibinfo {title} {Observation of chiral state transfer without encircling an exceptional point},\ }\href {https://doi.org/10.1038/s41586-022-04542-2} {\bibfield  {journal} {\bibinfo  {journal} {Nature}\ }\textbf {\bibinfo {volume} {605}},\ \bibinfo {pages} {256} (\bibinfo {year} {2022})}\BibitemShut {NoStop}%
\bibitem [{\citenamefont {Nguyen}\ \emph {et~al.}(2023)\citenamefont {Nguyen}, \citenamefont {Le}, \citenamefont {Sarelli}, \citenamefont {Malgrey}, \citenamefont {Luu}, \citenamefont {Chu}, \citenamefont {Vu}, \citenamefont {Tong}, \citenamefont {Vu}, \citenamefont {Seassal}, \citenamefont {Le-Van},\ and\ \citenamefont {Nguyen}}]{Nguyen2023}%
  \BibitemOpen
  \bibfield  {author} {\bibinfo {author} {\bibfnamefont {V.~A.}\ \bibnamefont {Nguyen}}, \bibinfo {author} {\bibfnamefont {V.~H.}\ \bibnamefont {Le}}, \bibinfo {author} {\bibfnamefont {E.}~\bibnamefont {Sarelli}}, \bibinfo {author} {\bibfnamefont {L.}~\bibnamefont {Malgrey}}, \bibinfo {author} {\bibfnamefont {D.-K.}\ \bibnamefont {Luu}}, \bibinfo {author} {\bibfnamefont {H.~L.}\ \bibnamefont {Chu}}, \bibinfo {author} {\bibfnamefont {T.~T.}\ \bibnamefont {Vu}}, \bibinfo {author} {\bibfnamefont {C.~Q.}\ \bibnamefont {Tong}}, \bibinfo {author} {\bibfnamefont {D.~L.}\ \bibnamefont {Vu}}, \bibinfo {author} {\bibfnamefont {C.}~\bibnamefont {Seassal}}, \bibinfo {author} {\bibfnamefont {Q.}~\bibnamefont {Le-Van}},\ and\ \bibinfo {author} {\bibfnamefont {H.~S.}\ \bibnamefont {Nguyen}},\ }\bibfield  {title} {\bibinfo {title} {Direct observation of exceptional points in photonic crystal by cross-polarization imaging in momentum space},\ }\bibfield  {journal} {\bibinfo  {journal} {Applied Physics Letters}\ }\textbf
  {\bibinfo {volume} {123}},\ \href {https://doi.org/10.1063/5.0175024} {10.1063/5.0175024} (\bibinfo {year} {2023})\BibitemShut {NoStop}%
\bibitem [{\citenamefont {Fan}\ \emph {et~al.}(2003)\citenamefont {Fan}, \citenamefont {Suh},\ and\ \citenamefont {Joannopoulos}}]{Fan2003}%
  \BibitemOpen
  \bibfield  {author} {\bibinfo {author} {\bibfnamefont {S.}~\bibnamefont {Fan}}, \bibinfo {author} {\bibfnamefont {W.}~\bibnamefont {Suh}},\ and\ \bibinfo {author} {\bibfnamefont {J.~D.}\ \bibnamefont {Joannopoulos}},\ }\bibfield  {title} {\bibinfo {title} {Temporal coupled-mode theory for the fano resonance in optical resonators},\ }\href {https://doi.org/10.1364/josaa.20.000569} {\bibfield  {journal} {\bibinfo  {journal} {Journal of the Optical Society of America A}\ }\textbf {\bibinfo {volume} {20}},\ \bibinfo {pages} {569} (\bibinfo {year} {2003})}\BibitemShut {NoStop}%
\bibitem [{\citenamefont {Suh}\ \emph {et~al.}(2004)\citenamefont {Suh}, \citenamefont {Wang},\ and\ \citenamefont {Fan}}]{WonjooSuh2004}%
  \BibitemOpen
  \bibfield  {author} {\bibinfo {author} {\bibfnamefont {W.}~\bibnamefont {Suh}}, \bibinfo {author} {\bibfnamefont {Z.}~\bibnamefont {Wang}},\ and\ \bibinfo {author} {\bibfnamefont {S.}~\bibnamefont {Fan}},\ }\bibfield  {title} {\bibinfo {title} {Temporal coupled-mode theory and the presence of non-orthogonal modes in lossless multimode cavities},\ }\href {https://doi.org/10.1109/jqe.2004.834773} {\bibfield  {journal} {\bibinfo  {journal} {IEEE Journal of Quantum Electronics}\ }\textbf {\bibinfo {volume} {40}},\ \bibinfo {pages} {1511–1518} (\bibinfo {year} {2004})}\BibitemShut {NoStop}%
\bibitem [{\citenamefont {Letartre}\ \emph {et~al.}(2022)\citenamefont {Letartre}, \citenamefont {Mazauric}, \citenamefont {Cueff}, \citenamefont {Benyattou}, \citenamefont {Nguyen},\ and\ \citenamefont {Viktorovitch}}]{Letartre2022}%
  \BibitemOpen
  \bibfield  {author} {\bibinfo {author} {\bibfnamefont {X.}~\bibnamefont {Letartre}}, \bibinfo {author} {\bibfnamefont {S.}~\bibnamefont {Mazauric}}, \bibinfo {author} {\bibfnamefont {S.}~\bibnamefont {Cueff}}, \bibinfo {author} {\bibfnamefont {T.}~\bibnamefont {Benyattou}}, \bibinfo {author} {\bibfnamefont {H.~S.}\ \bibnamefont {Nguyen}},\ and\ \bibinfo {author} {\bibfnamefont {P.}~\bibnamefont {Viktorovitch}},\ }\bibfield  {title} {\bibinfo {title} {Analytical non-hermitian description of photonic crystals with arbitrary lateral and transverse symmetry},\ }\bibfield  {journal} {\bibinfo  {journal} {Physical Review A}\ }\textbf {\bibinfo {volume} {106}},\ \href {https://doi.org/10.1103/physreva.106.033510} {10.1103/physreva.106.033510} (\bibinfo {year} {2022})\BibitemShut {NoStop}%
\bibitem [{\citenamefont {Streifer}\ \emph {et~al.}(1977)\citenamefont {Streifer}, \citenamefont {Scifres},\ and\ \citenamefont {Burnham}}]{Streifer_1977}%
  \BibitemOpen
  \bibfield  {author} {\bibinfo {author} {\bibfnamefont {W.}~\bibnamefont {Streifer}}, \bibinfo {author} {\bibfnamefont {D.}~\bibnamefont {Scifres}},\ and\ \bibinfo {author} {\bibfnamefont {R.}~\bibnamefont {Burnham}},\ }\bibfield  {title} {\bibinfo {title} {Coupled wave analysis of dfb and dbr lasers},\ }\href {https://doi.org/10.1109/jqe.1977.1069328} {\bibfield  {journal} {\bibinfo  {journal} {IEEE Journal of Quantum Electronics}\ }\textbf {\bibinfo {volume} {13}},\ \bibinfo {pages} {134–141} (\bibinfo {year} {1977})}\BibitemShut {NoStop}%
\bibitem [{\citenamefont {Kazarinov}\ and\ \citenamefont {Henry}(1985)}]{Kazarinov_1985}%
  \BibitemOpen
  \bibfield  {author} {\bibinfo {author} {\bibfnamefont {R.}~\bibnamefont {Kazarinov}}\ and\ \bibinfo {author} {\bibfnamefont {C.}~\bibnamefont {Henry}},\ }\bibfield  {title} {\bibinfo {title} {Second-order distributed feedback lasers with mode selection provided by first-order radiation losses},\ }\href {https://doi.org/10.1109/jqe.1985.1072627} {\bibfield  {journal} {\bibinfo  {journal} {IEEE Journal of Quantum Electronics}\ }\textbf {\bibinfo {volume} {21}},\ \bibinfo {pages} {144–150} (\bibinfo {year} {1985})}\BibitemShut {NoStop}%
\bibitem [{\citenamefont {Lee}\ and\ \citenamefont {Magnusson}(2019)}]{Lee_2019}%
  \BibitemOpen
  \bibfield  {author} {\bibinfo {author} {\bibfnamefont {S.-G.}\ \bibnamefont {Lee}}\ and\ \bibinfo {author} {\bibfnamefont {R.}~\bibnamefont {Magnusson}},\ }\bibfield  {title} {\bibinfo {title} {Band flips and bound-state transitions in leaky-mode photonic lattices},\ }\bibfield  {journal} {\bibinfo  {journal} {Physical Review B}\ }\textbf {\bibinfo {volume} {99}},\ \href {https://doi.org/10.1103/physrevb.99.045304} {10.1103/physrevb.99.045304} (\bibinfo {year} {2019})\BibitemShut {NoStop}%
\bibitem [{\citenamefont {Nasari}\ \emph {et~al.}(2023)\citenamefont {Nasari}, \citenamefont {Pyrialakos}, \citenamefont {Christodoulides},\ and\ \citenamefont {Khajavikhan}}]{Nasari2023}%
  \BibitemOpen
  \bibfield  {author} {\bibinfo {author} {\bibfnamefont {H.}~\bibnamefont {Nasari}}, \bibinfo {author} {\bibfnamefont {G.~G.}\ \bibnamefont {Pyrialakos}}, \bibinfo {author} {\bibfnamefont {D.~N.}\ \bibnamefont {Christodoulides}},\ and\ \bibinfo {author} {\bibfnamefont {M.}~\bibnamefont {Khajavikhan}},\ }\bibfield  {title} {\bibinfo {title} {Non-hermitian topological photonics},\ }\href {https://doi.org/10.1364/OME.483361} {\bibfield  {journal} {\bibinfo  {journal} {Opt. Mater. Express}\ }\textbf {\bibinfo {volume} {13}},\ \bibinfo {pages} {870} (\bibinfo {year} {2023})}\BibitemShut {NoStop}%
\bibitem [{\citenamefont {Liang}\ \emph {et~al.}(2011)\citenamefont {Liang}, \citenamefont {Peng}, \citenamefont {Sakai}, \citenamefont {Iwahashi},\ and\ \citenamefont {Noda}}]{Liang2011}%
  \BibitemOpen
  \bibfield  {author} {\bibinfo {author} {\bibfnamefont {Y.}~\bibnamefont {Liang}}, \bibinfo {author} {\bibfnamefont {C.}~\bibnamefont {Peng}}, \bibinfo {author} {\bibfnamefont {K.}~\bibnamefont {Sakai}}, \bibinfo {author} {\bibfnamefont {S.}~\bibnamefont {Iwahashi}},\ and\ \bibinfo {author} {\bibfnamefont {S.}~\bibnamefont {Noda}},\ }\bibfield  {title} {\bibinfo {title} {Three-dimensional coupled-wave model for square-lattice photonic crystal lasers with transverse electric polarization: A general approach},\ }\bibfield  {journal} {\bibinfo  {journal} {Physical Review B}\ }\textbf {\bibinfo {volume} {84}},\ \href {https://doi.org/10.1103/physrevb.84.195119} {10.1103/physrevb.84.195119} (\bibinfo {year} {2011})\BibitemShut {NoStop}%
\bibitem [{\citenamefont {Liang}\ \emph {et~al.}(2012)\citenamefont {Liang}, \citenamefont {Peng}, \citenamefont {Sakai}, \citenamefont {Iwahashi},\ and\ \citenamefont {Noda}}]{Liang_2012}%
  \BibitemOpen
  \bibfield  {author} {\bibinfo {author} {\bibfnamefont {Y.}~\bibnamefont {Liang}}, \bibinfo {author} {\bibfnamefont {C.}~\bibnamefont {Peng}}, \bibinfo {author} {\bibfnamefont {K.}~\bibnamefont {Sakai}}, \bibinfo {author} {\bibfnamefont {S.}~\bibnamefont {Iwahashi}},\ and\ \bibinfo {author} {\bibfnamefont {S.}~\bibnamefont {Noda}},\ }\bibfield  {title} {\bibinfo {title} {Three-dimensional coupled-wave analysis for square-lattice photonic crystal surface emitting lasers with transverse-electric polarization: finite-size effects},\ }\href {https://doi.org/10.1364/oe.20.015945} {\bibfield  {journal} {\bibinfo  {journal} {Optics Express}\ }\textbf {\bibinfo {volume} {20}},\ \bibinfo {pages} {15945} (\bibinfo {year} {2012})}\BibitemShut {NoStop}%
\bibitem [{\citenamefont {Peng}\ \emph {et~al.}(2012)\citenamefont {Peng}, \citenamefont {Liang}, \citenamefont {Sakai}, \citenamefont {Iwahashi},\ and\ \citenamefont {Noda}}]{Peng2012}%
  \BibitemOpen
  \bibfield  {author} {\bibinfo {author} {\bibfnamefont {C.}~\bibnamefont {Peng}}, \bibinfo {author} {\bibfnamefont {Y.}~\bibnamefont {Liang}}, \bibinfo {author} {\bibfnamefont {K.}~\bibnamefont {Sakai}}, \bibinfo {author} {\bibfnamefont {S.}~\bibnamefont {Iwahashi}},\ and\ \bibinfo {author} {\bibfnamefont {S.}~\bibnamefont {Noda}},\ }\bibfield  {title} {\bibinfo {title} {Three-dimensional coupled-wave theory analysis of a centered-rectangular lattice photonic crystal laser with a transverse-electric-like mode},\ }\bibfield  {journal} {\bibinfo  {journal} {Physical Review B}\ }\textbf {\bibinfo {volume} {86}},\ \href {https://doi.org/10.1103/physrevb.86.035108} {10.1103/physrevb.86.035108} (\bibinfo {year} {2012})\BibitemShut {NoStop}%
\bibitem [{\citenamefont {Liang}\ \emph {et~al.}(2013)\citenamefont {Liang}, \citenamefont {Peng}, \citenamefont {Ishizaki}, \citenamefont {Iwahashi}, \citenamefont {Sakai}, \citenamefont {Tanaka}, \citenamefont {Kitamura},\ and\ \citenamefont {Noda}}]{Liang_2013}%
  \BibitemOpen
  \bibfield  {author} {\bibinfo {author} {\bibfnamefont {Y.}~\bibnamefont {Liang}}, \bibinfo {author} {\bibfnamefont {C.}~\bibnamefont {Peng}}, \bibinfo {author} {\bibfnamefont {K.}~\bibnamefont {Ishizaki}}, \bibinfo {author} {\bibfnamefont {S.}~\bibnamefont {Iwahashi}}, \bibinfo {author} {\bibfnamefont {K.}~\bibnamefont {Sakai}}, \bibinfo {author} {\bibfnamefont {Y.}~\bibnamefont {Tanaka}}, \bibinfo {author} {\bibfnamefont {K.}~\bibnamefont {Kitamura}},\ and\ \bibinfo {author} {\bibfnamefont {S.}~\bibnamefont {Noda}},\ }\bibfield  {title} {\bibinfo {title} {Three-dimensional coupled-wave analysis for triangular-lattice photonic-crystal surface-emitting lasers with transverse-electric polarization},\ }\href {https://doi.org/10.1364/oe.21.000565} {\bibfield  {journal} {\bibinfo  {journal} {Optics Express}\ }\textbf {\bibinfo {volume} {21}},\ \bibinfo {pages} {565} (\bibinfo {year} {2013})}\BibitemShut {NoStop}%
\bibitem [{\citenamefont {Yang}\ \emph {et~al.}(2014)\citenamefont {Yang}, \citenamefont {Peng}, \citenamefont {Liang}, \citenamefont {Li},\ and\ \citenamefont {Noda}}]{Yang_2014}%
  \BibitemOpen
  \bibfield  {author} {\bibinfo {author} {\bibfnamefont {Y.}~\bibnamefont {Yang}}, \bibinfo {author} {\bibfnamefont {C.}~\bibnamefont {Peng}}, \bibinfo {author} {\bibfnamefont {Y.}~\bibnamefont {Liang}}, \bibinfo {author} {\bibfnamefont {Z.}~\bibnamefont {Li}},\ and\ \bibinfo {author} {\bibfnamefont {S.}~\bibnamefont {Noda}},\ }\bibfield  {title} {\bibinfo {title} {Analytical perspective for bound states in the continuum in photonic crystal slabs},\ }\bibfield  {journal} {\bibinfo  {journal} {Physical Review Letters}\ }\textbf {\bibinfo {volume} {113}},\ \href {https://doi.org/10.1103/physrevlett.113.037401} {10.1103/physrevlett.113.037401} (\bibinfo {year} {2014})\BibitemShut {NoStop}%
\bibitem [{\citenamefont {Hsu}\ \emph {et~al.}(2013)\citenamefont {Hsu}, \citenamefont {Zhen}, \citenamefont {Lee}, \citenamefont {Chua}, \citenamefont {Johnson}, \citenamefont {Joannopoulos},\ and\ \citenamefont {Soljačić}}]{Hsu_2013}%
  \BibitemOpen
  \bibfield  {author} {\bibinfo {author} {\bibfnamefont {C.~W.}\ \bibnamefont {Hsu}}, \bibinfo {author} {\bibfnamefont {B.}~\bibnamefont {Zhen}}, \bibinfo {author} {\bibfnamefont {J.}~\bibnamefont {Lee}}, \bibinfo {author} {\bibfnamefont {S.-L.}\ \bibnamefont {Chua}}, \bibinfo {author} {\bibfnamefont {S.~G.}\ \bibnamefont {Johnson}}, \bibinfo {author} {\bibfnamefont {J.~D.}\ \bibnamefont {Joannopoulos}},\ and\ \bibinfo {author} {\bibfnamefont {M.}~\bibnamefont {Soljačić}},\ }\bibfield  {title} {\bibinfo {title} {Observation of trapped light within the radiation continuum},\ }\href {https://doi.org/10.1038/nature12289} {\bibfield  {journal} {\bibinfo  {journal} {Nature}\ }\textbf {\bibinfo {volume} {499}},\ \bibinfo {pages} {188–191} (\bibinfo {year} {2013})}\BibitemShut {NoStop}%
\bibitem [{\citenamefont {Andreani}\ and\ \citenamefont {Gerace}(2006)}]{Andreani_2006}%
  \BibitemOpen
  \bibfield  {author} {\bibinfo {author} {\bibfnamefont {L.~C.}\ \bibnamefont {Andreani}}\ and\ \bibinfo {author} {\bibfnamefont {D.}~\bibnamefont {Gerace}},\ }\bibfield  {title} {\bibinfo {title} {Photonic-crystal slabs with a triangular lattice of triangular holes investigated using a guided-mode expansion method},\ }\bibfield  {journal} {\bibinfo  {journal} {Physical Review B}\ }\textbf {\bibinfo {volume} {73}},\ \href {https://doi.org/10.1103/physrevb.73.235114} {10.1103/physrevb.73.235114} (\bibinfo {year} {2006})\BibitemShut {NoStop}%
\bibitem [{\citenamefont {Zanotti}\ \emph {et~al.}(2024)\citenamefont {Zanotti}, \citenamefont {Minkov}, \citenamefont {Nigro}, \citenamefont {Gerace}, \citenamefont {Fan},\ and\ \citenamefont {Andreani}}]{Zanotti2024}%
  \BibitemOpen
  \bibfield  {author} {\bibinfo {author} {\bibfnamefont {S.}~\bibnamefont {Zanotti}}, \bibinfo {author} {\bibfnamefont {M.}~\bibnamefont {Minkov}}, \bibinfo {author} {\bibfnamefont {D.}~\bibnamefont {Nigro}}, \bibinfo {author} {\bibfnamefont {D.}~\bibnamefont {Gerace}}, \bibinfo {author} {\bibfnamefont {S.}~\bibnamefont {Fan}},\ and\ \bibinfo {author} {\bibfnamefont {L.~C.}\ \bibnamefont {Andreani}},\ }\bibfield  {title} {\bibinfo {title} {Legume: A free implementation of the guided-mode expansion method for photonic crystal slabs},\ }\href {https://doi.org/https://doi.org/10.1016/j.cpc.2024.109286} {\bibfield  {journal} {\bibinfo  {journal} {Computer Physics Communications}\ }\textbf {\bibinfo {volume} {304}},\ \bibinfo {pages} {109286} (\bibinfo {year} {2024})}\BibitemShut {NoStop}%
\bibitem [{\citenamefont {Leung}\ \emph {et~al.}(1994)\citenamefont {Leung}, \citenamefont {Liu},\ and\ \citenamefont {Young}}]{Leung1994}%
  \BibitemOpen
  \bibfield  {author} {\bibinfo {author} {\bibfnamefont {P.~T.}\ \bibnamefont {Leung}}, \bibinfo {author} {\bibfnamefont {S.~Y.}\ \bibnamefont {Liu}},\ and\ \bibinfo {author} {\bibfnamefont {K.}~\bibnamefont {Young}},\ }\bibfield  {title} {\bibinfo {title} {Completeness and orthogonality of quasinormal modes in leaky optical cavities},\ }\href {https://doi.org/10.1103/physreva.49.3057} {\bibfield  {journal} {\bibinfo  {journal} {Physical Review A}\ }\textbf {\bibinfo {volume} {49}},\ \bibinfo {pages} {3057–3067} (\bibinfo {year} {1994})}\BibitemShut {NoStop}%
\bibitem [{\citenamefont {Lu}\ \emph {et~al.}(2020)\citenamefont {Lu}, \citenamefont {Le-Van}, \citenamefont {Ferrier}, \citenamefont {Drouard}, \citenamefont {Seassal},\ and\ \citenamefont {Nguyen}}]{Lu2020}%
  \BibitemOpen
  \bibfield  {author} {\bibinfo {author} {\bibfnamefont {L.}~\bibnamefont {Lu}}, \bibinfo {author} {\bibfnamefont {Q.}~\bibnamefont {Le-Van}}, \bibinfo {author} {\bibfnamefont {L.}~\bibnamefont {Ferrier}}, \bibinfo {author} {\bibfnamefont {E.}~\bibnamefont {Drouard}}, \bibinfo {author} {\bibfnamefont {C.}~\bibnamefont {Seassal}},\ and\ \bibinfo {author} {\bibfnamefont {H.~S.}\ \bibnamefont {Nguyen}},\ }\bibfield  {title} {\bibinfo {title} {Engineering a light–matter strong coupling regime in perovskite-based plasmonic metasurface: quasi-bound state in the continuum and exceptional points},\ }\href {https://doi.org/10.1364/prj.404743} {\bibfield  {journal} {\bibinfo  {journal} {Photonics Research}\ }\textbf {\bibinfo {volume} {8}},\ \bibinfo {pages} {A91} (\bibinfo {year} {2020})}\BibitemShut {NoStop}%
\bibitem [{\citenamefont {Do}\ \emph {et~al.}(2025)\citenamefont {Do}, \citenamefont {Yuan}, \citenamefont {Durmusoglu}, \citenamefont {Shamkhi}, \citenamefont {Valuckas}, \citenamefont {Zhao}, \citenamefont {Kuznetsov}, \citenamefont {Demir}, \citenamefont {Dang}, \citenamefont {Nguyen},\ and\ \citenamefont {Ha}}]{Do2025}%
  \BibitemOpen
  \bibfield  {author} {\bibinfo {author} {\bibfnamefont {T.~T.~H.}\ \bibnamefont {Do}}, \bibinfo {author} {\bibfnamefont {Z.}~\bibnamefont {Yuan}}, \bibinfo {author} {\bibfnamefont {E.~G.}\ \bibnamefont {Durmusoglu}}, \bibinfo {author} {\bibfnamefont {H.~K.}\ \bibnamefont {Shamkhi}}, \bibinfo {author} {\bibfnamefont {V.}~\bibnamefont {Valuckas}}, \bibinfo {author} {\bibfnamefont {C.}~\bibnamefont {Zhao}}, \bibinfo {author} {\bibfnamefont {A.~I.}\ \bibnamefont {Kuznetsov}}, \bibinfo {author} {\bibfnamefont {H.~V.}\ \bibnamefont {Demir}}, \bibinfo {author} {\bibfnamefont {C.}~\bibnamefont {Dang}}, \bibinfo {author} {\bibfnamefont {H.~S.}\ \bibnamefont {Nguyen}},\ and\ \bibinfo {author} {\bibfnamefont {S.~T.}\ \bibnamefont {Ha}},\ }\bibfield  {title} {\bibinfo {title} {Room-temperature lasing at flatband bound states in the continuum},\ }\href {https://doi.org/10.1021/acsnano.5c01972} {\bibfield  {journal} {\bibinfo  {journal} {ACS Nano}\ }\textbf {\bibinfo {volume} {19}},\ \bibinfo {pages} {19287–19296}
  (\bibinfo {year} {2025})}\BibitemShut {NoStop}%
\bibitem [{\citenamefont {Blanchard}\ \emph {et~al.}(2014)\citenamefont {Blanchard}, \citenamefont {Viktorovitch},\ and\ \citenamefont {Letartre}}]{Blanchard2014}%
  \BibitemOpen
  \bibfield  {author} {\bibinfo {author} {\bibfnamefont {C.}~\bibnamefont {Blanchard}}, \bibinfo {author} {\bibfnamefont {P.}~\bibnamefont {Viktorovitch}},\ and\ \bibinfo {author} {\bibfnamefont {X.}~\bibnamefont {Letartre}},\ }\bibfield  {title} {\bibinfo {title} {Perturbation approach for the control of the quality factor in photonic crystal membranes: Application to selective absorbers},\ }\bibfield  {journal} {\bibinfo  {journal} {Physical Review A}\ }\textbf {\bibinfo {volume} {90}},\ \href {https://doi.org/10.1103/physreva.90.033824} {10.1103/physreva.90.033824} (\bibinfo {year} {2014})\BibitemShut {NoStop}%
\bibitem [{\citenamefont {Ovcharenko}\ \emph {et~al.}(2020)\citenamefont {Ovcharenko}, \citenamefont {Blanchard}, \citenamefont {Hugonin},\ and\ \citenamefont {Sauvan}}]{Ovcharenko2020}%
  \BibitemOpen
  \bibfield  {author} {\bibinfo {author} {\bibfnamefont {A.~I.}\ \bibnamefont {Ovcharenko}}, \bibinfo {author} {\bibfnamefont {C.}~\bibnamefont {Blanchard}}, \bibinfo {author} {\bibfnamefont {J.-P.}\ \bibnamefont {Hugonin}},\ and\ \bibinfo {author} {\bibfnamefont {C.}~\bibnamefont {Sauvan}},\ }\bibfield  {title} {\bibinfo {title} {Bound states in the continuum in symmetric and asymmetric photonic crystal slabs},\ }\bibfield  {journal} {\bibinfo  {journal} {Physical Review B}\ }\textbf {\bibinfo {volume} {101}},\ \href {https://doi.org/10.1103/physrevb.101.155303} {10.1103/physrevb.101.155303} (\bibinfo {year} {2020})\BibitemShut {NoStop}%
\bibitem [{\citenamefont {Contractor}\ \emph {et~al.}(2020)\citenamefont {Contractor}, \citenamefont {Noh}, \citenamefont {Le-Van},\ and\ \citenamefont {Kant\'{e}}}]{Contractor20}%
  \BibitemOpen
  \bibfield  {author} {\bibinfo {author} {\bibfnamefont {R.}~\bibnamefont {Contractor}}, \bibinfo {author} {\bibfnamefont {W.}~\bibnamefont {Noh}}, \bibinfo {author} {\bibfnamefont {Q.}~\bibnamefont {Le-Van}},\ and\ \bibinfo {author} {\bibfnamefont {B.}~\bibnamefont {Kant\'{e}}},\ }\bibfield  {title} {\bibinfo {title} {Doping-induced plateau of strong electromagnetic confinement in the momentum space},\ }\href {https://doi.org/10.1364/OL.395625} {\bibfield  {journal} {\bibinfo  {journal} {Opt. Lett.}\ }\textbf {\bibinfo {volume} {45}},\ \bibinfo {pages} {3653} (\bibinfo {year} {2020})}\BibitemShut {NoStop}%
\bibitem [{\citenamefont {Minkov}\ \emph {et~al.}(2018)\citenamefont {Minkov}, \citenamefont {Williamson}, \citenamefont {Xiao},\ and\ \citenamefont {Fan}}]{Minkov2018}%
  \BibitemOpen
  \bibfield  {author} {\bibinfo {author} {\bibfnamefont {M.}~\bibnamefont {Minkov}}, \bibinfo {author} {\bibfnamefont {I.~A.~D.}\ \bibnamefont {Williamson}}, \bibinfo {author} {\bibfnamefont {M.}~\bibnamefont {Xiao}},\ and\ \bibinfo {author} {\bibfnamefont {S.}~\bibnamefont {Fan}},\ }\bibfield  {title} {\bibinfo {title} {Zero-index bound states in the continuum},\ }\href {https://doi.org/10.1103/PhysRevLett.121.263901} {\bibfield  {journal} {\bibinfo  {journal} {Phys. Rev. Lett.}\ }\textbf {\bibinfo {volume} {121}},\ \bibinfo {pages} {263901} (\bibinfo {year} {2018})}\BibitemShut {NoStop}%
\bibitem [{\citenamefont {Dong}\ \emph {et~al.}(2021)\citenamefont {Dong}, \citenamefont {Liang}, \citenamefont {Camayd-Muñoz}, \citenamefont {Liu}, \citenamefont {Tang}, \citenamefont {Kita}, \citenamefont {Chen}, \citenamefont {Wu}, \citenamefont {Chu}, \citenamefont {Mazur},\ and\ \citenamefont {Li}}]{Dong2021}%
  \BibitemOpen
  \bibfield  {author} {\bibinfo {author} {\bibfnamefont {T.}~\bibnamefont {Dong}}, \bibinfo {author} {\bibfnamefont {J.}~\bibnamefont {Liang}}, \bibinfo {author} {\bibfnamefont {S.}~\bibnamefont {Camayd-Muñoz}}, \bibinfo {author} {\bibfnamefont {Y.}~\bibnamefont {Liu}}, \bibinfo {author} {\bibfnamefont {H.}~\bibnamefont {Tang}}, \bibinfo {author} {\bibfnamefont {S.}~\bibnamefont {Kita}}, \bibinfo {author} {\bibfnamefont {P.}~\bibnamefont {Chen}}, \bibinfo {author} {\bibfnamefont {X.}~\bibnamefont {Wu}}, \bibinfo {author} {\bibfnamefont {W.}~\bibnamefont {Chu}}, \bibinfo {author} {\bibfnamefont {E.}~\bibnamefont {Mazur}},\ and\ \bibinfo {author} {\bibfnamefont {Y.}~\bibnamefont {Li}},\ }\bibfield  {title} {\bibinfo {title} {Ultra-low-loss on-chip zero-index materials},\ }\bibfield  {journal} {\bibinfo  {journal} {Light: Science \& Applications}\ }\textbf {\bibinfo {volume} {10}},\ \href {https://doi.org/10.1038/s41377-020-00436-y} {10.1038/s41377-020-00436-y} (\bibinfo {year} {2021})\BibitemShut {NoStop}%
\bibitem [{\citenamefont {Contractor}\ \emph {et~al.}(2022)\citenamefont {Contractor}, \citenamefont {Noh}, \citenamefont {Redjem}, \citenamefont {Qarony}, \citenamefont {Martin}, \citenamefont {Dhuey}, \citenamefont {Schwartzberg},\ and\ \citenamefont {Kanté}}]{Contractor2022}%
  \BibitemOpen
  \bibfield  {author} {\bibinfo {author} {\bibfnamefont {R.}~\bibnamefont {Contractor}}, \bibinfo {author} {\bibfnamefont {W.}~\bibnamefont {Noh}}, \bibinfo {author} {\bibfnamefont {W.}~\bibnamefont {Redjem}}, \bibinfo {author} {\bibfnamefont {W.}~\bibnamefont {Qarony}}, \bibinfo {author} {\bibfnamefont {E.}~\bibnamefont {Martin}}, \bibinfo {author} {\bibfnamefont {S.}~\bibnamefont {Dhuey}}, \bibinfo {author} {\bibfnamefont {A.}~\bibnamefont {Schwartzberg}},\ and\ \bibinfo {author} {\bibfnamefont {B.}~\bibnamefont {Kanté}},\ }\bibfield  {title} {\bibinfo {title} {Scalable single-mode surface-emitting laser via open-dirac singularities},\ }\href {https://doi.org/10.1038/s41586-022-05021-4} {\bibfield  {journal} {\bibinfo  {journal} {Nature}\ }\textbf {\bibinfo {volume} {608}},\ \bibinfo {pages} {692–698} (\bibinfo {year} {2022})}\BibitemShut {NoStop}%
\bibitem [{\citenamefont {Kanté}(2024)}]{Kant2024}%
  \BibitemOpen
  \bibfield  {author} {\bibinfo {author} {\bibfnamefont {B.}~\bibnamefont {Kanté}},\ }\bibfield  {title} {\bibinfo {title} {Berksel: A scale-invariant laser beyond the schawlow-townes two-mirror strategy},\ }\bibfield  {journal} {\bibinfo  {journal} {Nature Communications}\ }\textbf {\bibinfo {volume} {15}},\ \href {https://doi.org/10.1038/s41467-024-46338-0} {10.1038/s41467-024-46338-0} (\bibinfo {year} {2024})\BibitemShut {NoStop}%
\bibitem [{\citenamefont {Frau}\ \emph {et~al.}(2025)\citenamefont {Frau}, \citenamefont {Zanotti}, \citenamefont {Ferrier}, \citenamefont {Gerace},\ and\ \citenamefont {Nguyen}}]{frau2025}%
  \BibitemOpen
  \bibfield  {author} {\bibinfo {author} {\bibfnamefont {L.}~\bibnamefont {Frau}}, \bibinfo {author} {\bibfnamefont {S.}~\bibnamefont {Zanotti}}, \bibinfo {author} {\bibfnamefont {L.}~\bibnamefont {Ferrier}}, \bibinfo {author} {\bibfnamefont {D.}~\bibnamefont {Gerace}},\ and\ \bibinfo {author} {\bibfnamefont {H.~S.}\ \bibnamefont {Nguyen}},\ }\href {https://arxiv.org/abs/2506.13698} {\bibinfo {title} {Shaping bulk fermi arcs in the momentum space of photonic crystal slabs}} (\bibinfo {year} {2025}),\ \Eprint {https://arxiv.org/abs/2506.13698} {arXiv:2506.13698 [physics.optics]} \BibitemShut {NoStop}%
\bibitem [{\citenamefont {Dong}\ \emph {et~al.}(2016)\citenamefont {Dong}, \citenamefont {Chen}, \citenamefont {Zhu}, \citenamefont {Wang},\ and\ \citenamefont {Zhang}}]{Dong2016}%
  \BibitemOpen
  \bibfield  {author} {\bibinfo {author} {\bibfnamefont {J.-W.}\ \bibnamefont {Dong}}, \bibinfo {author} {\bibfnamefont {X.-D.}\ \bibnamefont {Chen}}, \bibinfo {author} {\bibfnamefont {H.}~\bibnamefont {Zhu}}, \bibinfo {author} {\bibfnamefont {Y.}~\bibnamefont {Wang}},\ and\ \bibinfo {author} {\bibfnamefont {X.}~\bibnamefont {Zhang}},\ }\bibfield  {title} {\bibinfo {title} {Valley photonic crystals for control of spin and topology},\ }\href {https://doi.org/10.1038/nmat4807} {\bibfield  {journal} {\bibinfo  {journal} {Nature Materials}\ }\textbf {\bibinfo {volume} {16}},\ \bibinfo {pages} {298–302} (\bibinfo {year} {2016})}\BibitemShut {NoStop}%
\bibitem [{\citenamefont {Guo}\ \emph {et~al.}(2020)\citenamefont {Guo}, \citenamefont {Xiao}, \citenamefont {Guo}, \citenamefont {Yuan},\ and\ \citenamefont {Fan}}]{Guo2020}%
  \BibitemOpen
  \bibfield  {author} {\bibinfo {author} {\bibfnamefont {C.}~\bibnamefont {Guo}}, \bibinfo {author} {\bibfnamefont {M.}~\bibnamefont {Xiao}}, \bibinfo {author} {\bibfnamefont {Y.}~\bibnamefont {Guo}}, \bibinfo {author} {\bibfnamefont {L.}~\bibnamefont {Yuan}},\ and\ \bibinfo {author} {\bibfnamefont {S.}~\bibnamefont {Fan}},\ }\bibfield  {title} {\bibinfo {title} {Meron spin textures in momentum space},\ }\bibfield  {journal} {\bibinfo  {journal} {Physical Review Letters}\ }\textbf {\bibinfo {volume} {124}},\ \href {https://doi.org/10.1103/physrevlett.124.106103} {10.1103/physrevlett.124.106103} (\bibinfo {year} {2020})\BibitemShut {NoStop}%
\bibitem [{\citenamefont {Xue}\ \emph {et~al.}(2021)\citenamefont {Xue}, \citenamefont {Yang},\ and\ \citenamefont {Zhang}}]{Xue2021}%
  \BibitemOpen
  \bibfield  {author} {\bibinfo {author} {\bibfnamefont {H.}~\bibnamefont {Xue}}, \bibinfo {author} {\bibfnamefont {Y.}~\bibnamefont {Yang}},\ and\ \bibinfo {author} {\bibfnamefont {B.}~\bibnamefont {Zhang}},\ }\bibfield  {title} {\bibinfo {title} {Topological valley photonics: Physics and device applications},\ }\bibfield  {journal} {\bibinfo  {journal} {Advanced Photonics Research}\ }\textbf {\bibinfo {volume} {2}},\ \href {https://doi.org/10.1002/adpr.202100013} {10.1002/adpr.202100013} (\bibinfo {year} {2021})\BibitemShut {NoStop}%
\bibitem [{\citenamefont {Nguyen}\ \emph {et~al.}(2024)\citenamefont {Nguyen}, \citenamefont {Wu}, \citenamefont {Ha~Do}, \citenamefont {Dieu~Nguyen}, \citenamefont {Sergeev}, \citenamefont {Zhu}, \citenamefont {Valuckas}, \citenamefont {Pham}, \citenamefont {Son~Bui}, \citenamefont {Hoang}, \citenamefont {Tung}, \citenamefont {Khuyen}, \citenamefont {Nguyen}, \citenamefont {Nguyen}, \citenamefont {Lam}, \citenamefont {Rogach}, \citenamefont {Ha},\ and\ \citenamefont {Le-Van}}]{Nguyen2024}%
  \BibitemOpen
  \bibfield  {author} {\bibinfo {author} {\bibfnamefont {V.~A.}\ \bibnamefont {Nguyen}}, \bibinfo {author} {\bibfnamefont {Y.}~\bibnamefont {Wu}}, \bibinfo {author} {\bibfnamefont {T.~T.}\ \bibnamefont {Ha~Do}}, \bibinfo {author} {\bibfnamefont {L.~T.}\ \bibnamefont {Dieu~Nguyen}}, \bibinfo {author} {\bibfnamefont {A.~A.}\ \bibnamefont {Sergeev}}, \bibinfo {author} {\bibfnamefont {D.}~\bibnamefont {Zhu}}, \bibinfo {author} {\bibfnamefont {V.}~\bibnamefont {Valuckas}}, \bibinfo {author} {\bibfnamefont {D.}~\bibnamefont {Pham}}, \bibinfo {author} {\bibfnamefont {H.~X.}\ \bibnamefont {Son~Bui}}, \bibinfo {author} {\bibfnamefont {D.~M.}\ \bibnamefont {Hoang}}, \bibinfo {author} {\bibfnamefont {B.~S.}\ \bibnamefont {Tung}}, \bibinfo {author} {\bibfnamefont {B.~X.}\ \bibnamefont {Khuyen}}, \bibinfo {author} {\bibfnamefont {T.~B.}\ \bibnamefont {Nguyen}}, \bibinfo {author} {\bibfnamefont {H.~S.}\ \bibnamefont {Nguyen}}, \bibinfo {author} {\bibfnamefont {V.~D.}\ \bibnamefont {Lam}}, \bibinfo {author} {\bibfnamefont
  {A.~L.}\ \bibnamefont {Rogach}}, \bibinfo {author} {\bibfnamefont {S.~T.}\ \bibnamefont {Ha}},\ and\ \bibinfo {author} {\bibfnamefont {Q.}~\bibnamefont {Le-Van}},\ }\bibfield  {title} {\bibinfo {title} {Micrometer-resolution fluorescence and lifetime mappings of cspbbr3 nanocrystal films coupled with a tio2 grating},\ }\href {https://doi.org/10.1021/acs.jpclett.4c02546} {\bibfield  {journal} {\bibinfo  {journal} {The Journal of Physical Chemistry Letters}\ }\textbf {\bibinfo {volume} {15}},\ \bibinfo {pages} {11291} (\bibinfo {year} {2024})}\BibitemShut {NoStop}%
\bibitem [{\citenamefont {Yin}\ \emph {et~al.}(2025)\citenamefont {Yin}, \citenamefont {Chen}, \citenamefont {Zhang}, \citenamefont {Zhang}, \citenamefont {Noda},\ and\ \citenamefont {Peng}}]{Yin2025}%
  \BibitemOpen
  \bibfield  {author} {\bibinfo {author} {\bibfnamefont {X.}~\bibnamefont {Yin}}, \bibinfo {author} {\bibfnamefont {Y.}~\bibnamefont {Chen}}, \bibinfo {author} {\bibfnamefont {X.}~\bibnamefont {Zhang}}, \bibinfo {author} {\bibfnamefont {Z.}~\bibnamefont {Zhang}}, \bibinfo {author} {\bibfnamefont {S.}~\bibnamefont {Noda}},\ and\ \bibinfo {author} {\bibfnamefont {C.}~\bibnamefont {Peng}},\ }\bibfield  {title} {\bibinfo {title} {Observation of berry curvature in non-hermitian system from far-field radiation},\ }\bibfield  {journal} {\bibinfo  {journal} {Nature Communications}\ }\textbf {\bibinfo {volume} {16}},\ \href {https://doi.org/10.1038/s41467-025-58050-8} {10.1038/s41467-025-58050-8} (\bibinfo {year} {2025})\BibitemShut {NoStop}%
\bibitem [{\citenamefont {Yuan}\ \emph {et~al.}(2025)\citenamefont {Yuan}, \citenamefont {Malgrey}, \citenamefont {Sigurðsson}, \citenamefont {Nguyen},\ and\ \citenamefont {Salerno}}]{yuan2025}%
  \BibitemOpen
  \bibfield  {author} {\bibinfo {author} {\bibfnamefont {X.}~\bibnamefont {Yuan}}, \bibinfo {author} {\bibfnamefont {L.}~\bibnamefont {Malgrey}}, \bibinfo {author} {\bibfnamefont {H.}~\bibnamefont {Sigurðsson}}, \bibinfo {author} {\bibfnamefont {H.~S.}\ \bibnamefont {Nguyen}},\ and\ \bibinfo {author} {\bibfnamefont {G.}~\bibnamefont {Salerno}},\ }\href {https://arxiv.org/abs/2504.05188} {\bibinfo {title} {Breakdown of bulk-radiation correspondence in radiative photonic lattices}} (\bibinfo {year} {2025}),\ \Eprint {https://arxiv.org/abs/2504.05188} {arXiv:2504.05188 [physics.optics]} \BibitemShut {NoStop}%
\bibitem [{\citenamefont {Nguyen}\ \emph {et~al.}(2022)\citenamefont {Nguyen}, \citenamefont {Letartre}, \citenamefont {Drouard}, \citenamefont {Viktorovitch}, \citenamefont {Nguyen},\ and\ \citenamefont {Nguyen}}]{Nguyen2022}%
  \BibitemOpen
  \bibfield  {author} {\bibinfo {author} {\bibfnamefont {D.~X.}\ \bibnamefont {Nguyen}}, \bibinfo {author} {\bibfnamefont {X.}~\bibnamefont {Letartre}}, \bibinfo {author} {\bibfnamefont {E.}~\bibnamefont {Drouard}}, \bibinfo {author} {\bibfnamefont {P.}~\bibnamefont {Viktorovitch}}, \bibinfo {author} {\bibfnamefont {H.~C.}\ \bibnamefont {Nguyen}},\ and\ \bibinfo {author} {\bibfnamefont {H.~S.}\ \bibnamefont {Nguyen}},\ }\bibfield  {title} {\bibinfo {title} {Magic configurations in moiré superlattice of bilayer photonic crystals: Almost-perfect flatbands and unconventional localization},\ }\bibfield  {journal} {\bibinfo  {journal} {Physical Review Research}\ }\textbf {\bibinfo {volume} {4}},\ \href {https://doi.org/10.1103/physrevresearch.4.l032031} {10.1103/physrevresearch.4.l032031} (\bibinfo {year} {2022})\BibitemShut {NoStop}%
\bibitem [{\citenamefont {Saadi}\ \emph {et~al.}(2025)\citenamefont {Saadi}, \citenamefont {Cueff}, \citenamefont {Ferrier}, \citenamefont {Benamrouche}, \citenamefont {Gayrard}, \citenamefont {Drouard}, \citenamefont {Letartre}, \citenamefont {Nguyen},\ and\ \citenamefont {Callard}}]{Saadi2025}%
  \BibitemOpen
  \bibfield  {author} {\bibinfo {author} {\bibfnamefont {C.}~\bibnamefont {Saadi}}, \bibinfo {author} {\bibfnamefont {S.}~\bibnamefont {Cueff}}, \bibinfo {author} {\bibfnamefont {L.}~\bibnamefont {Ferrier}}, \bibinfo {author} {\bibfnamefont {A.}~\bibnamefont {Benamrouche}}, \bibinfo {author} {\bibfnamefont {M.}~\bibnamefont {Gayrard}}, \bibinfo {author} {\bibfnamefont {E.}~\bibnamefont {Drouard}}, \bibinfo {author} {\bibfnamefont {X.}~\bibnamefont {Letartre}}, \bibinfo {author} {\bibfnamefont {H.~S.}\ \bibnamefont {Nguyen}},\ and\ \bibinfo {author} {\bibfnamefont {S.}~\bibnamefont {Callard}},\ }\bibfield  {title} {\bibinfo {title} {Tailoring flatband dispersion in bilayer moiré photonic crystals},\ }\bibfield  {journal} {\bibinfo  {journal} {Laser and Photonics Reviews}\ }\href {https://doi.org/10.1002/lpor.202501038} {10.1002/lpor.202501038} (\bibinfo {year} {2025})\BibitemShut {NoStop}%
\bibitem [{\citenamefont {Zhang}\ \emph {et~al.}(2023)\citenamefont {Zhang}, \citenamefont {Dong}, \citenamefont {Li}, \citenamefont {Meng}, \citenamefont {Li}, \citenamefont {Munagavalasa}, \citenamefont {Grigoropoulos}, \citenamefont {Wu},\ and\ \citenamefont {Yao}}]{Zhang2023}%
  \BibitemOpen
  \bibfield  {author} {\bibinfo {author} {\bibfnamefont {T.}~\bibnamefont {Zhang}}, \bibinfo {author} {\bibfnamefont {K.}~\bibnamefont {Dong}}, \bibinfo {author} {\bibfnamefont {J.}~\bibnamefont {Li}}, \bibinfo {author} {\bibfnamefont {F.}~\bibnamefont {Meng}}, \bibinfo {author} {\bibfnamefont {J.}~\bibnamefont {Li}}, \bibinfo {author} {\bibfnamefont {S.}~\bibnamefont {Munagavalasa}}, \bibinfo {author} {\bibfnamefont {C.~P.}\ \bibnamefont {Grigoropoulos}}, \bibinfo {author} {\bibfnamefont {J.}~\bibnamefont {Wu}},\ and\ \bibinfo {author} {\bibfnamefont {J.}~\bibnamefont {Yao}},\ }\bibfield  {title} {\bibinfo {title} {Twisted moiré photonic crystal enabled optical vortex generation through bound states in the continuum},\ }\bibfield  {journal} {\bibinfo  {journal} {Nature Communications}\ }\textbf {\bibinfo {volume} {14}},\ \href {https://doi.org/10.1038/s41467-023-41068-1} {10.1038/s41467-023-41068-1} (\bibinfo {year} {2023})\BibitemShut {NoStop}%
\bibitem [{\citenamefont {Ni}\ \emph {et~al.}(2024)\citenamefont {Ni}, \citenamefont {Liu}, \citenamefont {Lou}, \citenamefont {Zhang}, \citenamefont {Hu}, \citenamefont {Fan}, \citenamefont {Mazur},\ and\ \citenamefont {Tang}}]{Ni2024}%
  \BibitemOpen
  \bibfield  {author} {\bibinfo {author} {\bibfnamefont {X.}~\bibnamefont {Ni}}, \bibinfo {author} {\bibfnamefont {Y.}~\bibnamefont {Liu}}, \bibinfo {author} {\bibfnamefont {B.}~\bibnamefont {Lou}}, \bibinfo {author} {\bibfnamefont {M.}~\bibnamefont {Zhang}}, \bibinfo {author} {\bibfnamefont {E.~L.}\ \bibnamefont {Hu}}, \bibinfo {author} {\bibfnamefont {S.}~\bibnamefont {Fan}}, \bibinfo {author} {\bibfnamefont {E.}~\bibnamefont {Mazur}},\ and\ \bibinfo {author} {\bibfnamefont {H.}~\bibnamefont {Tang}},\ }\bibfield  {title} {\bibinfo {title} {Three-dimensional reconfigurable optical singularities in bilayer photonic crystals},\ }\bibfield  {journal} {\bibinfo  {journal} {Physical Review Letters}\ }\textbf {\bibinfo {volume} {132}},\ \href {https://doi.org/10.1103/physrevlett.132.073804} {10.1103/physrevlett.132.073804} (\bibinfo {year} {2024})\BibitemShut {NoStop}%
\bibitem [{\citenamefont {Gromyko}\ \emph {et~al.}(2024)\citenamefont {Gromyko}, \citenamefont {An}, \citenamefont {Gorelik}, \citenamefont {Xu}, \citenamefont {Lim}, \citenamefont {Lee}, \citenamefont {Tjiptoharsono}, \citenamefont {Tan}, \citenamefont {Qiu}, \citenamefont {Dong},\ and\ \citenamefont {Wu}}]{Gromyko2024}%
  \BibitemOpen
  \bibfield  {author} {\bibinfo {author} {\bibfnamefont {D.}~\bibnamefont {Gromyko}}, \bibinfo {author} {\bibfnamefont {S.}~\bibnamefont {An}}, \bibinfo {author} {\bibfnamefont {S.}~\bibnamefont {Gorelik}}, \bibinfo {author} {\bibfnamefont {J.}~\bibnamefont {Xu}}, \bibinfo {author} {\bibfnamefont {L.~J.}\ \bibnamefont {Lim}}, \bibinfo {author} {\bibfnamefont {H.~Y.~L.}\ \bibnamefont {Lee}}, \bibinfo {author} {\bibfnamefont {F.}~\bibnamefont {Tjiptoharsono}}, \bibinfo {author} {\bibfnamefont {Z.-K.}\ \bibnamefont {Tan}}, \bibinfo {author} {\bibfnamefont {C.-W.}\ \bibnamefont {Qiu}}, \bibinfo {author} {\bibfnamefont {Z.}~\bibnamefont {Dong}},\ and\ \bibinfo {author} {\bibfnamefont {L.}~\bibnamefont {Wu}},\ }\bibfield  {title} {\bibinfo {title} {Unidirectional chiral emission via twisted bi-layer metasurfaces},\ }\bibfield  {journal} {\bibinfo  {journal} {Nature Communications}\ }\textbf {\bibinfo {volume} {15}},\ \href {https://doi.org/10.1038/s41467-024-54262-6} {10.1038/s41467-024-54262-6} (\bibinfo {year}
  {2024})\BibitemShut {NoStop}%
\bibitem [{\citenamefont {Dennis}(2002)}]{Dennis2002}%
  \BibitemOpen
  \bibfield  {author} {\bibinfo {author} {\bibfnamefont {M.}~\bibnamefont {Dennis}},\ }\bibfield  {title} {\bibinfo {title} {Polarization singularities in paraxial vector fields: morphology and statistics},\ }\href {https://doi.org/https://doi.org/10.1016/S0030-4018(02)02088-6} {\bibfield  {journal} {\bibinfo  {journal} {Optics Communications}\ }\textbf {\bibinfo {volume} {213}},\ \bibinfo {pages} {201} (\bibinfo {year} {2002})}\BibitemShut {NoStop}%
\end{thebibliography}%
\end{document}